\documentclass[a4paper,11pt]{article}
\pdfoutput=1 

\usepackage{jheppub} 

\usepackage[T1]{fontenc} 

\usepackage{amsmath,epsfig}
\usepackage{amssymb,amsfonts}
\usepackage{latexsym}
\usepackage[latin1]{inputenc}
\usepackage{subeqnarray}
\usepackage{xcolor}

\usepackage{graphicx}
\usepackage{longtable}
\usepackage{enumerate}
\usepackage[export]{adjustbox}

\relax
\renewcommand{\theequation}{\arabic{section}.\arabic{equation}}
\def\be{\begin{equation}}
\def\ee{\end{equation}}
\def\bea{\begin{eqnarray}}
\def\eea{\end{eqnarray}}

\newcommand\fverb{\setbox\pippobox=\hbox\bgroup\verb}
\newcommand\fverbdo{\egroup\medskip\noindent%
                        \fbox{\unhbox\pippobox}\ }
\newcommand\fverbit{\egroup\item[\fbox{\unhbox\pippobox}]}

\newcommand{\bear}{\begin{eqnarray}}

\newcommand{\eear}{\end{eqnarray}}

\newcommand{\bsea}{\begin{subeqnarray}}
\newcommand{\esea}{\end{subeqnarray}}
\newbox\pippobox

\def\6{\partial}

\def\a{\alpha}

\def\m{\mu}
\def\n{\nu}

\def\sp{\;\;\;,\;\;\;}

\def\sq
\def\a{\alpha}

\def\hri#1{\href{http://arxiv.org/abs/#1}{arXiv:#1}}
\def\hri#1#2{\href{http://arxiv.org/abs/#1}{arXiv:#1}}
\def\hre#1#2{\href{http://arxiv.org/abs/#1/#2}{#1/#2}}

\newcommand{\e}{\mathrm{e}}

\def\e{\epsilon}

\title{On Holographic Insulators and Supersolids}

\author[a]{Elias Kiritsis}
\author[a,b]{and Jie Ren}

\affiliation[a]{\href{http://hep.physics.uoc.gr}{Crete Center for Theoretical Physics},\\ Department of Physics, University of Crete, 71003 Heraklion, Greece}
\affiliation[b]{Univ Paris Diderot, Sorbonne Paris Cit\'e, \href{http://www.apc.univ-paris7.fr/APC_CS/}{APC},
UMR 7164 CNRS, F-75205 Paris, France}

\emailAdd{kiritsis@physics.uoc.gr}
\emailAdd{jren@physics.uoc.gr}

\preprint{\begin{tabular}{r}
CCTP-2015-07\\
CCQCN-2015-55
\end{tabular}}

\abstract{We obtain holographic realizations for systems that have strong similarities to Mott insulators and supersolids, after examining the ground states of Einstein-Maxwell-scalar systems. The real part of the AC conductivity has a hard gap and a discrete spectrum only. We add momentum dissipation to resolve the $\delta$-function in the conductivity due to  translational invariance. We develop tools to directly calculate the Drude weight for a large class of solutions and to support our claims. Numerical RG flows are also constructed to verify that such saddle points are IR fixed points of asymptotically AdS$_4$  geometries.}

\arxivnumber{1503.03481}

\begin{document}
\maketitle
\flushbottom

\section{Introduction}
\label{intro}

Insulators and superconductors are central concepts in strongly correlated electron materials giving rise to high $T_c$ superconductivity, \cite{zaanen}. The less understood region is the region in which the insulator-metal transition happens, \cite{mit,mit2}, characterized by novel features of the electronic conductivity, \cite{gap}.

The mechanisms that create insulators can be broadly classified as follows:
\begin{enumerate}

\item Band gap insulators, where the conduction band is empty, and there is therefore a gap that prevents current transport.

\item Anderson localization, \cite{anderson},  where strong disorder inhibits conduction.

\item  Mott localization, where strong on-site interactions localize electrons, \cite{mott}. This has been argued  to be at work in insulating anti-ferromagnets, \cite{mott1} as was first pointed out by Mott and van Vleck.

\end{enumerate}

Lately a new further mechanism was argued to lead to insulating behavior:

\begin{enumerate}

\item[4] Momentum-dissipating interactions become relevant (strong) in the IR, and they inhibit conduction, \cite{hd}. The model used as an example was a holographic model with a saddle-point of helical symmetry. The mechanism was generalized to Einstein-Maxwell-Dilaton (EMD) theories in \cite{dgk} and many saddle points corresponding to insulators, bad metals and conventional metals were found.

\end{enumerate}

Holography has provided a new paradigm and a new arena for theoretical models that address physics at strong coupling, but in a semiclassical setup (that is typically present because of a large-$N$ limit involved).\footnote{Useful reviews for condensed matter physics applications can be found in \cite{sachdev}, \cite{rev,rev2} and section 2 of \cite{kkp}.}
It is a natural framework to describe quantum critical systems at zero and finite density. It is convenient to describe out of equilibrium effects as well as conductivity, a major observable in condensed matter. A study of holographic ground states has indicated that they are very diverse in their properties, \cite{Hartnoll:2008kx}, \cite{Horowitz:2009ij,Gubser:2009cg}, \cite{cgkkm,gk1,gk2}, \cite{helical,helical2,helical3,helical4,Lippert:2014jma, Erdmenger:2015qqa}.

Most holographic systems analyzed at finite density are translationally invariant.
Exceptions also exist, using D-brane defects and magnetic vortices, \cite{def,def2,def3,def4}, but such systems have been much more difficult to analyze so far, although exceptions exist, \cite{Fujita:2014mqa}. The standard symmetry argument indicates that the real part of the AC conductivity will have a $\delta(\omega)$ contribution as in a translationally invariant system a constant electric field generates an infinite current. The $\delta$-function can be argued to relate via causality to a ${1\over \omega}$ pole in the imaginary part of the conductivity, via  standard dispersion relations. In the context of holographic correlators this was analyzed recently in \cite{Hartnoll:2008kx}. This $\delta$-function is distinct from the one appearing in superfluid/superconducting phases.

The interaction with momentum dissipation agents has been discussed in rather general terms in \cite{impurities,impurities2,Hartnoll:2012rj}. When the interaction with dissipators is IR irrelevant with respect to RG fixed points, a perturbative IR calculation can determine the scaling of the IR DC conductivity. When the dissipation is IR relevant, it can change the nature of the saddle point, turning the system into an insulator as was first argued in \cite{hd}. This is the mechanism 4 above that leads to insulating behavior.

There have been several lines of research addressing the breaking of translational invariance in holographic saddle points at finite density and its impact on conductivity.
A first line of research introduced a holographic lattice imprinted by boundary conditions on the bulk charge contribution, \cite{lattice1,lattice12,lattice13,lattice2,Ling:2013nxa}. In the regimes accessible to the numerical calculations, the lattice perturbation is irrelevant in the IR and it controls to leading order in the IR the DC conductivity as predicted in \cite{Hartnoll:2012rj} on general principles.

Another line of approach, \cite{massive,massive2,massive3,Blake:2013bqa,Amoretti:2014mma,bp}, was to assume an effective action  treatment for momentum dissipation associated to the breaking of translational invariance and this was provided by a massive gravity action.

More recently a third approach to breaking translational invariance and therefore introducing momentum dissipation was introduced in  \cite{DG2} and \cite{AW}. In \cite{DG2} the phase of UV relevant, charge neutral, complex scalars was used in order to break translational symmetry on the boundary.
The construction of \cite{AW} used bulk fields which are perturbatively without a potential on AdS. In order to relax momentum, the massless scalars  were given a linear dependence on the spatial coordinates of the boundary.

The formula obtained for the DC conductivity in all the studies above is a sum of two contributions:
\be
\sigma_{DC}=\sigma_{DC}^{ccs}+\sigma_{DC}^{diss}
\label{1}\ee
The first term, $\sigma_{DC}^{ccs}$, has been interpreted, \cite{KO}, as a pair creation contribution as it persists at zero charge density. For the RN black hole it is a constant proportional to the inverse of the bulk gauge coupling constant that counts the relative density of charge-carrying degrees of freedom to the neutral ones in the strongly-coupled plasma.
More recently, it was verified in \cite{DG1} that the first term in (\ref{1}) is the electric conductivity in the absence of a heat current. The interpretation as a pair creation term is then natural since charged pairs are created with zero total momentum and therefore not contributing to a net matter flow. It is fair to say however that this pair production interpretation, although appealing is not in agreement with other related expectations from weak coupling, like the fact that we expect it to be exponentially suppressed in the chemical potential as this gaps free fermions. At finite density, this term persists and can be powerlike  in the temperature in scaling IR geometries.  We will henceforth call this term the charge conjugation symmetric (ccs) term following J. Zaanen's suggestion.

The second contribution in (\ref{1}) is due to the effects of dissipating momentum. When translation-breaking operators are irrelevant, the system is expected to be metallic and this term is expected to give the leading contribution to the DC conductivity. Then, a description of momentum relaxation in terms of the memory matrix formalism is appropriate and shows that the conductivity takes a Drude-like form, though no quasi-particle description is assumed \cite{Hartnoll:2012rj}.

This general form of the DC conductivity was seen already in pure metric backgrounds in \cite{KO} and was generalized to dilatonic backgrounds in \cite{cgkkm}. In both cases, as the gauge field action is the DBI action, (\ref{1}) is replaced by
\be
\sigma_{DC}=\sqrt{(\sigma_{DC}^{ccs})^2+(\sigma_{DC}^{diss})^2}
\label{a2}\ee
giving results compatible with (\ref{1}) in the regimes where pair creation or diffusion dominates the conductivity.
In general, we expect a nonlinear formula that reflects the bulk action of the gauge field. In the probe DBI cases the momentum dissipation is due to the fact that charge degrees of freedom are subleading compared to uncharged ones. This means that there is a momentum conserving $\delta$-function but its coefficient is hierarchically suppressed (by ${N_f\over N_c}$).

In \cite{cgkkm} it was observed, based on (\ref{a2}), that for running scalars other than the dilaton and in 2+1 boundary dimensions, the drag DC resistivity, when it dominates,  is proportional to the electronic entropy. This is a general property of strange metals where both the measured electronic entropy and resistivity   are linear in temperature. This  was extended in \cite{DSZ} to more general cases using the massive graviton theory, and most importantly provided a kinetic explanation for the correlation suggesting a more general validity.

The general properties of holographic conductivity were further corroborated recently by a careful study of the holographic current-current correlators, and the associated properties of their poles in the complex plane, \cite{dg}.

In most holographic cases, the translationally invariant systems are good conductors and we will call them ``holographic metals''. There have been also systems that have a discrete  spectrum in the current-current correlator and they are therefore candidates for insulators.
At zero density such systems were described in \cite{witten,ihqcd,ihqcd2,taka}.

More interestingly, at finite density a large class of systems were found in \cite{cgkkm,gk1} that had a discrete spectrum for the current-current correlator (as well as the stress-tensor correlators). One particular case in a different supergravity was independently observed in \cite{McGreevy} and a new one (the two-charge STU solution in $AdS_4$) is presented in this paper. Such holographic saddle points have very peculiar properties. In particular, the spectra of the stress tensor and current-current correlators were discrete. The current-current correlator had a $\delta$-function at zero frequency because of translational invariance. Finally, the IR endpoint of the bulk geometry does not affect IR physics but UV physics. Moreover, although the analysis of \cite{cgkkm} was based on ``bottom-up'' theories, the example of \cite{McGreevy} and the one studied here are found in M-theory compactifications and are both therefore ``top-down".

Interestingly, the thermodynamics for such systems is reminiscent of  that of Yang-Mills in four dimensions, \cite{gkmn,gkmn2}.
At finite temperature, up to a transition temperature $T_c$, the dynamics is temperature-independent to leading order in the large-$N$ limit. However there is a first-order transition at $T_c$ to a new phase that is strongly coupled, liquid-like and gapless.  However, in these theories, the  saddle points are translationally invariant and the DC conductivity is strictly infinite because of the $\delta$-function.

The purpose of this work will be to introduce momentum dissipation in the holographic theories in question and study the ensuing holographic saddle points. We will model momentum dissipation by using massless scalars with translationally non-invariant backgrounds, \cite{DG2,AW} and choose it to be irrelevant in the IR. Our results are as follows:

\begin{itemize}

\item In EMD theories, as in (\ref{eq:action}) with no U(1) symmetry breaking ($W(\phi)=0$) and with  momentum dissipation we analyze the nature of the zero-temperature current-current correlator for theories that without momentum dissipation had a discrete spectrum. We
    find that the zero-temperature spectrum remains discrete but as expected the zero frequency $\delta$-function disappears.  Such saddle points are insulators with a hard gap and a discrete spectrum. They share similarities with insulators of type 1 above, with the difference that there,  electrons are weakly-coupled and the underlying spectrum of possible states is not-only gapped but has bands. Therefore the only similarity with type 1 insulators is the presence of a gap. In our case this is driven by strong interactions in the underlying theory, and in that respect it resembles type 3 (Mott) insulators although the detailed mechanism seems different.

Such ground states share many similarities with nuclear matter at finite charge density, \cite{ajkkrt}.

We believe that this mechanism of insulation is new and it should be number five on the previous list. To substantiate this claim however further work is needed and we comment on this in the outlook section.

\item To corroborate the previous claims a subtle and detailed analysis of the $T=0$ current-current correlator, and the conductivity are required. The reason is that at $T=0$ a zero frequency $\delta$-functions can re-emerge as it happens in metals, because the momentum dissipating effects are IR irrelevant and therefore shut-off at $T=0$.
Our analysis indicates that in the saddle-points in question this does not happen. It is interestingly correlated with another feature of non-trivial quantum critical saddle points: the presence of the mild IR (resolvable) singularity and its resolution. As was first argued in detail in \cite{cgkkm}, sometimes the calculation of a correlator in the presence of a resolvable singularity is well defined and the result independent of its resolution. We called such cases, ``repulsive" or ``holographically well-defined". We show here that there is a clear correlation between such holographically well defined cases and the absence of the zero frequency $\delta$-function.

\item We construct examples of such saddle points numerically using a concrete bulk gravitational theory and a complete flow from the UV to the IR. As the zero temperature geometry is numerically challenging we approach it from the unstable black hole branch. Our findings corroborate the asymptotic analytical tools used to analyze the zero
    temperature saddle points.

\item We further consider effective holographic actions with U(1) symmetry breaking (spontaneous or explicit) as in (\ref{eq:action}) with ($W(\phi)\not= 0$). We consider fractionalized IR fixed points of the type found in \cite{gk2}, with a discrete spectrum for the current-current correlator.
We add momentum dissipation as in the previous case and verify analytically that the nature of the spectrum of the current-current correlator remains similar: it is discrete.
In the case where the U(1) symmetry breaking is spontaneous, there is always a $\delta$-function at zero frequency due to the superfluid pole.
Such saddle points describe superfluids in the presence of the breaking of translation invariance, with a discrete spectrum and a hard gap above the superfluid pole.
Their behavior in this respect resembles supersolids,\footnote{Supersolids (see \cite{review} for a review) have been anticipated theoretically, \cite{Legget} and have been described in several contexts, \cite{fisher,anderson2,anderson3,anderson4,anderson5,son,nicolis}. Experimental evidence for supersolids has been claimed in \cite{kim}. Realizations in cold atoms have been also put forward, \cite{cirac}. Most recent experiments however cast doubt on the first experimental observation, \cite{kim2} although there are further experimental claims for verifications.} although in the supersolids discussed in the literature the translation invariance breaking is expected to be spontaneous. We may call them ``charge supersolids'' although the spectral density of the spin-2 (shear) part and the spin-0 (bulk) part of the stress tensor is also discrete and gapped in these saddle points (as detailed in section~\ref{sec:tensor-scalar}).

\end{itemize}

The rest of this paper is organized as follows:

\begin{itemize}

\item In the next subsection the main gravitational (bulk) Lagrangians are introduced and the procedure for constructing the extremal IR geometries (along the lines of \cite{cgkkm,gk1,gk2}) is described.

\item In section~\ref{sec:geometry}, the equations of motion for the holographic saddle points are analyzed and an analytic forms of the asymptotic solutions are obtained and studied.

\item In section~\ref{sec:conductivity}, the conductivity is studied analytically for scaling solutions of the EMD theories introduced in section~\ref{sec:review}. Such solutions contains both the charged and neutral hyperscaling violating solutions of \cite{cgkkm,gk1}.

\item In section~\ref{sec:conductivity2}, the conductivity is analyzed in the presence of momentum dissipation. In particular it is shown that the nature of the charged spectra do not change in the relevant theories.

\item In section~\ref{sec:DC-conductivity}, a further analysis of the DC conductivity substantiates the claim that such systems have properties reminiscent of strongly coupled insulators and supersolids.

\item Section~\ref{sec:finite-temp} contains the numerical results in concrete holographic models that substantiate the analytical conclusions obtained in previous sections.

\item In section~\ref{sec:tensor-scalar}, we consider the spin-2 and spin-0 parts of the stress-tenror two-point function.

\item In section~\ref{sec:conclusion}, we present our conclusions and outlook for the present work.

\item In appendix~\ref{sec:eom}, we analyze in detail the equations of motion.

\item In appendix~\ref{sec:potential}, the potential for the scalar, used for the numerical evaluations is analyzed in detail.

\item Appendix~\ref{sec:SUGRA} contains an overview of the properties of charged black holes known from supergravity, and compares them with the systems described in this paper.

\item Finally in appendix~\ref{sec:zero-density}, we analyze zero density systems that are gapped, by applying the same method.

\end{itemize}

\subsection{Holographic finite density systems at quantum criticality}
\label{sec:review}
We will consider a general Einstein-Maxwell-scalar action in four dimensions with a massive gauge field (mEMD)
\begin{equation}
S=\int d^{4}x\sqrt{-g}\left[R-\frac{1}{2}(\partial\phi)^2-V(\phi)-\frac{Z(\phi)}{4}F^2-
\frac{W(\phi)}{2}A^2-\frac{Y(\phi)}{2}\sum_{i=1}^2(\partial\psi_i)^2\right],\label{eq:action}
\end{equation}
where $\psi_i$ ($i=1$, $2$) are massless scalars  responsible for the breaking the translational invariance. We will eventually require the last term to be irrelevant in the IR. Moreover these scalars will take a simple form, namely $\psi_i=k_ix_i$ (no sum on $i$) in order to break translational invariance.

The scalar $\phi$ is dual to the most important scalar relevant operator of the system that drives a non-trivial RG flow from the UV to the IR.
We assume that $V(\phi)$, $Z(\phi)$, $W(\phi)$, and $Y(\phi)$  have exponential asymptotics as in supergravity systems. To obtain an asymptotically AdS geometry, $V(\phi)$ must have a finite extremum. The detailed analysis of the IR geometries from \eqref{eq:action} without the axions was  given in \cite{gk2}.

\begin{figure}
  \centering
  \includegraphics[width=\textwidth]{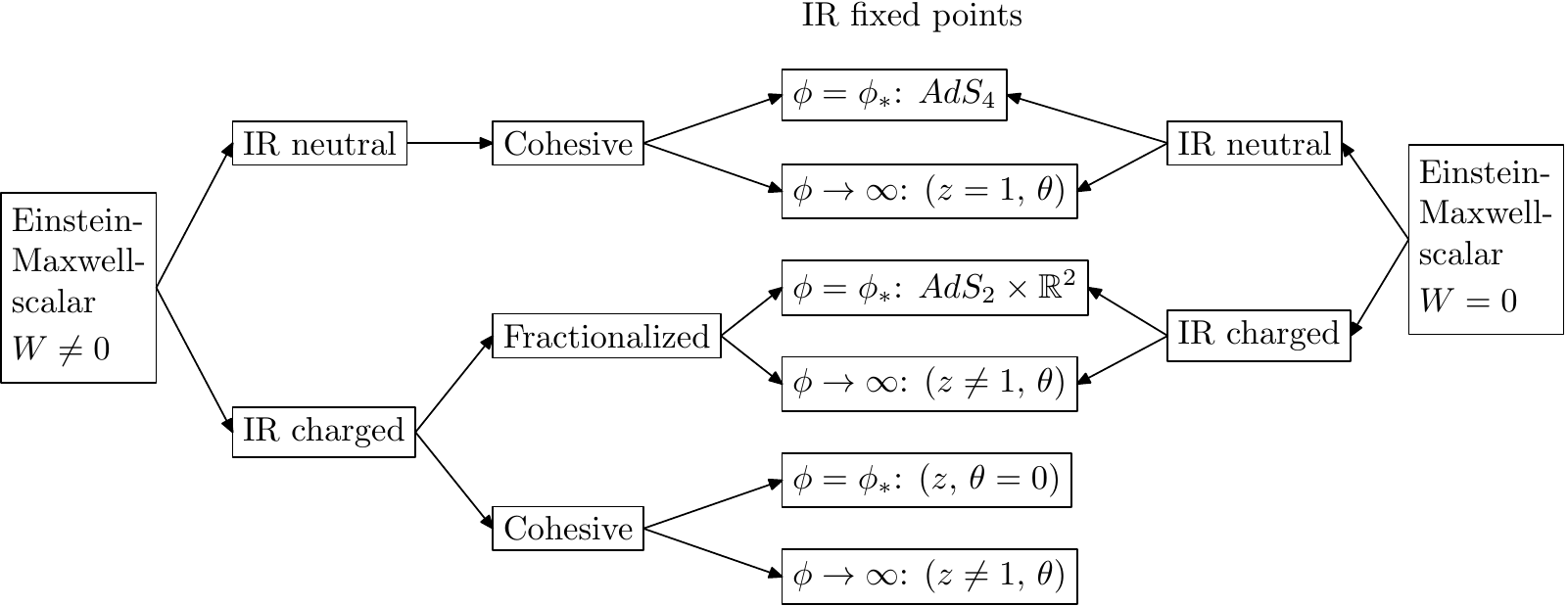}
  \caption{\label{fig:whole} The possible IR geometries from systems with a U(1) symmetry which may be broken or unbroken. The running scalar cases correspond to those that may have a gapped geometry.}
\end{figure}

The possible IR scaling geometries are summarized in figure~\ref{fig:whole} where we use the terminology of reference \cite{gk2} to characterize the holographic critical scaling geometries. In particular, the two sides of the diagram refer to the bulk action used. On the right-hand side, the bulk action has the U(1) symmetry unbroken (or the dual gauge boson is massless). On the left-hand side the symmetry is broken (either spontaneously or explicitly) and the gauge boson is massive.
As we study a finite density system, the gauge field is always non-trivial in the bulk background solution. There are two possibilities:
\begin{itemize}
  \item The {\it IR-neutral solution} is a solution where the gauge field vanishes fast enough and therefore does not affect the background solution in the IR. This does not necessarily imply that the electric flux vanishes in the IR.

  \item On the other hand in {\it IR-charged solutions} the gauge field is affecting non-trivially the solution in the IR.
\end{itemize}

The scalar $\phi$ can either be a constant or running in the IR. We are interested in the case when $\phi$ is running, which leads to the IR geometry
\begin{equation}
ds^2=\tilde{r}^\theta\left(-\frac{dt^2}{\tilde{r}^{2z}}+\frac{d\tilde{r}^2+dx^2+dy^2}{\tilde{r}^2}\right),\label{eq:ztheta1}
\end{equation}
where $z$ is the Lifshitz scaling exponent, and $\theta$ is the hyperscaling violation exponent. We will change $\tilde{r}$ to $r$ later. The exponents $(z,\theta)$ do not  uniquely characterize the solution. Beyond the metric the behavior of the gauge field is important to characterize the solution and there is another critical exponent $\zeta$ that was introduced in \cite{gk2} and characterized in more detail in \cite{Gouteraux:2013oca}.\footnote{A somewhat similar exponent was also introduced for different reasons in \cite{ho} and also in \cite{karch}.}

\begin{figure}
  \centering
  \includegraphics[width=0.5\textwidth]{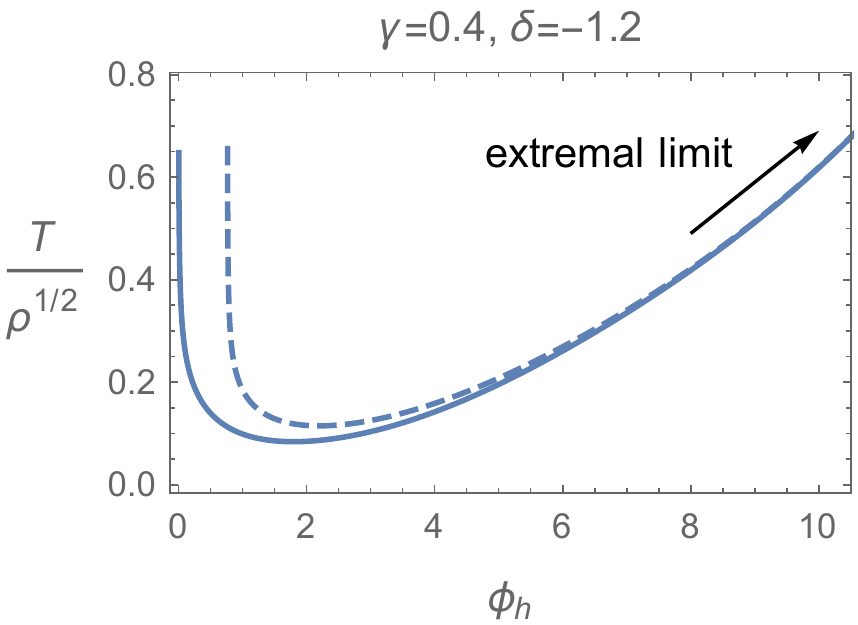}
  \caption{\label{fig:Tphia_2} A typical relation between the temperature and $\phi_h\equiv\phi(r_h)$ for the gapped geometry. There is a minimal temperature, and the ground system can be obtain by taking the extremal limit $T\to\infty$. The dashed line shows the relation between $T$ and $\phi_h$ when the axions are added, while the solid line shows the case without axions. The parameters are $k=1/2$ and $\lambda=1/2$.}
\end{figure}

\begin{figure}
  \centering
  \includegraphics[]{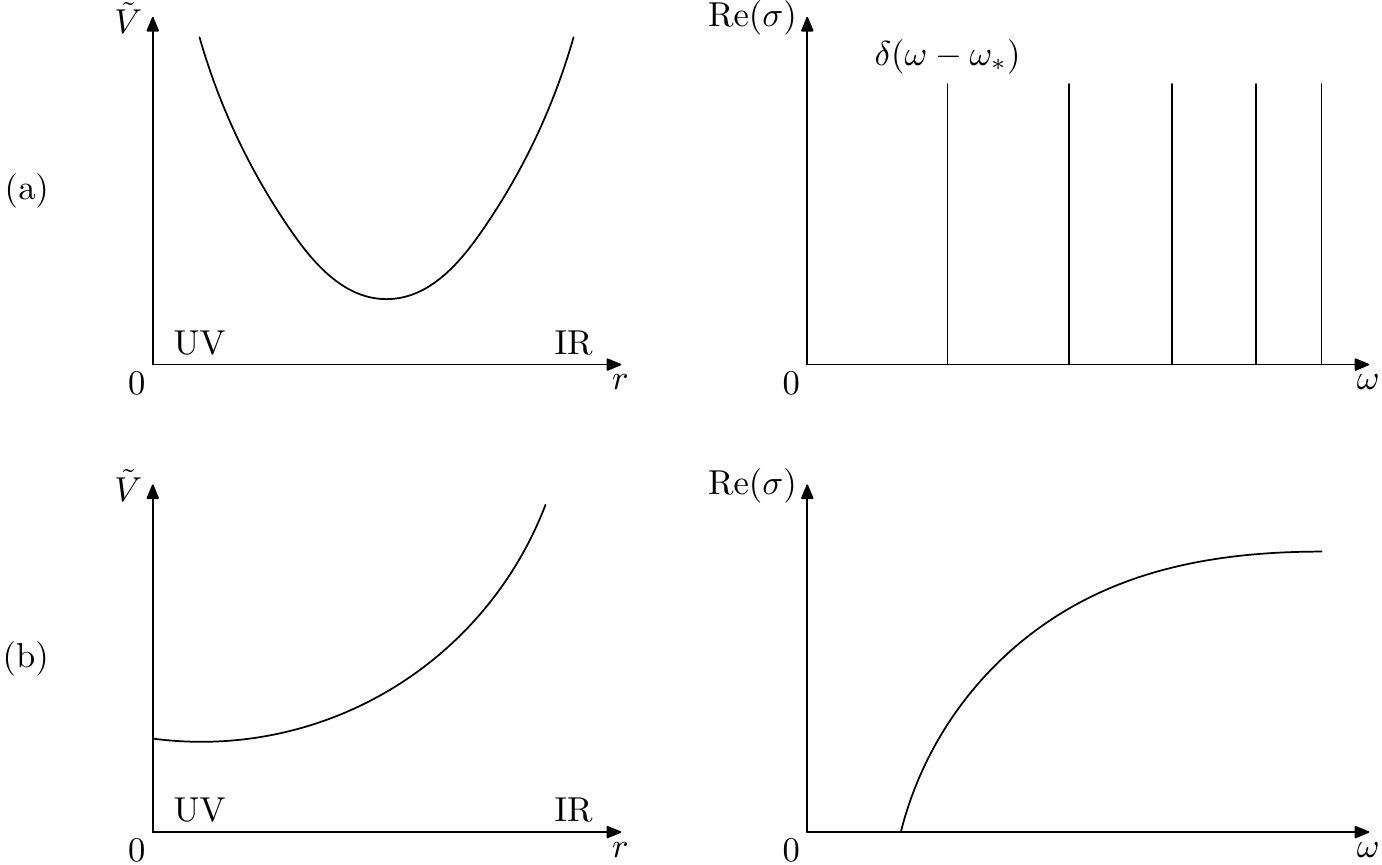}
  \caption{\label{fig:Veff} Schematic plots of the effective potential and the conductivity with a gap. (a) The spectrum is discrete. (b) The spectrum is continuous with a gap.}
\end{figure}

 In the case where the U(1) symmetry is broken and therefore there is a nontrivial mass term  $W(\phi)A^2$ for the gauge field in the bulk, we need to use two more concepts.
 \begin{itemize}
\item  If the symmetry breaking term does not contribute to the leading order of the IR geometry, then the electric flux in the IR
\begin{equation}
\int\star[Z(\phi)F]\sim r^\xi
\end{equation}
is constant, and the system is in a {\bf fractionalized} phase.

\item If the symmetry breaking term contributes to leading order to the IR geometry, then the electric flux in the IR is zero, and the system is in a {\bf cohesive} phase.
\end{itemize}

We will mainly study the fractionalized phase in this work. See \cite{Hartnoll:2011pp,Hartnoll:2012ux} for details about the fractionalized and cohesive phases.

We now focus on the IR charged, fractionalized phase, and briefly mention the other two hyperscaling violating cases, which are cohesive phases. In the fractionalized phase, the metric exponents $(z,\theta)$ can be obtained  in terms of two parameters in the Lagrangian $(\gamma,\delta)$.
These are defined in terms of the asymptotic form of the potential functions in (\ref{eq:action}) in the far IR of the solution as
\be
V(\phi)\sim e^{-\delta \phi}\sp Z(\phi)\sim e^{\gamma\phi}\sp W(\phi)\sim e^{\e \phi}.
\label{asy}\ee

The Gubser criterion for the singularity that is present in these backgrounds is implying \cite{Gubser:2000nd,cgkkm}\footnote{This is stronger than the null energy condition: $(2-\theta)(2z-2-\theta)\geq 0$, $(z-1)(2+z-\theta)\geq 0$ \cite{Dong:2012se}. The Gubser criterion in terms of $z$ and $\theta$ is \cite{gk1}
  \begin{equation}
  \frac{z+2-\theta}{2z-2-\theta}>0,\qquad \frac{z-\theta+1}{2z-2-\theta}>0,\qquad \frac{z-1}{2z-2-\theta}>0.
  \end{equation}}
\begin{equation}
3\gamma^2-2\gamma\delta-\delta^2+4>0,\qquad \gamma^2-\gamma\delta+2>0,\qquad \delta^2-\gamma\delta-2<0.
\end{equation}
When these inequalities are satisfied, the naked singularity is expected to be resolvable. Moreover, we require that the singularity is ``repulsive". In practice this means that  the Sturm-Liouville problem (associated to the conductivity calculation)  be well defined, i.e., there is only one normalizable solution in the IR limit, among the two linearly independent solutions. Under the above conditions, the naked singularity is expected to be resolvable\footnote{Either by embedding in a higher-dimensional geometry or by a stringy resolution.}  and moreover the conductivity does not depend on how this resolution is done.

The conductivity for holographic models satisfies a Schr\"{o}dinger-like equation. If the Schr\"{o}dinger potential $\tilde{V}$ only allows bound states, then the spectrum is discrete. This implies that  the poles of the conductivity have a discrete pattern and the conductivity is given by a discrete sum over the poles. See figures~\ref{fig:Tphia_2} and \ref{fig:Veff}.

The two-dimensional parameter space $(\gamma,\delta)$ will be  characterised  on the basis of conductivity. We plot the parameter space for $(\gamma,\delta)$ when $\tilde{V}\to\infty$ in the IR in figure~\ref{fig:par} below. If $\tilde{V}\to\infty$ in the UV also, then $\text{Re}[\sigma(\omega)]$ has discrete spectrum only; if $\tilde{V}$ is a positive constant in the UV, then $\text{Re}[\sigma(\omega)]$ has a hard gap, may have a few low-lying poles and continuous spectrum above this. The parameter space when the conductivity has a gap will be identified, and the corresponding geometry at finite temperature will be  numerically constructed.

The DC conductivity for a translationally invariant system at finite density has a $\delta$-function at $\omega=0$. This is because in a translationally invariant system charge at rest can be boosted to arbitrary velocities implying infinite DC conductivity.

Momentum dissipation can resolve the $\delta$-function. In this work we will choose as the mechanism for momentum dissipation a distribution of string charge, described by an inhomogeneous solution of scalars without potentials,\footnote{These are known as axions in string theory.} \cite{DG2,AW}. The fluctuation equations that determine the conductivity involve two coupled equations that couple the gauge field fluctuation to the axion fluctuation responsible for the momentum dissipation.
When the effects of momentum dissipation on the background solution vanish in the IR,\footnote{This is tantamount to the dissipation dynamics being IR irrelevant.} we show that the qualitative behavior of the conductivity is the same as without momentum dissipation, except that now the $\delta$-function peak has disappeared. The gap in the conductivity, and the discrete spectrum of poles persists.

To determine whether the system is a metal or insulator, we must calculate the DC conductivity. The DC conductivity of holographic systems has two types of contributions. One exists and is finite without the momentum dissipation, and is also non-trivial at zero charge density. It has been interpreted as due to charge pair-production albeit in a critical massless system, \cite{KO} and this interpretation was reinforced in \cite{cgkkm} and \cite{DG1}, although this interpretation is still tentative.

 The other term originates from the momentum dissipation. If there are several sources of momentum dissipation then there are several such contributions that add-up. As seen in special cases in \cite{dgk}, this second term has the structure ${Q^2\over \Gamma_1+\Gamma_2+\cdots}$, where $Q$ is the charge density and $\Gamma_i$ are the various diffusion rates.

In the class of theories that we study the dissipative term always dominates. This is true for both gapless and gapped geometries. In each case the system can be either a conductor or an insulator. We will later obtain the parameter space for which the system is a conductor or insulator.

We find theories that in the presence of momentum dissipation have a current two-point function whose imaginary part is a discrete sum of  poles on the real axis. This gives a conductivity whose real part is a sum of $\delta$-functions.
This behavior persists at all non-zero temperatures $T<T_c$ up to a fist order
phase transition in which the system jumps into a ``plasma" phase described by a hairy black hole. The current correlator at leading order in $1/N_c$ is independent of temperature in the low temperature phase $0<T<T_c$. The system in this phase transport-wise resembles a Mott insulator.\footnote{Mott insulators are defined as insulators due to strong repulsive on-site interactions. In this paper we use a more liberal and more general definition: insulators due to strong interactions.} This behavior was conjectured first in \cite{cgkkm} after translationally-invariant solutions with a discrete spectrum in the current-current channel was found.\footnote{Another example of a translationally invariant holographic finite density system with a discrete spectrum was found independently in \cite{McGreevy}.}

Alternatively, we analyze holographic systems that are effective theories with a broken U(1) symmetry in the presence of momentum dissipation (arising from broken translational invariance), along the lines of \cite{gk2}. We again use axion solutions that break translation invariance.
We find holographic superfluid holographic ground states, whose two-point function of the current has a discrete set of poles including the pole at zero frequency guaranteed by Goldstone's theorem. These systems are superconductors with a discrete spectrum. In many respects they resemble supersolids, \cite{review,cirac}: Translational invariance is broken, there is a pole in the conductivity at zero frequency, and they have a discrete spectrum of lattice vibrations. A main difference is that supersolids are crystalline while the holographic system is (not-surprisingly) continuous.
Like the Mott-insulator-like ground states described above, here also there is a low temperature supersolid phase up to a finite transitional temperature, $T_c$. To leading order in $1/N_c$ the spectrum and conductivity are temperature independent. The system enters a liquid plasma phase (black hole) above $T_c$.

To determine whether such IR ground states can be reached by  RG flows from the UV  we solve the finite temperature equations of motion and then take the near-extremal limit. We explicitly construct two systems with hyperscaling violating geometries in the IR. The near-extremal scaling behaviors are obtained and matched with the analytic results from the near-extremal black hole. This gives rise to complete holographic ground states realizing the behaviors (Mott-insulator and supersolid) described above.

\section{Constructing the extremal geometry}
\label{sec:geometry}
The detailed analysis of the IR geometries from \eqref{eq:action} is given by \cite{gk2}. We will construct the hyperscaling violating geometries from a Einstein-Maxwell-scalar system with $W=0$. This is sufficient for us to discuss the fractionalized phase, in which the symmetry breaking term is subleading in the IR.

We want to obtain the extremal geometry of the following system
\be
S=\int d^{4}x\sqrt{-g}\left[R-{1\over 2}(\partial\phi)^2-V(\phi)-{Z(\phi)\over 4}F_{\m\n}F^{\m\n}\right],\label{eq:LVZ}
\ee
where
\be
Z(\phi)=e^{\gamma\phi},\qquad V(\phi)=V_0e^{-\delta\phi}+V_1e^{\delta_1\phi}+\cdots.\label{eq:Vexp}
\ee
We are looking for the extremal geometry when $\phi\to\infty$ in the IR. We assume that the leading term in $V(\phi)$ in the IR is the first term and $V_0<0$ is required.

The metric ansatz is
\begin{equation}
ds^2=\frac{1}{r^2}\left(-\frac{g(r)}{h(r)}dt^2+\frac{dr^2}{g(r)}+dx^2+dy^2\right).\label{eq:ansatz}
\end{equation}
The AdS boundary is at $r=0$, and the IR limit is $r\to\infty$. The IR geometry is
\begin{equation}
ds^2=\frac{1}{r^2}\left(-\frac{g_0}{h_0}r^{-\frac{4(z-1)}{2-\theta}}dt^2
+r^\frac{2\theta}{2-\theta}\frac{dr^2}{g_0}+dx^2+dy^2\right),\label{eq:ztheta}
\end{equation}
where $z$ is the Lifshitz scaling exponent, and $\theta$ is the hyperscaling violation exponent.\footnote{
The relation between $r$ and $\tilde{r}$ in \eqref{eq:ztheta1} is $r^2=\tilde{r}^{2-\theta}$. The advantage of the coordinates \eqref{eq:ansatz} is that it is the Poincar\'{e} coordinates in the UV, and thus more convenient for numerical calculations.}

The equations of motion are
\begin{subequations}
\begin{gather}
2r^2gh\phi''-4rgh\phi'+2r^2hg'\phi'-r^2gh'\phi'-2hV'(\phi)+r^4h^2A_t'^2Z'(\phi)=0,\label{eq:phi}\\
(\sqrt{h}Z(\phi)A_t')'=0,\label{eq:At}\\
4rg'-12g-2V(\phi)-r^4hZ(\phi)A_t'^2-r^2g\phi'^2=0,\label{eq:g}\\
2h'-rh\phi'^2=0.\label{eq:h}
\end{gather}\label{eq:eoms}
\end{subequations}
There is a conserved (extremality) charge
\begin{equation}
\mathcal{Q}=\sqrt{h}\Bigl[Z(\phi)A_tA_t'-\frac{1}{r^2}\Bigl(\frac{g}{h}\Bigr)'\Bigr],\label{eq:calQ}
\end{equation}
which is zero for the extremal geometry \cite{Gubser:2009cg}. Equation \eqref{eq:At} can be solved as
\be
\sqrt{h}Z(\phi)A_t'=-\rho\;,
 \ee
 where $\rho$ is the charge density. We may use this equation to eliminate $A_t$ in \eqref{eq:phi}, \eqref{eq:g}, and \eqref{eq:h}, and obtain three coupled equations for $\phi$, $g$, and $h$.\footnote{We can use the conserved charge $\mathcal{Q}$ to determine the chemical potential $\mu$. At the horizon, $\mathcal{Q}=Ts$, where $T$ is the temperature, and $s$ is the entropy density; at the AdS boundary, $\mathcal{Q}=-\mu\rho+3\epsilon/2$, where $\epsilon$ is the energy density determined by \eqref{eq:epsilon} below. However, if we work in the grand canonical ensemble (fixing $\mu$), we need to solve the four equations.}

The near horizon expansion for the functions is as follows
\begin{subequations}
\begin{align}
e^\phi &=r^{\bar{\kappa}}(f_0+{\color{purple}\tilde{f}_{1}r^{\beta}+\tilde{f}_{2}r^{2\beta}\cdots}+f_1r^\alpha+f_2r^{2\alpha}+\cdots),\\
A_t &=r^{\bar{\zeta}-z}(a_0+{\color{purple}\tilde{a}_{1}r^{\beta}+\tilde{a}_{2}r^{2\beta}\cdots}+a_1r^\alpha+a_2r^{2\alpha}+\cdots),\\
g &=r^{-\frac{2\theta}{2-\theta}}(g_0+{\color{purple}\tilde{g}_{1}r^{\beta}+\tilde{g}_{2}r^{2\beta}\cdots}
+g_1r^\alpha+g_2r^{2\alpha}+\cdots),\\
h &=r^{\frac{2(2z-2-\theta)}{2-\theta}}(h_0+{\color{purple}\tilde{h}_{1}r^{\beta}+\tilde{h}_{2}r^{2\beta}\cdots}
+h_1r^\alpha+g_2r^{2\alpha}+\cdots),
\end{align}\label{eq:hor}
\end{subequations}
where the leading terms are from the hyperscaling violation geometry, the terms involving $\beta$ are the modes of the linear perturbations and the terms involving $\alpha$ are from the second exponential in the scalar potential. There can be more powers if the potential has more exponentials.

The exponents $\kappa$ and $\zeta$ are defined in terms of the coordinates \eqref{eq:ztheta1} \cite{cgkkm}:
\be
\phi\sim\tilde{r}^\kappa\sp A_t\sim\tilde{r}^{\zeta-z}\;.
 \ee
 The relation between $\bar{\kappa}$, $\bar{\zeta}$ and $\kappa$, $\zeta$ is
\be
\bar{\kappa}=\frac{2\kappa}{2-\theta},\qquad \bar{\zeta}-z=\frac{2(\zeta-z)}{2-\theta}.
\label{zeta}\ee

After we substitute  the above expansion to the equations of motion, we obtain equations order by order in the perturbations. At the zeroth order, we can solve $\bar{\kappa}$, $\zeta$, $z$, $\theta$, $g_0$, and $a_0$ as follows. The parameter $h_0$ is fixed by requiring $h=1$ in the UV. Thus, the only free parameter in the IR is $f_0$. There are two sets of consistent solutions.

\begin{itemize}
  \item IR charged solution, \cite{cgkkm}:
  \begin{equation}
  \begin{split}
  z &=\frac{\gamma^2+2\gamma\delta-3\delta^2+4}{\gamma^2-\delta^2},\qquad \theta=\frac{4\delta}{\gamma+\delta},\\
  \bar{\kappa} &=\frac{4}{\gamma-\delta},\qquad \kappa=\frac{4}{\gamma+\delta},\qquad \zeta=\theta-2,\\
  g_0 &=-\frac{f_0^{-\delta}(\gamma-\delta)^4V_0}{2(3\gamma^2-2\gamma\delta-\delta^2+4)(\gamma^2-\gamma\delta+2)},\\ a_0^2 &=\frac{2f_0^{-\gamma-\delta}(\gamma-\delta)^4(\delta^2-\gamma\delta-2)V_0}{h_0(3\gamma^2-2\gamma\delta-\delta^2+4)^2(\gamma^2-\gamma\delta+2)}.
  \end{split}\label{eq:zeroth1}
  \end{equation}
  The powers $\beta_i$ arise from perturbing the hyperscaling violating  geometry,
  \begin{footnotesize}
  \begin{align}
  \beta_0 &=0,\qquad \beta_\text{u}=\frac{4+(\gamma-\delta)(3\gamma+\delta)}{2(\gamma-\delta)^2},\nonumber\\
  \beta_\pm &=\frac{4+(\gamma-\delta)(3\gamma+\delta)\pm\sqrt{(4+(\gamma-\delta)(3\gamma+\delta))
  (36+(\gamma-\delta)(17\delta+\gamma(19+8(\gamma-\delta)\delta)))}}{2(\gamma-\delta)^2}.
  \end{align}
  \end{footnotesize}
  In the IR expansion, $\beta_i$ can be taken as either $\beta_0$ or $\beta_-$.\footnote{If we write the numerator of $\beta_\pm$ as $a\pm\sqrt{b}$, we have $a>0$ and $a^2-b=8(3\gamma^2-2\gamma\delta-\delta^2+4)(\gamma^2-\gamma\delta+2)(\delta^2-\gamma\delta-2)<0$, according to the Gubser criterion; consequently, we always have $\beta_-<0$ and $\beta_+>0$.}

  \item IR neutral solution:
  \begin{equation}
  \begin{split}
  z &=1,\qquad \theta=\frac{2\delta^2}{\delta^2-1},\\
  \bar{\kappa} &=-2\delta,\qquad \kappa=\frac{2\delta}{\delta^2-1},\qquad
  \zeta=\frac{2(\delta^2-\gamma\delta-1)}{\delta^2-1},\\
  g_0 &=-\frac{f_0^{-\delta}V_0}{2(3-\delta^2)}.
  \end{split}\label{eq:zeroth2}
  \end{equation}
  We will only focus on the IR charged solution in this work.
\end{itemize}

The terms involves $\alpha$ are from other exponentials in the potential of the scalar. At the first order, we can solve $f_1$, $a_1$, $g_1$, $h_1$, and $\alpha$.  For the first set of solutions \eqref{eq:zeroth1}, the solution for $\alpha$ is
\begin{equation}
\alpha=\frac{4(\delta+\delta_1)}{\gamma-\delta}.
\end{equation}
For the second set of solutions \eqref{eq:zeroth2}, the solution for $\alpha$ is
\begin{equation}
\alpha=-2\delta(\delta+\delta_1).
\end{equation}
Note that $\alpha$ depends on the UV completion of the system. The value of $\alpha$ gives a further constraint for $z$ and $\theta$ by requiring $\alpha<0$.

The asymptotic behavior of the scalar field $\phi$ near the AdS boundary is
\begin{equation}
\phi=\phi_ar^\Delta(1+\cdots)+\phi_br^{3-\Delta}(1+\cdots),\label{eq:phibry}
\end{equation}
where $\phi_a$ is the expectation of the scalar operator dual to $\phi$, and $\phi_b$ is its source. When $\Delta$ is an integer, there may be log terms. We discuss the choice of the potential $V(\phi)$ to avoid the log terms in appendix~\ref{sec:potential}.

\section{Conductivity with translational invariance}
\label{sec:conductivity}
\subsection{The IR charged solution}
\label{sec:IR-charged}
We will focus on the conductivity calculated from the extremal geometries. We are interested in the case when the conductivity has a discrete spectrum for the hyperscaling violating geometries. To obtain the conductivity, we perturb the system \eqref{eq:LVZ} around the solution to \eqref{eq:eoms} by $\delta A_x=a_x(r)e^{-i\omega t}$ and $\delta g_{tx}=g_{tx}(r)e^{-i\omega t}$. Generically, for the metric
\begin{equation}
ds^2=-D(r)dt^2+B(r)dr^2+C(r)(dx^2+dy^2),\label{eq:BCD}
\end{equation}
the equation for $a_x$ after eliminating $g_{tx}$ is, \cite{cgkkm}
\begin{equation}
\left(Z\sqrt{\frac{D}{B}}a_x'\right)'+\left(Z\sqrt{\frac{B}{D}}\omega^2-\frac{Z^2A_t'^2}{\sqrt{BD}}\right)a_x=0.\label{eq:axtrans}
\end{equation}
After we impose an appropriate boundary condition in the IR (to be discussed below), the asymptotic behavior in the UV is
\begin{equation}
a_x(r)=a_x^{(0)}+a_x^{(1)}r+\cdots,
\end{equation}
and the conductivity is calculated from\footnote{If $Z=Z_0$ at the AdS boundary, then there will be an extra factor in $G$ so that $G=Z_0\frac{a_x^{(1)}}{a_x^{(0)}}$. We can set $Z_0=1$ by rescale $A_\mu$. This amounts to a suitable change of units of measuring the boundary charge density.}
\begin{equation}
G=\frac{a_x^{(1)}}{a_x^{(0)}},\qquad \sigma(\omega)=\frac{G}{i\omega}.
\end{equation}
We do not need to know  $a_x$ explicitly to decide whether the spectrum of $\text{Re}(\omega)$ is gapped/gapless and/or discrete/continuous.

After a change of variables by
\begin{equation}
\frac{d\xi}{dr}=\sqrt{\frac{B}{D}},\qquad \tilde{a}_x=\sqrt{Z}a_x,\label{eq:schrx}
\end{equation}
we can obtain a Schr\"{o}dinger equation
\begin{equation}
-\frac{d^2\tilde{a}_x}{d\xi^2}+\tilde{V}(\xi)\tilde{a}_x=\omega^2\tilde{a}_x.\label{eq:schr}
\end{equation}
The potential is given by
\begin{equation}
\tilde{V}=\frac{ZA_t'^2}{B}-\frac{DB'Z'}{4B^2Z}+\frac{D'Z'}{4BZ}-\frac{DZ'^2}{4BZ^2}+\frac{DZ''}{2BZ},
\end{equation}
where the prime is denoting a derivative with respect to $r$. If $\tilde{V}$ asymptotes to infinity in both the IR and the UV, then the Schr\"{o}dinger equation can only have bound states.

To analyze the behavior of $\tilde{V}$ in the IR, we use the hyperscaling violating geometry \eqref{eq:ztheta} as background:
\begin{equation}
D=r^{-\frac{4(z-1)}{2-\theta}-2},\qquad B=r^{\frac{2\theta}{2-\theta}-2},\qquad C=\frac{1}{r^2}.
\end{equation}
We change the variable $r$ to $\xi$ as
\begin{equation}
\xi=\frac{\sqrt{h_0}}{g_0}\frac{2-\theta}{2z}r^\frac{2z}{2-\theta}\;.\label{eq:xr}
\end{equation}
The asymptotic Schr\"{o}dinger potential is given by
\begin{equation}
\tilde{V}(\xi)=\frac{\nu^2-1/4}{\xi^2},\label{eq:c}
\end{equation}
where
\begin{equation}
\nu=\frac{(\gamma-\delta)(3\gamma+5\delta)+12}{2[(\gamma-\delta)(\gamma+3\delta)+4]}=\frac{3z-\theta}{2z}.
\label{eq:nu}\end{equation}
The expression of $\nu$ in terms of $z$ and $\theta$ holds for both the fractionalized phase and the cohesive phase when the current is dominant in the IR.\footnote{We use two different terms to describe two different effects. We call the current ``dominant" in the IR if the gauge field is both non-trivial in the IR and contributes to the leading order IR background solution. This is distinct from the term ``relevant" current that indicates a current perturbation of the IR solution that is relevant in the RG sense.}

The boundary condition in the IR depends on whether the IR limit is at $\xi\to\infty$ or $\xi\to 0$. This is directly related to whether in the extremal limit the black hole solution has  $T\to 0$ or $T\to\infty$ in the near-extremal geometry. The relation between the entropy and temperature is
\begin{equation}
S\sim T^\frac{2-\theta}{z}.\label{eq:ST}
\end{equation}
Comparing \eqref{eq:xr} and \eqref{eq:ST}, we can distinguish  three different cases as follows:
\begin{enumerate}[(1)]
  \item The IR limit is at $\xi\to\infty$, which happens when $\frac{z}{2-\theta}>0$. In this case, the extremal limit of the small black hole branch is at $T\to 0$. Moreover, the Gubser criterion implies that we have $\nu>{1\over 2}$. To leading order in the IR expansion, the solution for $\tilde{a}_x$ with the in-falling wave boundary condition is
  \begin{equation}
  \tilde{a}_x\sim\sqrt{\xi}H^{(1)}_\nu(\omega\xi)\sim e^{i\omega\xi}.
  \end{equation}
  The current-current correlator is gapless in this case.
  \item The IR limit is at $\xi\to 0$, which happens when $\frac{z}{2-\theta}<0$. In this case, the extremal limit of the small black hole branch (that is now thermodynamically unstable) is at $T\to\infty$. Moreover,  the Gubser criterion implies that $\nu<0$. To leading order in the IR expansion, the general solution for $\tilde{a}_x$ is
  \begin{equation}
  \tilde{a}_x ~=~ C_1\sqrt{\xi}J_{-\nu}(\omega\xi)+C_2\sqrt{\xi}J_{\nu}(\omega\xi)
  ~\sim~ C_1\,\xi^{1/2-\nu} +C_2\,\xi^{1/2+\nu}.
\label{li1}\ee
  The first linearly independent solution in (\ref{li1}) is always normalizable:
  \begin{equation}
  \int_0^\infty |\tilde{a}_x|^2d\xi<\infty.
  \end{equation}
  The second linearly independent solution in (\ref{li1}) is normalizable when $|\nu|<1$, and is non-normalizable when $|\nu|>1$. If there are two normalizable solutions, the boundary condition in the IR depends on how the singularity is resolved, and thus the calculation of the correlator is unreliable and can only be fixed when the singularity is properly resolved.

  When $|\nu|<1/2$, the current-current correlator is gapless; when $|\nu|>1/2$, the current-current correlator is gapped, if the Schr\"{o}dinger potential in the UV is also positive, which happens for scaling dimension $1/2<\Delta<2$.\footnote{In \cite{cgkkm} where an analogous condition was first derived it was assumed that near the UV fixed point at $\phi=0$, $Z(\phi)=1+{\cal O}(\phi^2)$. Here we assume the more general case  $Z(\phi)=1+{\cal O}(\phi)$.}
\item The IR limit is at a constant non-zero  $\xi$, which happens when $\frac{z}{2-\theta}\to 0$. In this case, the extremal limit is at a constant $T$, and the Schr\"{o}dinger potential is a constant $V_0$. The general solution for $a_x$ is
\begin{equation}
a_x=C_1e^{\sqrt{V_0-\omega^2}\,\xi}+C_2e^{-\sqrt{V_0-\omega^2}\,\xi}.
\end{equation}
When $\omega^2>V_0$, the first solution describes the in-falling wave with the $\omega\to\omega+i\epsilon$ prescription. When $\omega^2<V_0$, both solutions are real and normalizable. As the background extremal solution is singular, this situation seems to be in the holographically non-well defined class we have discussed earlier: we need to impose an extra boundary condition at the singularity in order to obtain a unique solution. However, in this case, analyticity (in $\omega$) of the correlator makes the solution unique. Indeed, if we analytically continue the $\omega^2>V_0$ solution to $\omega^2<V_0$, the solution for $a_x$ is unambiguous for all $\omega$.

When $V_0<0$, the current-current correlator is gapless; when $V_0>0$, the current-current correlator is gapped, if the Schr\"{o}dinger potential in the UV is also positive.\footnote{It is fair to say that such cases are degenerate and generalize a similar case that was discussed in \cite{gkmn,gkmn2}.  Further and deeper analysis is necessary in order to find out waht happens in such cases.}

 \end{enumerate}

If the IR is at $\xi\to\infty$ and $|\nu|>1/2$, the potential $\tilde{V}$ is divergent in the IR. To summarize, the conditions for a gapped background charged spectrum  are
\begin{itemize}
  \item Gubser criterion: $(\gamma-\delta)(3\gamma+\delta)+4>0$, $(\gamma-\delta)\gamma+2>0$, $(\gamma-\delta)\delta+2>0$ \cite{cgkkm}.
  \item Thermodynamical instability of the small black hole branch ($\frac{2-\theta}{z}<0$): $(\gamma-\delta)(\gamma+3\delta)+4<0$.
  \item The Sturm-Liouville problem is well-defined: $|\nu|>1$ (stronger than $|\nu|>1/2$).
\end{itemize}

\begin{figure}
  \centering
  \includegraphics[width=0.47\textwidth]{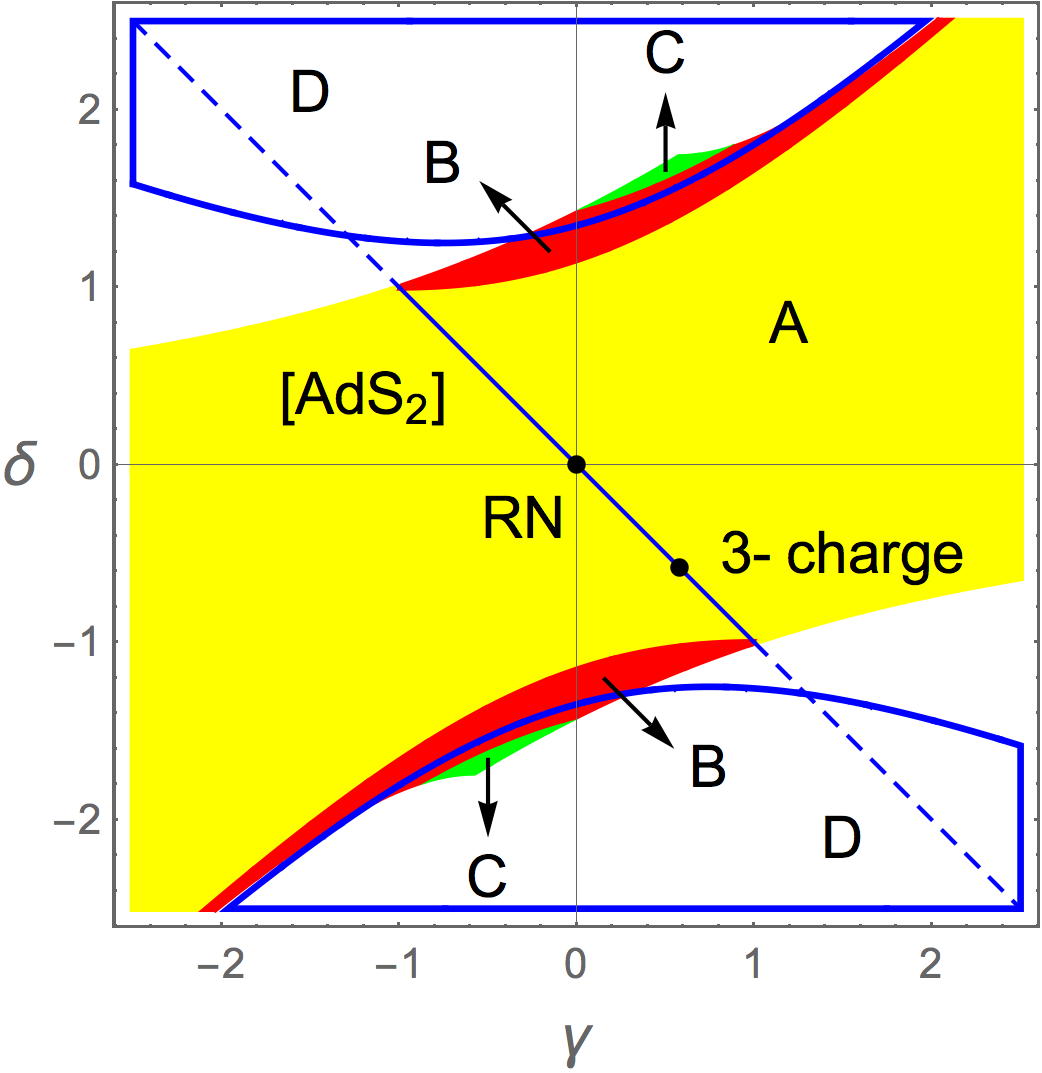}\qquad
  \includegraphics[width=0.47\textwidth]{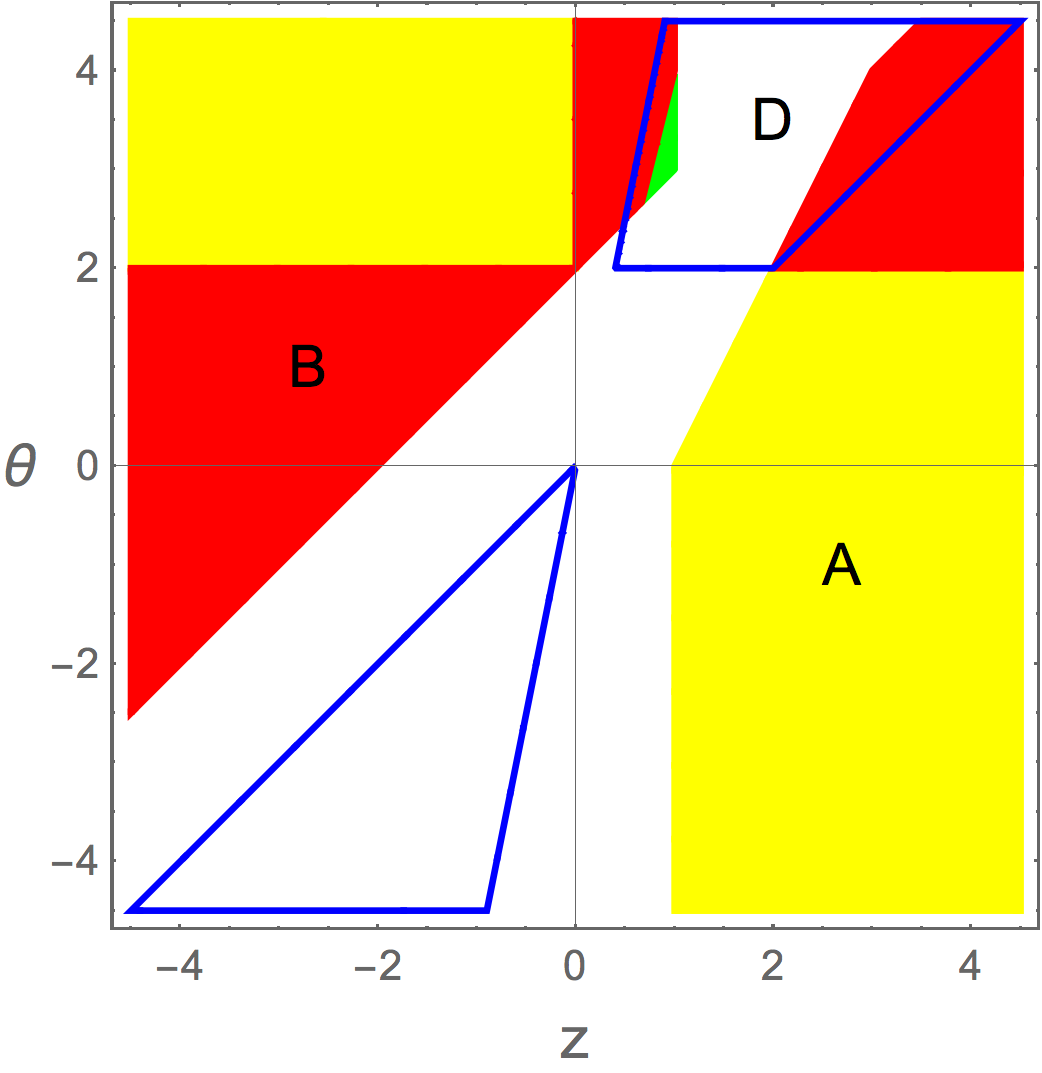}\qquad
  \caption{\label{fig:par} This figure refers to the IR charged solutions. The regions A, B, and C are the parameter space constrained by the Gubser criterion. We assume that  the conditions detailed at the end of section \protect\ref{sec:IR-charged} in the UV for a gapped geometry are satisfied. In the left plot the axes are given by the gravity action parameters ($\gamma,\delta$). In the right plot the axes are given by the metric critical
exponents ($z,\theta$) that are related to ($\gamma,\delta$) as in (\protect\ref{eq:zeroth1}). In region A (yellow), the extremal limit is at $T\to 0$, and the current-current correlator is gapless. In region B (red), the extremal limit is at $T\to\infty$, and the current-current correlator is gapped. In region C (green), the extremal limit is at $T\to\infty$, and the current-current correlator is gapless. Region D (enclosed by blue boundaries) is holographically unreliable.}
\end{figure}
Figure~\ref{fig:par} is a summary of the above discussion. The parameter space when the current-current correlator can have a gap is region B.

The Schr\"{o}dinger potential $\tilde{V}$ near the AdS boundary was analyzed in the Appendix of \cite{Bhattacharya:2014dea} and their conclusions are as follows. The asymptotic behavior of $\phi$ near the AdS boundary is \eqref{eq:phibry}. When both $\phi_a$ and $\phi_b$ are nonzero, $\tilde{V}$ in the UV is divergent when $\Delta\neq 1$ and $\Delta\neq 2$, and is constant when $\Delta=1$ or $2$. When the source is zero ($\phi_b=0$), $\tilde{V}$ in the UV diverges when $1/2<\Delta<1$ and $1<\Delta<2$, is constant when $\Delta=1$ or $2$, and vanishes when $\Delta>2$ (the unitarity bound is $\Delta>1/2$).

\subsection{The IR neutral solution}
When the current is subdominant in the IR we must use the IR neutral solution solution \eqref{eq:zeroth2}. To leading order in the IR geometry, the effective potential for the Schr\"{o}dinger-like equation is
\begin{equation}
V(x)=\frac{\nu_0^2-1/4}{\xi^2},
\end{equation}
where
\begin{equation}
\nu_0=\frac{\delta^2-2\gamma\delta-1}{2(\delta^2-1)}=\frac{\zeta-1}{2}.
\label{li2}\end{equation}
The analysis is similar to the case with a dominant current, except that the values $z$, $\theta$ and $\zeta$ are given by eq.~\eqref{eq:zeroth2}.

When the IR is at $\xi\to 0$, the solution for $\tilde{a}_x$ is
\begin{equation}
\tilde{a}_x=C_1\xi^{1/2-\nu_0}+C_2\xi^{1/2+\nu_0},
\end{equation}
where $\n_0$ is defined in (\ref{li2}). When $|\nu_0|<1$, both solutions are normalizable, and this is a holographically unreliable case as argued above in the sense that an extra boundary condition is necessary at the IR singularity. Therefore we require $\nu_0>1$ or $\nu_0<-1$.

The conditions for a gapped background solution are
\begin{itemize}
  \item Validity of the Gubser criterion $\delta^2<3$.
  \item Thermodynamical instability of the small black hole branch ($\frac{2-\theta}{z}<0$): $\delta^2>1$
  \item The Sturm-Liouville problem is well-defined: $|\nu_0|>1$ (stronger than $|\nu_0|>1/2$).
\end{itemize}

\begin{figure}
  \centering
  \includegraphics[width=0.47\textwidth]{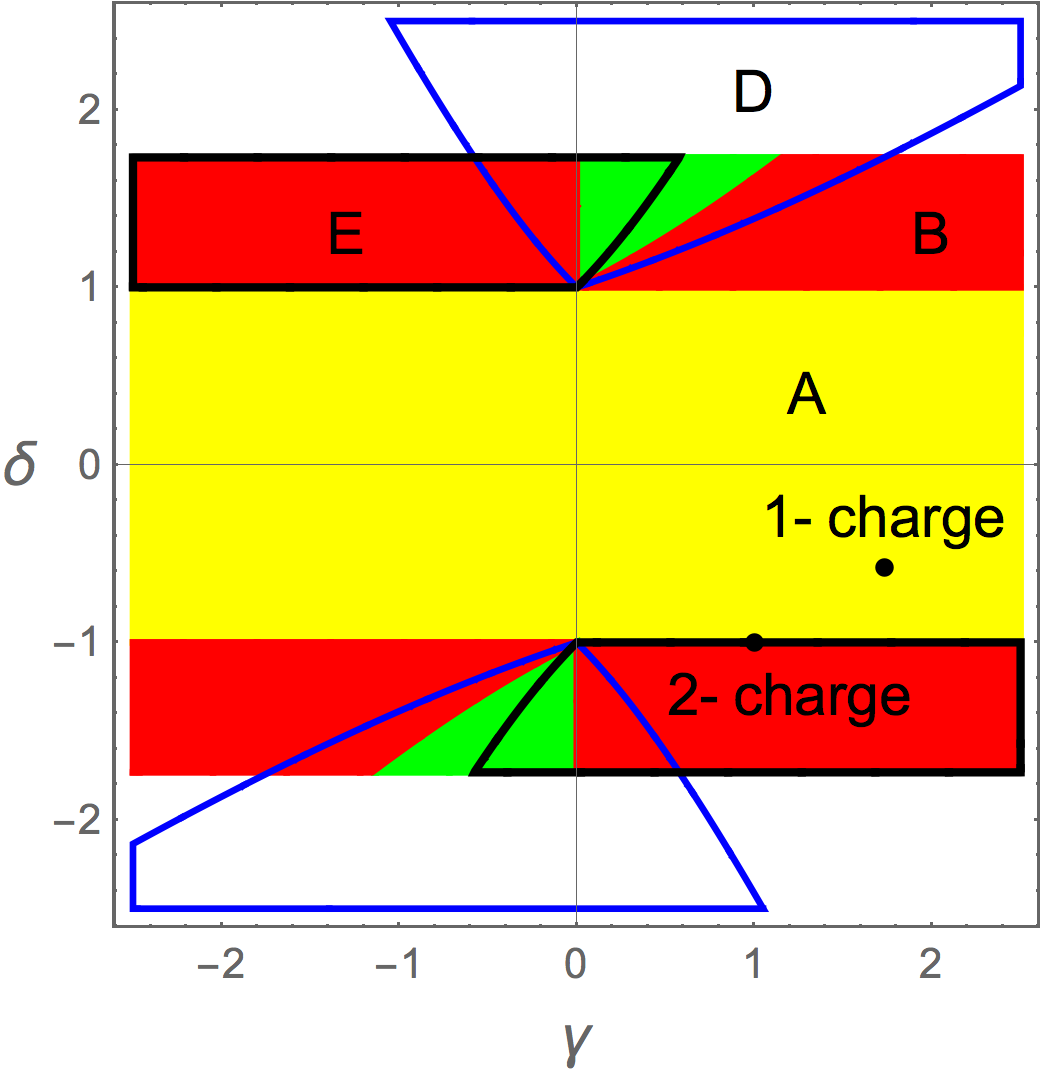}\qquad
  \includegraphics[width=0.47\textwidth]{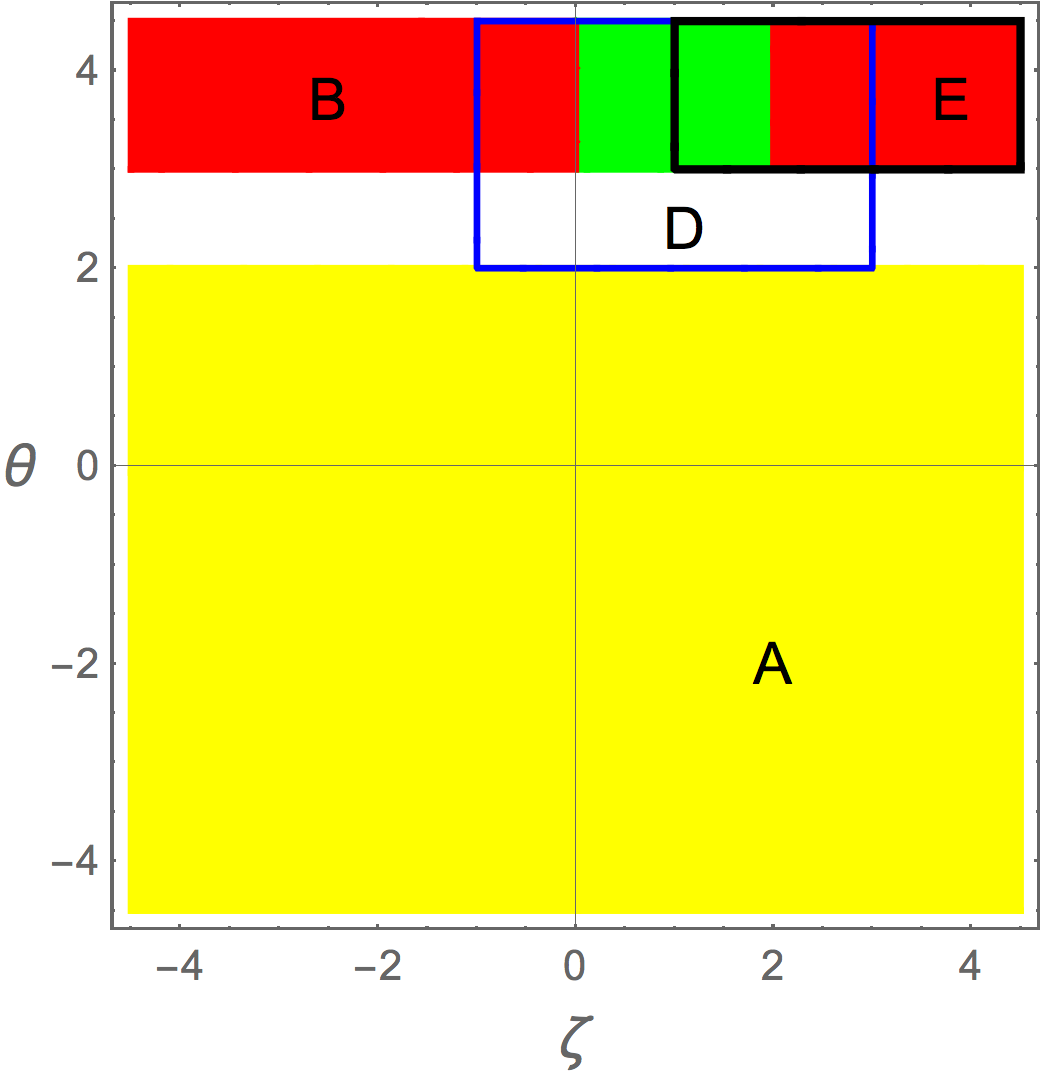}\qquad
  \caption{\label{fig:par0} This figure refers to the IR neutral solutions. The regions A, B, and C are the parameter space constrained by the Gubser criterion. We assume that  the conditions detailed at the end of section \protect\ref{sec:IR-charged} in the UV for a gapped geometry are satisfied. In the left plot the axes are given by the gravity action parameters ($\gamma,\delta$).  In the right plot the axes are given by the critical
exponents of the metric $\theta$ and  gauge field $\zeta$ defined in (\protect\ref{zeta}) and given in (\protect\ref{eq:zeroth2}). In region A (yellow), the extremal limit is at $T\to 0$, and the current-current correlator is gapless. In region B (red), the extremal limit is at $T\to\infty$, and the current-current correlator is gapped. In region C (green), the extremal limit is at $T\to\infty$, and the current-current correlator is gapless. Region D (enclosed by blue boundaries) is holographically unreliable. In region E (enclosed by black boundaries), the conductivity has a $\delta$-function at $\omega=0$.}
\end{figure}
Figure~\ref{fig:par0} is a summary of the above discussion. The parameter space when the current-current correlator has a gap is region B.

\section{Conductivity in the presence of momentum dissipation}
\label{sec:conductivity2}
To add momentum dissipation and at the same time preserve the rotational symmetry we will consider the following bulk theory
\begin{equation}
S=\int d^{4}x\sqrt{-g}\left[R-\frac{1}{2}(\partial\phi)^2-V(\phi)-\frac{Z(\phi)}{4}F^2-\frac{W(\phi)}{2}A^2-\frac{Y(\phi)}{2}\sum_{i=1}^2(\partial\psi_i)^2\right],
\label{li3}\end{equation}
where the leading IR behavior of $V$, $Z$, $W$, and $Y$ are
\begin{equation}
V(\phi)=V_0e^{-\delta\phi},\qquad Z(\phi)=e^{\gamma\phi},\qquad W(\phi)=W_0e^{\eta\phi},\qquad Y(\phi)=e^{\lambda\phi}.
\label{li5}\end{equation}
This is a simple way to add momentum dissipation \cite{DG2,AW} as the scalars $\psi_i$ take the form $\psi_i=kx_i$. Having the same slope for the axions guarantees a rotationally invariant solution.

To calculate the conductivity, we perturb the system by
\begin{equation}
\delta A_x=a_x(r)e^{-i\omega t},\qquad \delta g_{tx}=g_{tx}(r)e^{-i\omega t},\qquad \delta\psi_1=\chi(r)e^{-i\omega t}.
\end{equation}
The independent linearized equations are
\begin{align}
& a_x''+\left((\log Z)'-\frac{B'}{2B}+\frac{D'}{2D}\right)a_x'+\left(\frac{\omega^2}{D}-\frac{W}{Z}\right)Ba_x
+\frac{A_t'}{D}\left(g_{tx}'-\frac{C'}{C}g_{tx}\right)=0,\\
& \chi''+\left((\log Y)'-\frac{B'}{2B}+\frac{C'}{C}+\frac{D'}{2D}\right)\chi'+\frac{\omega^2B}{D}\chi-\frac{ik\omega Bg_{tx}}{CD}=0,\\
& g_{tx}'-\frac{C'}{C}g_{tx}+A_t'Za_x+\frac{ik}{\omega}DY\chi'=0.
\end{align}

We define
\begin{equation}
b_x\equiv\frac{ik}{\omega}\sqrt{\frac{D}{B}}CY\chi',\qquad q\equiv\frac{CZ}{\sqrt{BD}}A_t'.
\end{equation}
From \eqref{eq:Maxwell}, we observe that $q$ is a constant if $W=0$. If $W\neq 0$, $q$ is the minus charge density in the UV, but not a constant in the bulk. In the fractionalized phase, $q$ is a nonzero constant in the IR, while in the cohesive phase, $q=0$ in the IR. After eliminating $g_{tx}(r)$, two coupled equations are derived
\begin{align}
& \left(Z\sqrt{\frac{D}{B}}a_x'\right)'+\left(Z\sqrt{\frac{B}{D}}\omega^2-W\sqrt{BD}-q^2\frac{\sqrt{BD}}{C^2}\right)a_x-q\frac{\sqrt{BD}}{C^2}b_x=0,\\
& \left(\frac{1}{CY}\sqrt{\frac{D}{B}}b_x'\right)'+\left(\frac{\omega^2}{CY}\sqrt{\frac{B}{D}}-k^2\frac{\sqrt{BD}}{C^2}\right)b_x-k^2q\frac{\sqrt{BD}}{C^2}a_x=0.
\end{align}

The equations for $a_x$ and $b_x$ can be written in a more symmetric form. Define $Z_1:=Z=r^{\kappa\gamma}$ and $Z_2:=\frac{1}{CY}=r^{2-\kappa\lambda}$. The relation between the metric \eqref{eq:BCD} and \eqref{eq:ansatz} is
\begin{equation}
D=\frac{g}{r^2h},\qquad B=\frac{1}{r^2g},\qquad C=\frac{1}{r^2}.
\end{equation}
Also we define $f=g/\sqrt{h}=\sqrt{D/B}$. The coupled equations for $a_x$ and $b_x$ are
\begin{align}
\frac{1}{fZ_1}(fZ_1a_x')'+\frac{\omega^2}{f^2}a_x &=\frac{r^2}{gZ_1}\Bigl(\Bigl(q^2+\frac{W}{r^4}\Bigr)a_x+qb_x\Bigr),\\
\frac{1}{fZ_2}(fZ_2b_x')'+\frac{\omega^2}{f^2}b_x &=\frac{r^2}{gZ_2}(k^2b_x+k^2qa_x).
\end{align}
Define $\tilde{a}_x=\sqrt{Z_1}a_x$ and $\tilde{b}_x=\sqrt{Z_2}b_x$, and then we obtain
\begin{align}
-\frac{d^2\tilde{a}_x}{d\xi^2}+V_1(\xi)\tilde{a}_x &=\omega^2\tilde{a}_x+\frac{qgr^2}{h\sqrt{Z_1Z_2}}\tilde{b}_x,\label{eq:ax}\\
-\frac{d^2\tilde{b}_x}{d\xi^2}+V_2(\xi)\tilde{b}_x &=\omega^2\tilde{b}_x+\frac{k^2qgr^2}{h\sqrt{Z_1Z_2}}\tilde{a}_x,\label{eq:bx}
\end{align}
where $x$ is the same as in \eqref{eq:schrx} and
\begin{equation}
\begin{split}
V_i(\xi) &=\frac{1}{F^2}\left(Q_i+\frac{1}{4}P_i^2-\frac{1}{2}P_i'-\frac{3}{4}\frac{F'^2}{F^2}+\frac{1}{2}\frac{F''}{F}\right),\\
F &=\frac{1}{f},\qquad P_i= -\frac{(fZ_i)'}{fZ_i},\qquad Q_1=\frac{q^2r^2}{gZ_1}+\frac{W}{r^2gZ_1},\qquad Q_2=\frac{k^2r^2}{gZ_2}.
\end{split}\label{eq:V12}
\end{equation}
After inserting the hyperscaling-violating geometry to the equations, we obtain
\begin{align}
-\frac{d^2\tilde{a}_x}{d\xi^2}+\left(\frac{c_1}{\xi^2}+\frac{c_w}{\xi^\beta}\right)\tilde{a}_x &=\omega^2\tilde{a}_x+\frac{d_1}{\xi^\alpha}\tilde{b}_x,\label{eq:tilax}\\
-\frac{d^2\tilde{b}_x}{d\xi^2}+\left(\frac{c_2}{\xi^2}+\frac{c_3}{\xi^{2\alpha-2}}\right)\tilde{b}_x &=\omega^2\tilde{b}_x+\frac{d_2}{\xi^\alpha}\tilde{a}_x,\label{eq:tilbx}
\end{align}
where
\begin{alignat}{2}
\alpha &=2-\frac{(\gamma-\delta)(\gamma+\delta+2\lambda)}{(\gamma-\delta)(\gamma+3\delta)+4} \,&&=2-\frac{\kappa\lambda-2}{2z},\\
\beta &=2+\frac{4(\gamma-\delta)(\gamma-\delta-\eta)}{(\gamma-\delta)(\gamma+3\delta)+4} \,&&=2-\frac{\kappa\eta+2\theta-4}{z},
\end{alignat}
and $c_1$, $c_2$, $c_3$, $c_w$, $d_1$, and $d_2$ are constants.\footnote{The coefficients $c_3$, $c_w$, $d_1$, and $d_2$ are obtained from $$\frac{c_3}{\xi^{2\alpha-2}}=\frac{k^2r^2}{gZ_2},\qquad \frac{c_w}{\xi^\beta}=\frac{W}{r^2gZ_1},\qquad \frac{d_1}{\xi^\alpha}=\frac{qgr^2}{h\sqrt{Z_1Z_2}},\qquad \frac{d_2}{\xi^\alpha}=\frac{k^2qgr^2}{h\sqrt{Z_1Z_2}}.$$} The coefficients $c_1$ and $c_2$ are given by $c_1=\nu_1^2-1/4$ and $c_2=\nu_2^2-1/4$, where
\begin{alignat}{2}
\nu_1 &=\frac{(\gamma-\delta)(3\gamma+5\delta)+12}{2[(\gamma-\delta)(\gamma+3\delta)+4]} &&=\frac{3z-\theta}{2z},\\
\nu_2 &=\frac{(\gamma-\delta)(\gamma-5\delta-4\lambda)-4}{2[(\gamma-\delta)(\gamma+3\delta)+4]} &&=\frac{2-z-\theta-\kappa\lambda}{2z}.
\end{alignat}

The terms involving $\alpha$ are from the momentum dissipation sector and the term involving $\beta$  comes from the U(1) symmetry breaking. We  impose two conditions:

\begin{itemize}
\item The axions do not change the IR geometry to leading order.
After we include the axions, the effective potential for the scalar is
\begin{equation}
V_\text{tot} =V(\phi)+\frac{Y(\phi)}{2}g^{xx}k^2
\sim e^{-\delta\phi}+\frac{k^2}{2}r^2e^{\lambda\phi}
\sim r^{-\bar{\kappa}\delta}+\frac{k^2}{2}r^{2+\bar{\kappa}\lambda},\qquad \text{as }r\to\infty.
\label{vt}
\end{equation}
For the axions to not change the IR geometry, the first term must be dominant. Therefore we require
\begin{equation}
-\bar{\kappa}\delta>2+\bar{\kappa}\lambda\quad \Rightarrow\quad (\gamma-\delta)(\gamma+\delta+2\lambda)<0.
\end{equation}

\item The system is in the fractionalized phase, in which the symmetry breaking term is subleading in the IR.\footnote{In the cohesive phase, the symmetry breaking term is at the same order, which requires $\eta=\gamma-\delta$.} To implement this we demand that in \eqref{eq:Maxwell} the right hand side is subleading to the left hand side. This implies
\begin{equation}
(\gamma-\delta)(\gamma-\delta-\eta)>0.
\end{equation}
\end{itemize}

If the IR is at $\xi\to\infty$, i.e., the extremal limit is at $T\to 0$, we have $\alpha>2$ and $\beta>2$. If the IR is at $\xi\to 0$, i.e., the extremal limit is at $T\to\infty$, we have $\alpha<2$ and $\beta<2$. We will show that the terms involving $\alpha$ and $\beta$ are all subleading to the solution without those terms. This will imply that the qualitative features of the charged spectrum is not changed by the presence of the momentum-dissipating sector nor by the U(1) symmetry breaking.

We take first $c_w=0$ ($W=0$) for simplicity. We can obtain a fourth-order equation for $\tilde{a}_x$ after eliminating $\tilde{b}_x$:
\begin{multline}
\frac{d^4\tilde{a}_x}{d\xi^4}+\frac{2\alpha}{\xi}\frac{d^3\tilde{a}_x}{d\xi^3}+\left(\frac{\alpha(\alpha-1)-c_1-c_2}{\xi^3}-\frac{\epsilon c_3}{\xi^{2\alpha-2}}\right)\frac{d^2\tilde{a}_x}{d\xi^2}
+\frac{2(2-\alpha)c_1}{\xi^2}\frac{d\tilde{a}_x}{d\xi}\\
+\left(\frac{c_1(c_2-(2-\alpha)(3-\alpha))}{\xi^4}+\frac{\epsilon (c_1c_3-d_1d_2)}{\xi^{2\alpha}}\right)\tilde{a}_x=0,
\end{multline}
where $\epsilon$ is a bookkeeping parameter that keeps track of the order of perturbation and which we will eventually set to 1. The asymptotic behavior of $\tilde{a}_x$ at $\xi\to 0$ is
\begin{equation}
\tilde{a}_x=\xi^{1/2-\nu_1}\left(a_0+\epsilon a_1\xi^{2(2-\alpha)}+\epsilon^2 a_2\xi^{4(2-\alpha)}+\cdots\right).
\end{equation}
Since $\alpha<2$, the above expansion is valid and reliable. The coefficients $a_1$, $a_2$, $\cdots$ can be solved order by order. We observe that the coupled equations for $\tilde{a}_x$ and $\tilde{b}_x$ do not change the leading exponent of $\tilde{a}_x$. Therefore, if $\tilde{a}_x$ is normalizable at $\xi\to 0$ without the momentum dissipation, it is still normalizable with the momentum dissipation. The hard gap remains in the conductivity.

When the U(1) symmetry breaking term is included, the fourth-order equation for $\tilde{a}_x$ is
\begin{footnotesize}
\begin{align}
&\frac{d^4\tilde{a}_x}{d\xi^4}+\frac{2\alpha}{\xi}\frac{d^3\tilde{a}_x}{d\xi^3}+\left(\frac{\alpha(\alpha-1)-c_1-c_2}{\xi^2}-\frac{\epsilon_\alpha c_3}{\xi^{2\alpha-2}}-\frac{\epsilon_\beta c_w}{\xi^\beta}\right)\frac{d^2\tilde{a}_x}{d\xi^2}
+\left(\frac{2(2-\alpha)c_1}{\xi^3}+\frac{2\epsilon_\beta(\beta-\alpha)c_w}{\xi^{\beta+1}}\right)\frac{d\tilde{a}_x}{d\xi}\nonumber\\
&+\left(\frac{c_1(c_2-(2-\alpha)(3-\alpha))}{\xi^4}
+\frac{\epsilon_\alpha(c_1c_3-d_1d_2)}{\xi^{2\alpha}}+\frac{\epsilon_\beta c_w(c_2-(\beta-\alpha)(\beta+1-\alpha))}{\xi^{\beta+2}}+\frac{\epsilon_\alpha\epsilon_\beta c_3c_w}{\xi^{2\alpha+\beta-2}}\right)\tilde{a}_x=0,
\end{align}
\end{footnotesize}
where there are two subleading terms multiplied by $\epsilon_\alpha=\epsilon_\beta$ that in the end will be set to 1. The solution is a double expansion
\begin{equation}
\tilde{a}_x=\xi^{1/2-\nu_1}\left(a_0+\epsilon_\alpha a_{\alpha_1}\xi^{2(2-\alpha)}+\epsilon_\beta a_{\beta_1}\xi^{2-\beta}+\epsilon_\alpha\epsilon_\beta a_{\alpha_1\beta_1}\xi^{6-2\alpha-\beta}+\cdots\right),
\end{equation}
where all terms involving $\alpha$ and $\beta$ are subleading. Therefore, we confirm that in this case, the U(1) symmetry breaking term and the momentum dissipation effects do not charge the main characteristics of the spectrum: the spectrum remains discrete. The only difference is that the zero mode (responsible for the $\delta$-function at zero frequency) has now disappeared and therefore now the spectrum is gapped. We conclude that the parameter space for the gapped geometry in figure~\ref{fig:par}, for the fractionalized phase has not changed.

In the following we will discuss the solutions of $a_x$ and $b_x$ in the IR. We will impose the Gubser criterion so that the extremal limit is at the $T\to\infty$ end of the small black-hole branch ($\xi\to 0$). The general solutions of $\tilde{a}_x$ and $\tilde{b}_x$ are
\begin{align}
\tilde{a}_x &=C_1\xi^{1/2-\nu_1}+C_2\xi^{1/2+\nu_1},\\
\tilde{b}_x &=D_1\xi^{1/2-\nu_2}+D_2\xi^{1/2+\nu_2}.
\end{align}
For the system to be holographically reliable  we must impose the condition that only one normalizable solution exists for each of $\tilde{a}_x$ and $\tilde{b}_x$. This requires $|\nu_1|>1$ and $|\nu_2|>1$. We have the following observations:
\begin{itemize}
\item $\nu_1<0$ is always satisfied, but $\nu_1<-1$ gives nontrivial constraints for $(\gamma,\delta)$, as discussed before.
\item $\nu_2$ can be either positive or negative, and $|\nu_2|>1$ gives nontrivial constrains for $(\gamma,\delta)$.
\end{itemize}

In the UV, substituting in \eqref{eq:phibry}, the near-boundary expansion  of the potentials in \eqref{eq:V12}  is
\begin{align}
V_1= &\,\frac{1}{2}\Delta(\Delta-1)\gamma\phi_ar^{\Delta-2}+\frac{1}{4}(\Delta^2\gamma^2+4W_0\eta^2)\phi_a^2 r^{2(\Delta-1)},\nonumber\\
&+\frac{1}{2}(2-\Delta)(3-\Delta)\gamma\phi_br^{1-\Delta}+\frac{1}{4}((3-\Delta)^2\gamma^2+4W_0\eta^2)\phi_b^2r^{2(2-\Delta)},\\
V_2= &-\frac{1}{2}\Delta(\Delta+1)\lambda\phi_ar^{\Delta-2}+\frac{1}{4}\Delta^2\lambda^2\phi_a^2r^{2(\Delta-1)}\nonumber\\
&-\frac{1}{2}(3-\Delta)(4-\Delta)\lambda\phi_br^{1-\Delta}+\frac{1}{4}(3-\Delta)^2\lambda^2\phi_b^2r^{2(2-\Delta)}.
\end{align}
The UV behavior of $a_x$ and $b_x$ is
\begin{equation}
a_x=a_x^{(0)}+a_x^{(1)}r+\cdots,\qquad b_x=b_x^{(1)}r+\cdots.
\end{equation}
The Schr\"{o}dinger potentials $V_1$ and $V_2$ are divergent when $1/2<\Delta<2$, except for $\Delta=1$ or $2$, for which they asymptote to constants. The analysis is similar to the last paragraph of section~\ref{sec:IR-charged}.

\section{On holographic Mott-like insulators and supersolids}
\label{sec:DC-conductivity}
We will first consider the EMD bulk action with no U(1) symmetry breaking ( $W=0$ in (\ref{li3})). To decide whether the system is a metal or insulator, we must calculate its DC conductivity. With momentum dissipation, a formula for the DC conductivity was derived in \cite{Blake:2013bqa}. We will briefly review the procedure, and use it for the near-extremal geometries. Then we will apply this formulation to extremal geometries. We define for latter convenience
\begin{equation}
H(r)=Z_1+\frac{q^2}{k^2}Z_2\equiv Z+\frac{q^2}{k^2CY}.
\end{equation}
The two eigenmodes of the equations for $a_x$ and $b_x$ are
\begin{equation}
\lambda_1=\frac{Z_1}{H}\left(a_x-\frac{q}{k^2}\frac{Z_2}{Z_1}b_x\right),\qquad
\lambda_2=\frac{Z_2}{H}\left(qa_x+b_x\right).
\end{equation}
Following \cite{Blake:2013bqa,Gouteraux:2014hca}, we can derive an equation
\begin{equation}
\left[fH\lambda_1'-\frac{q}{k^2}fZ_2\left(\frac{Z_1}{Z_2}\right)'\lambda_2\right]'+\frac{\omega^2}{f}H\lambda_1=0.\label{eq:lambda1}
\end{equation}
From this we observe that the equations for $a_x$ and $b_x$ have a massless eigenmode.
Therefore, we define
\begin{equation}
\Pi=fH\lambda_1'-\frac{q}{k^2}fZ_2\left(\frac{Z_1}{Z_2}\right)'\lambda_2,
\end{equation}
and $\Pi$ is a radially conserved quantity at $\omega=0$. At the UV, we have $H=1$,\footnote{As we set $Z_0=1$.} $\lambda_1=a_x^{(0)}$, $\lambda_2=0$, and $\Pi=a_x^{(1)}$ in the leading order. The conductivity is
\begin{equation}
\sigma(\omega)=\frac{a_x^{(1)}}{i\omega a_x^{(0)}}=\left.\frac{\Pi}{i\omega\lambda_1}\right|_{r\to 0}.\label{eq:dcdef}
\end{equation}
At $\omega=0$, $\Pi$ can be evaluated at the horizon,
\begin{equation}
\sigma_\text{DC}=\left.\frac{\Pi}{i\omega\lambda_1}\right|_{r=r_h}.\label{eq:sigmaPi}
\end{equation}
Refs.~\cite{Gouteraux:2014hca,Donos:2014uba} derived the DC conductivity at finite temperature as
\begin{equation}
\sigma_\text{DC}=Z_h+\frac{q^2}{k^2C_hY_h}=e^{\gamma\phi_h}+\frac{q^2}{k^2C_he^{\lambda\phi_h}},\label{eq:sigmaDC}
\end{equation}
where $\phi_h\equiv\phi(r_h)$ and $C_h\equiv C(r_h)$ are the values of the scalar (dilaton) and the scale factor on the black-hole horizon. At finite temperature, $\sigma_\text{DC}$ is finite. The formula in (\ref{eq:sigmaDC}) however cannot be directly applied to extremal geometries. We will discuss the finite temperature, near-extremal geometries first, and then the extremal geometries.

\subsection{DC conductivity for near-extremal black holes}
We now return to the analysis of the nature of the conductivity in the hyperscaling violating Lifshitz geometries perturbed by U(1) breaking as well as by the momentum dissipating sector.

\begin{figure}
  \centering
  \includegraphics[width=0.29\textwidth]{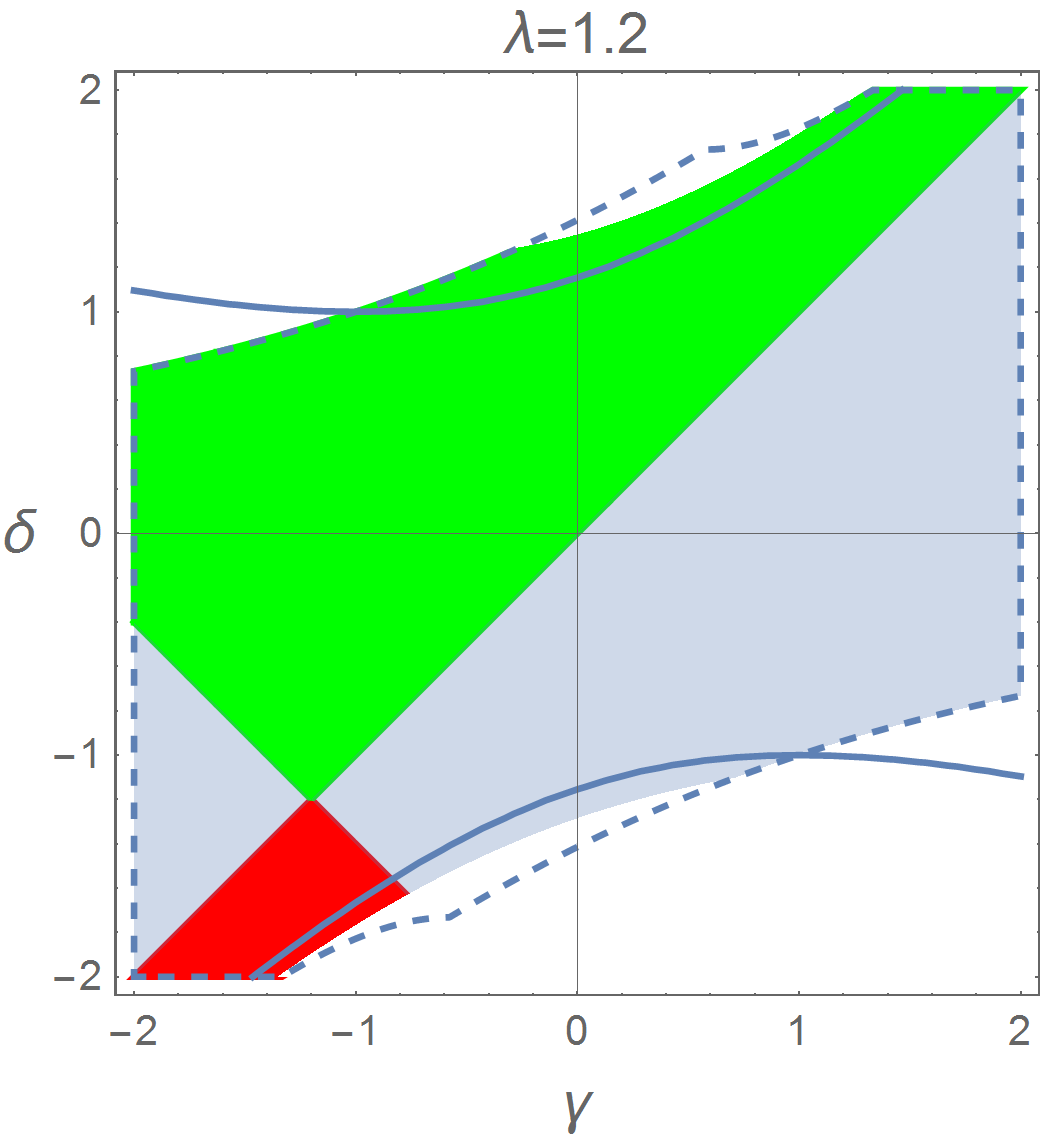}\qquad
  \includegraphics[width=0.29\textwidth]{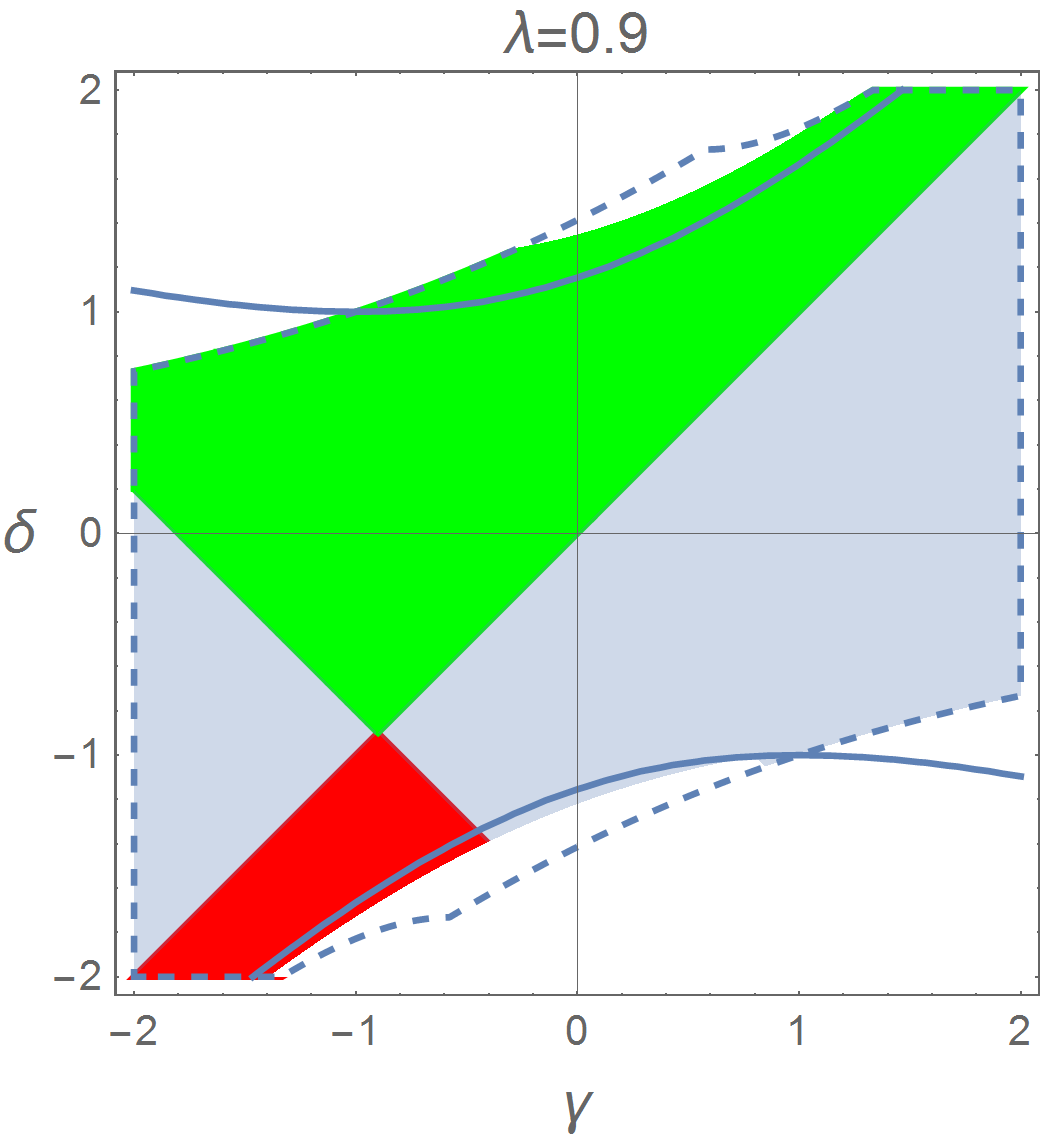}\qquad
  \includegraphics[width=0.29\textwidth]{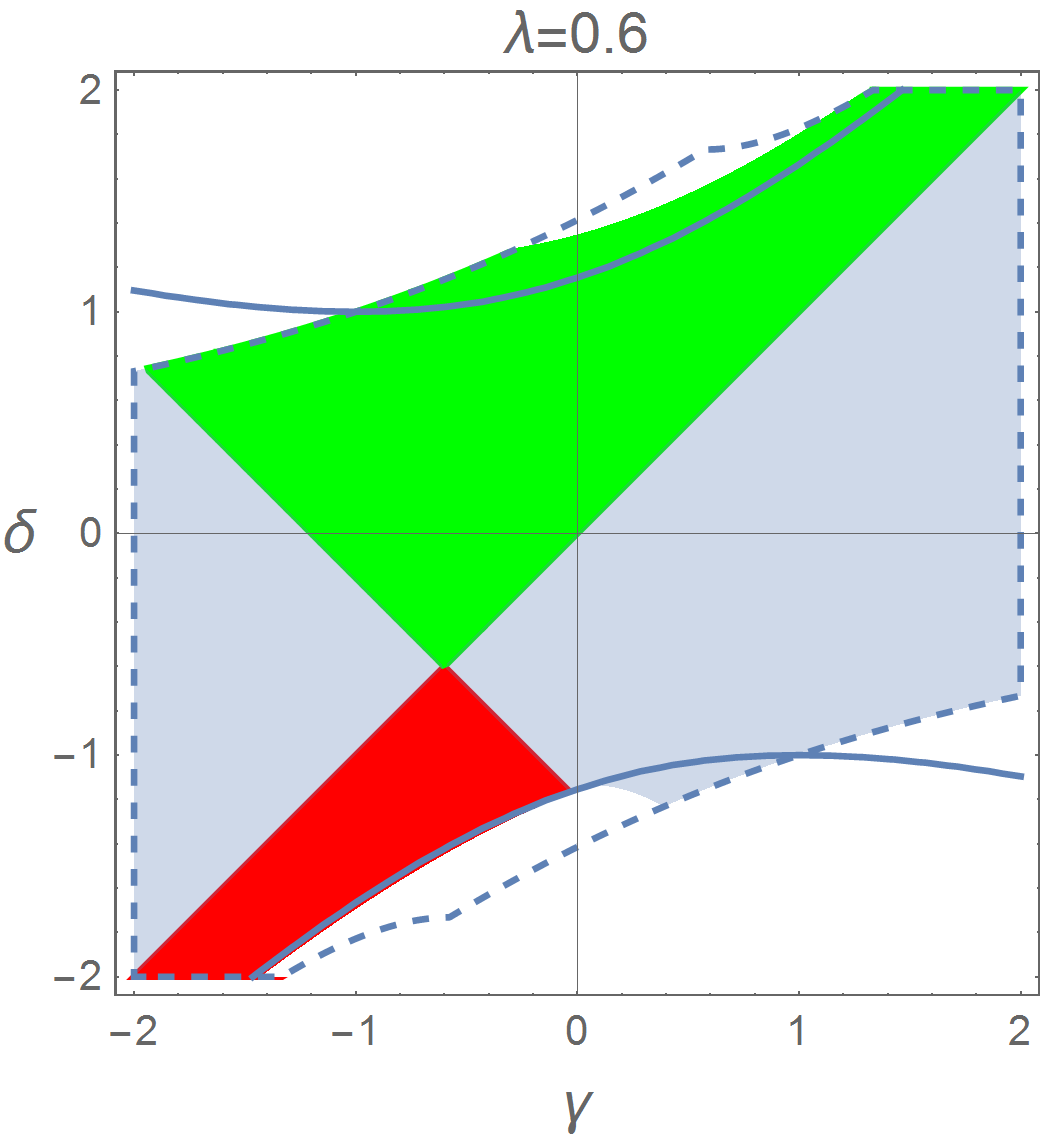}\\
  \vspace{10pt}
  \includegraphics[width=0.29\textwidth]{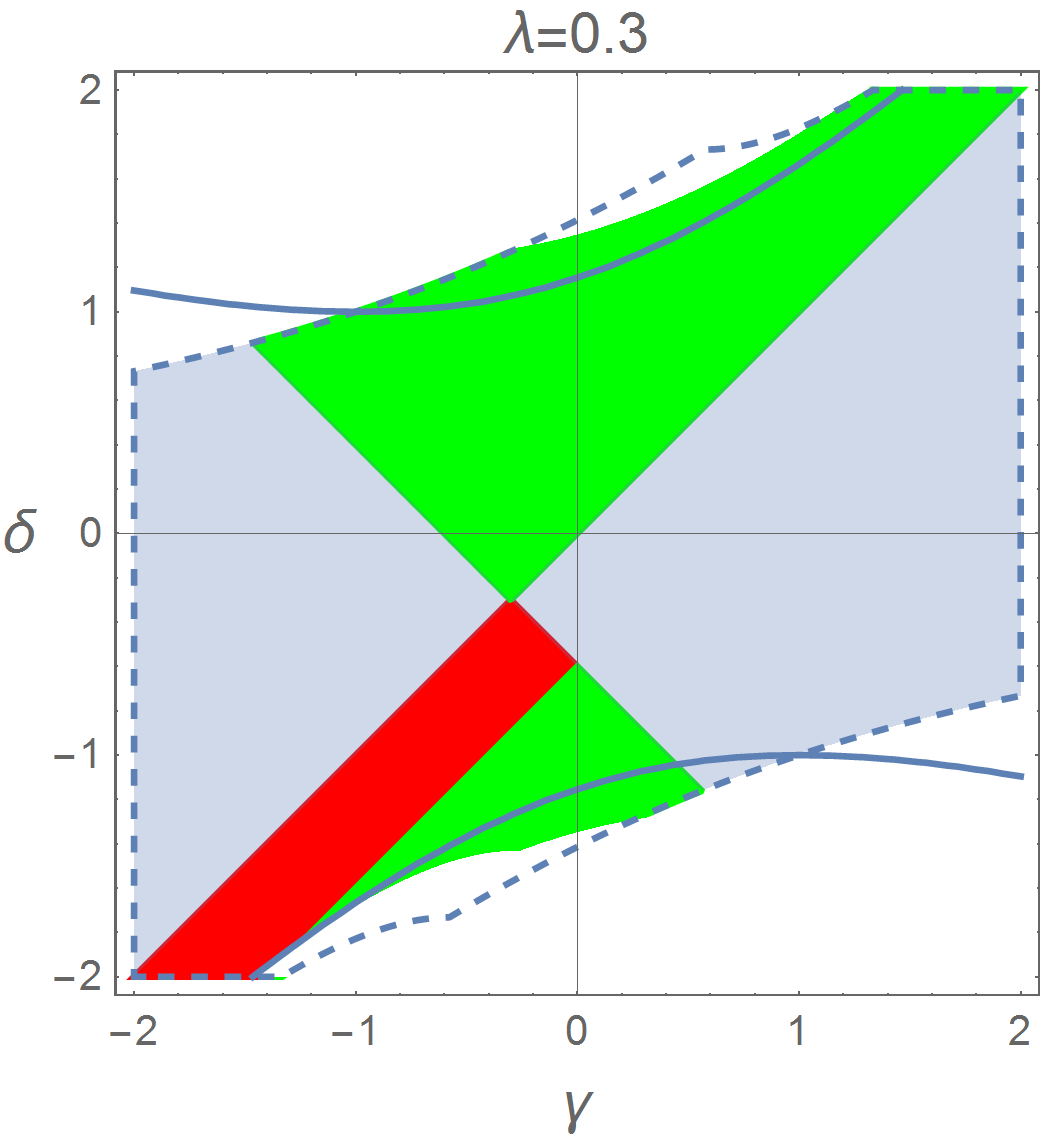}\qquad
  \includegraphics[width=0.29\textwidth]{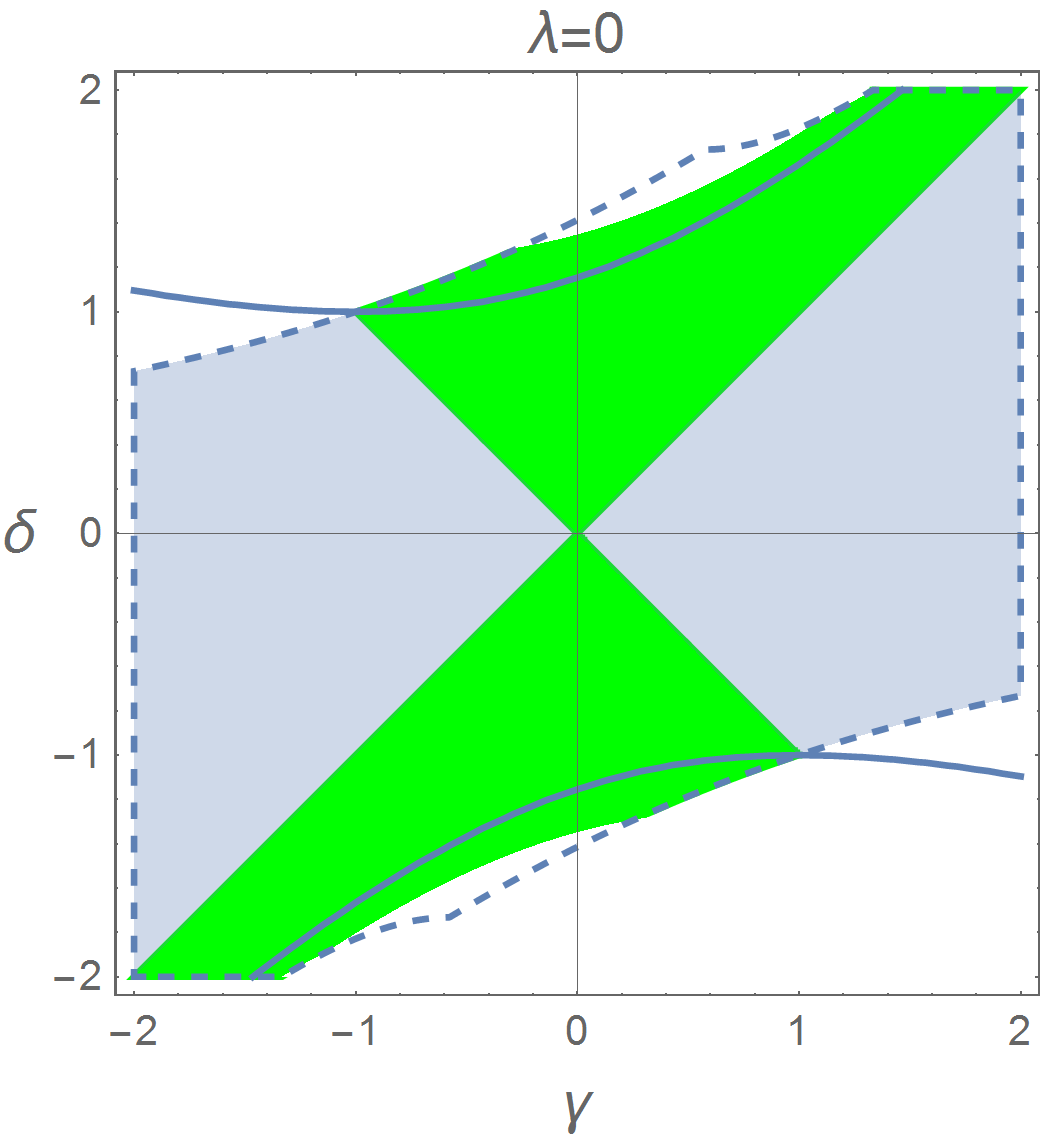}\qquad
  \includegraphics[width=0.29\textwidth]{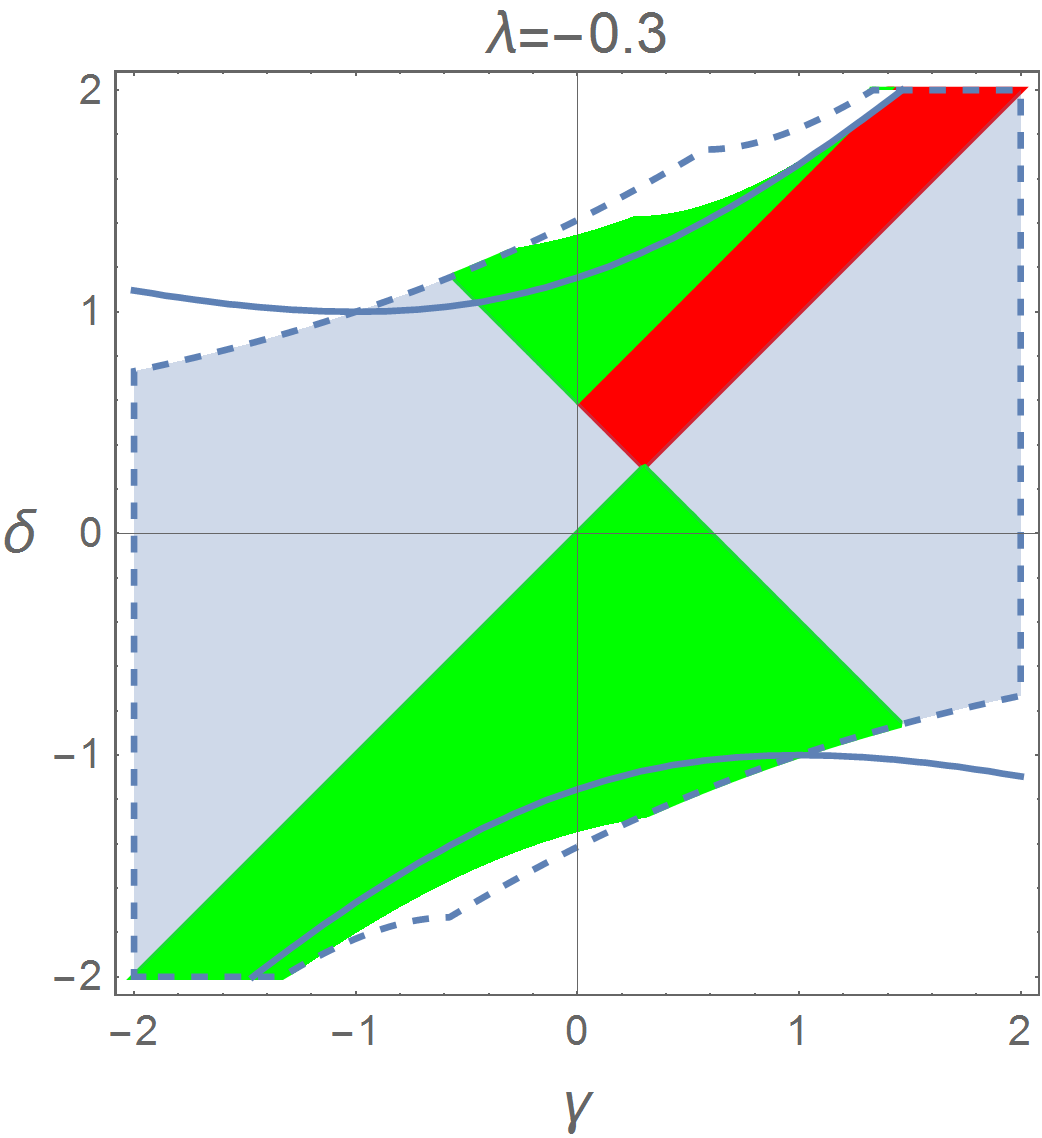}\\
  \vspace{10pt}
  \includegraphics[width=0.29\textwidth]{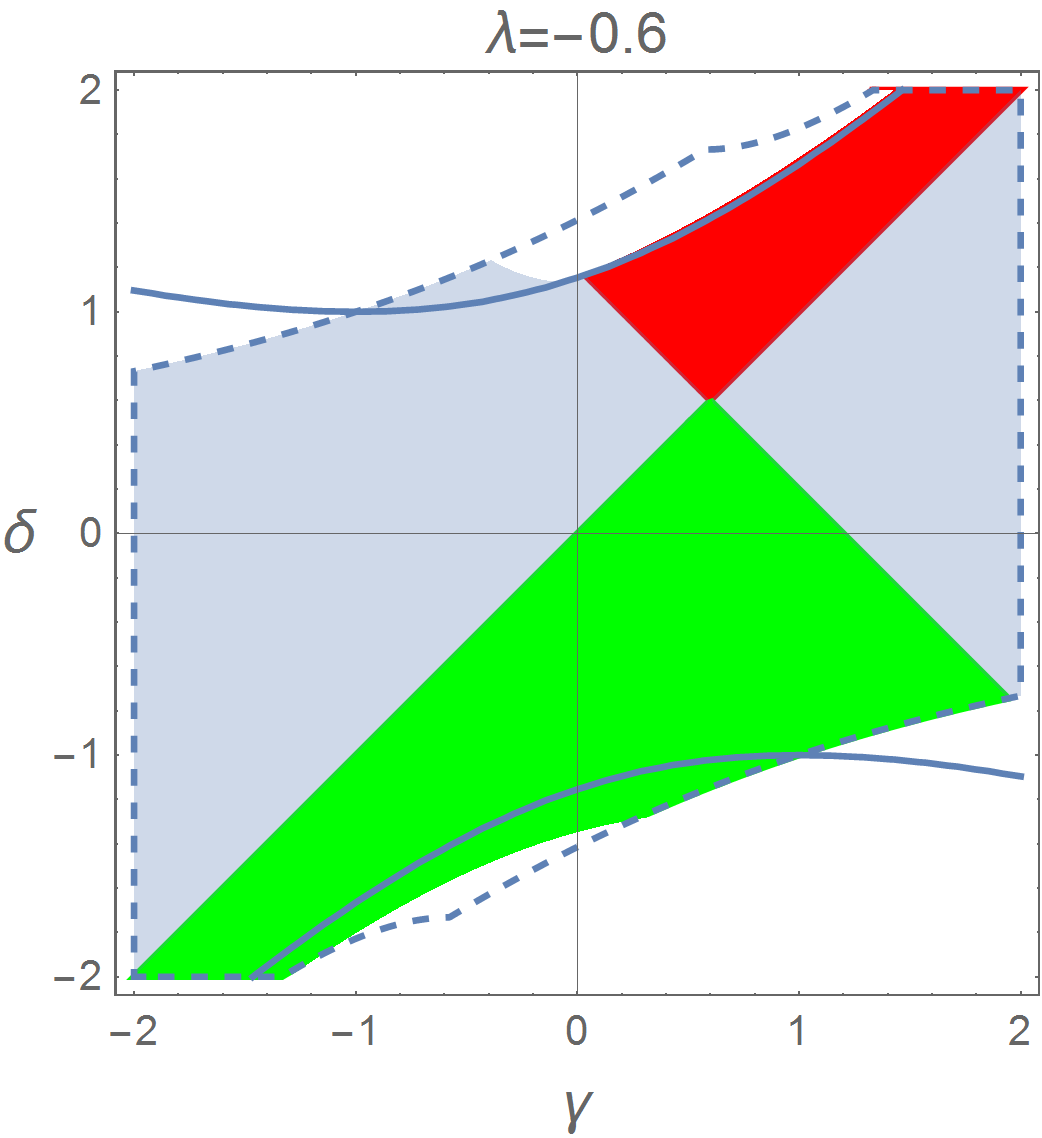}\qquad
  \includegraphics[width=0.29\textwidth]{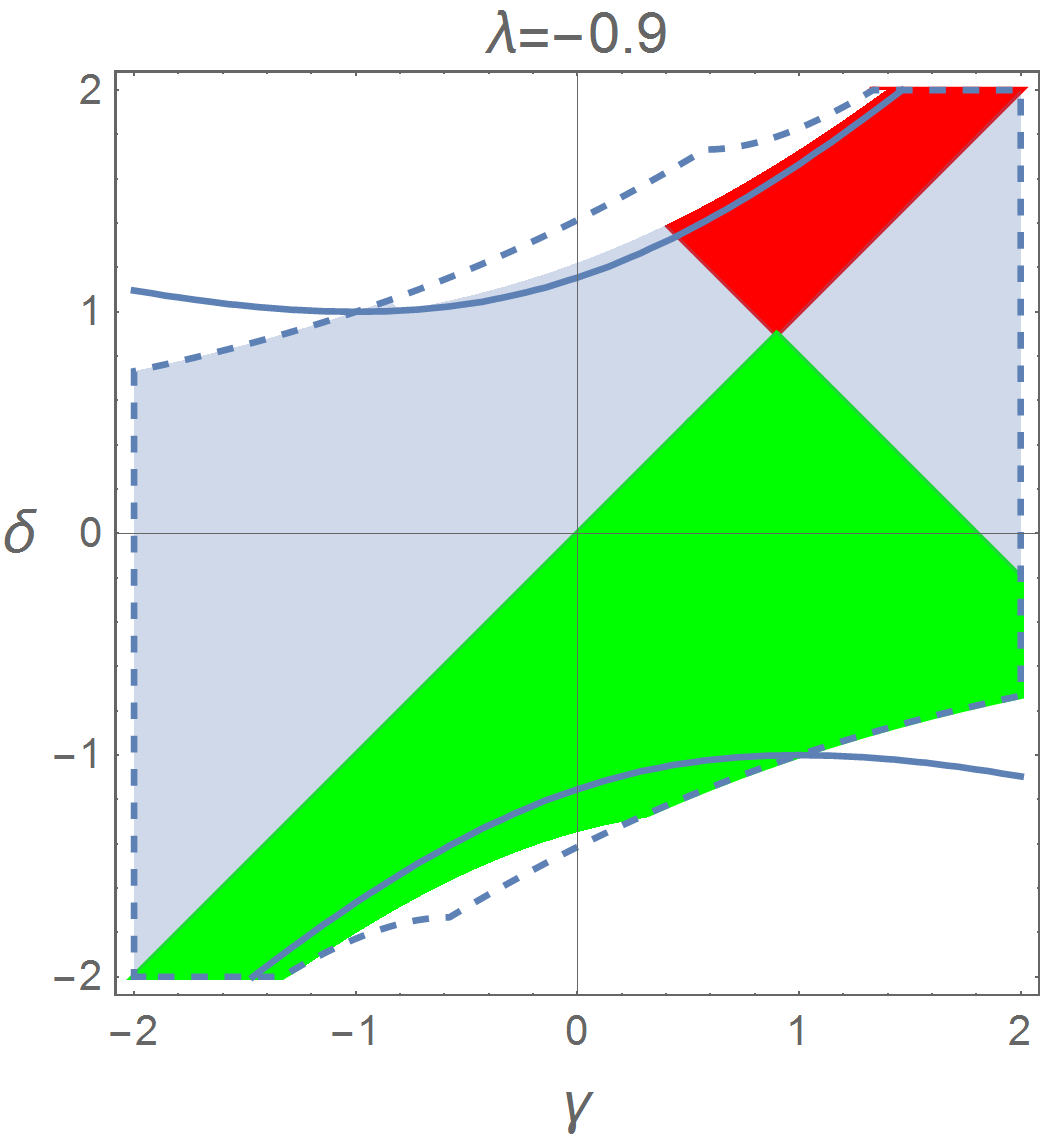}\qquad
  \includegraphics[width=0.29\textwidth]{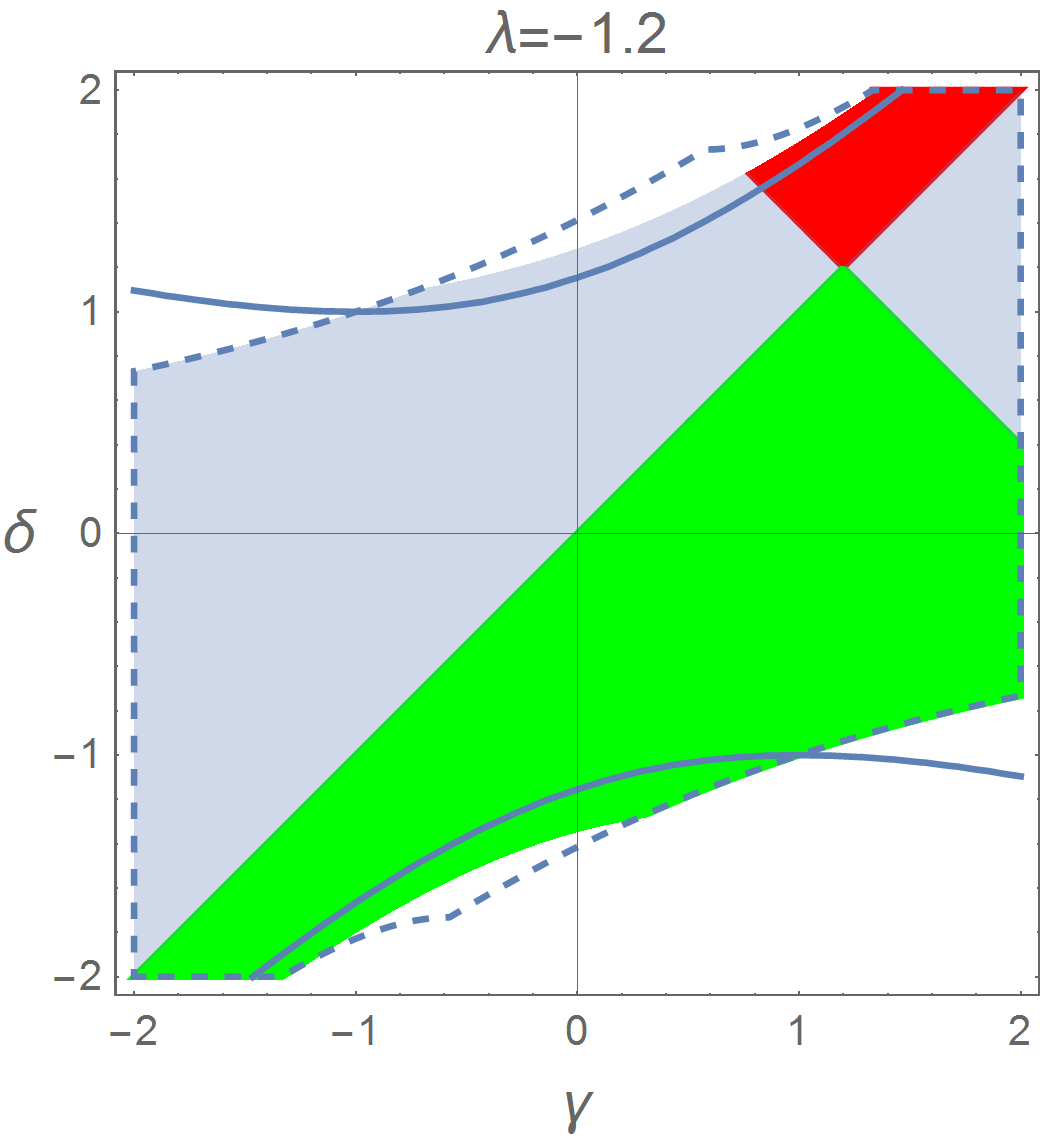}
  \caption{\label{fig:lambda} Parameter space for the conducting and insulating phases. The parameter space constrained by the Gubser criterion is enclosed by dashed lines; moreover, we have excluded the holographically unreliable region in which there are two normalizable solutions in the IR. In the gray region, the axions term is relevant in the IR. In the green region, the system is a conductor; in the red region, the system is an insulator. The parameter space for the gapped geometry is above and below the two curves $(\gamma-\delta)(\gamma+3\delta)+4=0$.}
\end{figure}

We apply the formula \eqref{eq:sigmaDC} for the DC conductivity to the near-extremal geometries, and then take the extremal limit. The near-extremal solution from \cite{cgkkm} is\footnote{The $AdS_4$ near-extremal solution in coordinates \eqref{eq:ansatz} is (we need to add appropriate coefficients before $dt^2$ and $dr^2$, as in eq.~\eqref{eq:ztheta})
\begin{equation}
ds^2=\frac{1}{r^2}\left(-fr^{-\frac{4(z-1)}{2-\theta}}dt^2+r^\frac{2\theta}{2-\theta}\frac{dr^2}{f}+d\mathbf{x}^2\right),\qquad f=1-\Bigl(\frac{r}{r_h}\Bigr)^\frac{2(2+z-\theta)}{2-\theta}.
\end{equation}
The power in the blackness function $f(r)$ is always positive as required by the Gubser criterion. The extremal solution can be obtained by sending $r_h\to\infty$.}
\begin{equation}
\begin{split}
ds^2 &=-V(\tilde{r})\tilde{r}^{-4\frac{\gamma(\gamma-\delta)}{wu}}dt^2+\frac{e^{\delta\phi}d\tilde{r}^2}{-w\Lambda V(\tilde{r})}+\tilde{r}^{2\frac{(\gamma-\delta)^2}{wu}}(dx^2+dy^2),\\
e^\phi &=e^{\phi_0}\tilde{r}^{-4\frac{(\gamma-\delta)}{wu}},\qquad
A_t =2\sqrt{\frac{-v}{wu}}e^{-\frac{\gamma}{2}\phi_0}(\tilde{r}-\tilde{r}_h)\quad
V(\tilde{r}) =\tilde{r}(\tilde{r}-\tilde{r}_h),\\
wu &=3\gamma^2-2\gamma\delta-\delta^2+4,\qquad u=\gamma^2-\gamma\delta+2,\qquad v=\delta^2-\gamma\delta-2.
\end{split}
\end{equation}
Note that this $\tilde{r}$ is different from the $r$ in \eqref{eq:ansatz}. Thus we obtain
\begin{equation}
\sigma_\text{DC}\sim \tilde{r}_h^{-\frac{4\gamma(\gamma-\delta)}{3\gamma^2-2\gamma\delta-\delta^2+4}}
+\frac{q^2}{k^2}\tilde{r}_h^{-\frac{2(\gamma-\delta)(\gamma-\delta-2\lambda)}{3\gamma^2-2\gamma\delta-\delta^2+4}}
\equiv \tilde{r}_h^{\alpha_1}+\frac{q^2}{k^2}\tilde{r}_h^{\alpha_2}.
\end{equation}
In terms of $z$ and $\theta$,
\begin{equation}
\alpha_1=\frac{\theta-4}{2+z-\theta},\qquad \alpha_2=\frac{\theta-2+\kappa\lambda}{2+z-\theta}.
\end{equation}
The extremal limit is $r_h\to 0$ keeping the other coordinates fixed. The Gubser criterion and the condition that the axions are irrelevant in the IR imply that the second term is always dominant:
\begin{equation}
\alpha_1-\alpha_2=-\frac{2(\gamma-\delta)(\gamma+\delta+2\lambda)}{3\gamma^2-2\gamma\delta-\delta^2+4}>0.
\end{equation}
Therefore, the DC conductivity is always dissipation-dominated in the extremal limit. If $\sigma_\text{DC}\to 0$ as $\tilde{r}_h\to 0$, the system is an insulator; if $\sigma_\text{DC}\to\infty$, the system is a (perfect) conductor. The condition that the system is an insulator is $\alpha_2<0$, i.e.,
\begin{equation}
(\gamma-\delta)(\gamma-\delta-2\lambda)<0.
\end{equation}
For different values of the parameter $\lambda$ that controls the IR strength of momentum dissipation, the parameter spaces for insulators and conductors are plotted in figure~\ref{fig:lambda}. In the green region, the system is a conductor, and in the red region, the system is an insulator. Finally in the light blue region the axions are relevant in the IR, a case that was analyzed further in \cite{Donos:2014uba,Gouteraux:2014hca}, but we do not analyze it further here.

The relation between the temperature $T$ and the horizon radius $\tilde{r}_h$ is \cite{cgkkm}
\begin{equation}
T=\frac{1}{4\pi}\sqrt{-w\Lambda}e^{-\frac{\delta}{2}\phi_0}\tilde{r}_h^{1-2\frac{(\gamma-\delta)^2}{wu}}
\sim \tilde{r}_h^\frac{\gamma^2+2\gamma\delta-3\delta^2+4}{3\gamma^2-2\gamma\delta-\delta^2+4}.
\end{equation}
Thus, the DC conductivity for the near-extremal black hole is
\begin{equation}
\sigma_\text{DC}\sim T^{-\frac{4\gamma(\gamma-\delta)}{\gamma^2+2\gamma\delta-3\delta^2+4}}
+\frac{q^2}{k^2}T^{-\frac{2(\gamma-\delta)(\gamma-\delta-2\lambda)}{\gamma^2+2\gamma\delta-3\delta^2+4}}
\equiv T^{\beta_1}+\frac{q^2}{k^2}T^{\beta_2}.
\end{equation}
In terms of $z$ and $\theta$,
\begin{equation}
\beta_1=\frac{\theta-4}{z}=\frac{\zeta-2}{z},\qquad \beta_2=\frac{\theta-2+\kappa\lambda}{z},
\end{equation}
where $\zeta$ is the conduction exponent given in \eqref{eq:zeroth1}, \cite{gk2,Gouteraux:2013oca}. As mentioned above, the second term is always dominant no matter whether the extremal limit is $T\to 0$ or $T\to\infty$.

We are especially interested in Mott-like insulators. In figure~\ref{fig:lambda}, the parameter space in which the conductivity can have a hard gap and a discrete spectrum is above and below the two curves $(\gamma-\delta)(\gamma+3\delta)+4=0$ (which gives $\frac{2-\theta}{z}=0$). This is in the region B in figure~\ref{fig:par}. From figure~\ref{fig:lambda}, we can see that this region can be either a perfect conductor or an insulator, depending on the value of $\lambda$. In sum, in the red region of the parameter space for the gapped geometry, the system is a Mott insulator, if the scaling dimension of the scalar operator in the UV satisfies $1/2<\Delta<2$.

If the U(1) symmetry is spontaneously broken, there will be a $\delta$-function at $\omega=0$ even in the presence of momentum dissipation (the superfluid  mode) provided that the breaking is spontaneous and we will have therefore a superfluid (or superconductor if the U(1) is weakly gauged)  with a discrete spectrum.
This will provide a system that it is in the broken phase ($ \phi$ there will be interpreted as the modulus of a charged scalar), and will therefore have a $\delta$-function at zero frequency in the conductivity but will have discrete spectrum otherwise!
This characteristic ressembles what is expected from the conductivity in supersolids (although there are also differences that were expanded upon in the introductory sections).

\subsection{DC conductivity for gapped, extremal geometries}
For extremal geometries, the formula for the DC conductivity \eqref{eq:sigmaDC} cannot be directly applied. When the extremal limit is at $T\to 0$, the $\omega^2$ terms in the perturbation equations can not be neglected to calculate the $\omega\to 0$ behavior of the conductivity; we need to use the asymptotic match. On the contrary, when the extremal limit is at $T\to\infty$, we can neglect the $\omega^2$ terms. Moreover, for gapped geometries, the IR solution is independent of $\omega$, and there is no in-falling wave boundary condition. From \eqref{eq:dcdef}, we can evaluate the radially conserved quantity $\Pi$ at the IR, and the $\omega\to 0$ limit of the conductivity is
\begin{equation}
\sigma_\text{DC}=\frac{\Pi(r\to\infty)}{i\omega\lambda_1^{(0)}},\label{eq:sigmaPi}
\end{equation}
where $\lambda_1^{(0)}=a_x^{(0)}$ is the source of the perturbation, and $\Pi$ can be regarded as the strength (coefficient) of the zero-frequency $\delta$-function.

The condition that the presence of axion-generated  dissipation do not change the original IR solution is equivalent to
\begin{equation}
Z_1/Z_2\to 0\qquad\text{as}\qquad r\to\infty.
\end{equation}
Another important observation is that we always have\footnote{We do not always have $a_x/b_x\to 0$ as $\xi\to 0$.}
\begin{equation}
\frac{Z_1}{Z_2}\frac{a_x}{b_x}\sim\sqrt{\frac{Z_1}{Z_2}}\,\xi^{|\nu_1|-|\nu_2|}\to 0,
\end{equation}
no matter $\nu_2>0$ or $\nu_2<0$. Consequently, the leading behavior of $\Pi$ in the IR is
\begin{equation}
\Pi=-\frac{q}{k^2}fZ_2b_x'+\cdots
\end{equation}
If $\Pi\to 0$ as $r\to\infty$, the conductivity does not have a $\delta$-function at $\omega=0$. If $\Pi$ is constant as $r\to\infty$, the conductivity has a $\delta$-function at $\omega=0$ and its coefficient is proportional to $\Pi(\omega=0)$. That there is only one normalizable solution for $b_x$ requires $|\nu_2|>1$. There are two cases to consider as follows:

\begin{itemize}
\item Case 1: $\nu_2>1$, for which we have $(\gamma-\delta)(\gamma-5\delta-4\lambda)-4<0$. The solution for $b_x$ is $b_x\sim\sqrt{Z_2}\xi^{1/2+\nu_2}$
The radially conserved quantity $\Pi$ evaluated at the IR is
\begin{equation}
\Pi\sim fZ_2b_x'\sim r^\frac{(\gamma-\delta)(\gamma-5\delta-4\lambda)-4}{(\gamma-\delta)^2}\to 0.
\end{equation}

\item Case 2: $\nu_2<-1$, for which we have $(\gamma-\delta)(\gamma-5\delta-4\lambda)-4>0$. The solution for $b_x$ is $b_x\sim\sqrt{Z_2}\xi^{1/2-\nu_2}$. The radially conserved quantity evaluated in the IR is
\begin{equation}
\Pi\sim fZ_2b_x'\to\text{constant}.
\end{equation}
\end{itemize}

We expect that the insulating phase corresponds to the case 1, and the conducting phase corresponds to the case 2. We will show that this is true after we excluded the holographically unacceptable region $|\nu_2|<1$ and have imposed Gubser's criterion.

{\it Gapped, insulating phase.} The condition that the extremal geometries are at $T\to\infty$ limit of the small black-hole branch is $(\gamma-\delta)(\gamma+3\delta)+4<0$. The condition that the system is an insulator is $(\gamma-\delta)(\gamma-\delta-2\lambda)<0$. These two inequalities give $(\gamma-\delta)(3\gamma+\delta-4\lambda)+4<0$. Consequently,
\begin{equation}
\nu_2+1=\frac{(\gamma-\delta)(3\gamma+\delta-4\lambda)+4}{(\gamma-\delta)(\gamma+3\delta)+4}>0.
\end{equation}
If we further impose that there is only one normalizable solution, which is $|\nu_2|>1$, we obtain $\nu_2>1$. According to the discussion above, this corresponds to the case 1. The conductivity does not have a $\delta$-function at $\omega=0$ in the gapped, insulating phase. Note that this is not always true without the constraint $|\nu_2|>1$.

{\it Gapped, conducting phase.} The condition that the extremal geometries are at $T\to\infty$ is $(\gamma-\delta)(\gamma+3\delta)+4<0$. The condition that the system is a conductor is $(\gamma-\delta)(\gamma-\delta-2\lambda)>0$. These two inequalities give $(\gamma-\delta)(\gamma+11\delta+4\lambda)+12<0$. Consequently,
\begin{equation}
\nu_2-1=-\frac{(\gamma-\delta)(\gamma+11\delta+4\lambda)+12}{(\gamma-\delta)(\gamma+3\delta)+4}<0.
\end{equation}
If we further impose that there is only one normalizable solution, which is $|\nu_2|>1$, we obtain $\nu_2<-1$. According to the discussion above, this corresponds to the case 2. The conductivity has a $\delta$-function at $\omega=0$ in the gapped conducting phase.

\section{Constructing the finite temperature geometry}
\label{sec:finite-temp}
To determine which is the ground state of the system \eqref{eq:LVZ}, we need to solve for the finite temperature geometry, and lower the temperature to obtain the near-extremal geometry. Note that the RN black hole is not a solution of the system \eqref{eq:LVZ} when $Z'(\phi)|_{\phi=0}\neq 0$.

If we want to choose $\Delta=1$ for general values of $\delta$ without $\log r$ terms in the boundary, we can use a three-exponential potential. The simplest potential is
\begin{equation}
V(\phi)=-6-\frac{4}{\delta^2}\sinh^2\frac{\delta\phi}{2}.\label{eq:3exp}
\end{equation}
The reason why we choose this potential is explained in appendix~\ref{sec:potential}.

After eliminating $A_t'$ by $\sqrt{h}Z(\phi)A_t'=-\rho$, where $\rho$ is the charge density, we solve the three coupled equations for $\phi$, $g$, and $h$. We work in the canonical ensemble. At finite temperature, we can set the horizon be at $r_h=1$ by a rescaling of coordinates. The asymptotic behavior of the scalar field $\phi$ near the AdS boundary is
\begin{equation}
\phi=\phi_ar+\phi_br^2+\cdots.
\end{equation}
We consider the UV scaling dimension $\Delta=1$, so $\phi_a$ is the expectation value, and $\phi_b$ is the source. The way to impose the boundary condition in the UV is not unique, for example,
\begin{itemize}

  \item $\phi_b=\phi_0$. This is the standard Dirichlet case. Here $\phi_0$ has the dimension of energy. We have a two-parameter family of the solutions $(\rho,\phi_0)$.

  \item $\phi_b/\phi_a=\kappa$. This corresponds to a double-trace perturbation by the scalar operator dual to $\phi$. In this case, $\kappa$ has the dimension of energy. We have a two-parameter family of the solutions $(\rho,\kappa)$.

  \item $\phi_b/\phi_a^2=\tau$. This corresponds to a triple-trace perturbation by the scalar operator dual to $\phi$. In this case, the only quantity that has a dimension is the charge density. We have a two-parameter family of the solutions $(\rho,\tau)$.
\end{itemize}
On the scalar we will use boundary conditions of the third type.\footnote{The bulk action that has the gapped solution, has also an AdS$_2$ IR critical solution. It is RG-unstable. The bulk action that has the gapless solution has also an AdS$_2$ IR critical solution which  is RG-stable. However we do not reach it for the values of the sources we explored in our numerics.
 We should also note that the parameters of the action with the gapped solution were chosen slightly outside the region determined in the previous section for reasons of numerical reliability.} We further need four boundary conditions as in the following table (``X" means no boundary condition).

\begin{center}
\renewcommand{\arraystretch}{1.2}
\begin{tabular}{c|c|c}
\hline
& horizon ($r=1$) & AdS boundary ($r=0$)\\\hline
$\phi$ & regularity & $\phi''-2\tau\phi'^2=0$ ($\Delta=1$)\\\hline
$A_t$ & $A_t=0$ & $A_t'=-\rho$\\\hline
$g$ & $g=0$ & X\\\hline
$h$ & X & $h=1$\\\hline
\end{tabular}
\end{center}
\vskip 1cm

The entropy is $S=4\pi r_h^2V=4\pi V$, as we have set $r_h=1$, where $V$ is the volume (area) in the spatial dimensions. The dimensionless temperature and entropy density are
\begin{equation}
\tilde{T}=\frac{T}{\sqrt{\rho}}=-\frac{g'_h}{4\pi\sqrt{h_h\rho}},\qquad \tilde{s}=\frac{s}{\rho}=\frac{4\pi}{\rho}.
\end{equation}
The extremal limit is $T\to 0$ or $T\to\infty$. Although  the RN black hole has finite entropy at zero temperature, the other cases studied here have $S\to 0$ in the extremal limit, which is obtained by sending $\rho\to\infty$. When the geometry is near-extremal, we obtain
\begin{equation}
\tilde{S}\propto\tilde{T}^\frac{2-\theta}{z}.
\end{equation}
The free energy in the canonical ensemble is
\begin{equation}
\frac{F}{V}=\epsilon-Ts,
\end{equation}
where $\epsilon$ is the energy density from
\begin{equation}
\left.\frac{g}{h}\right|_{r\to 0}=1-\frac{\epsilon}{2}r^3+\cdots.\label{eq:epsilon}
\end{equation}

\begin{figure}
  \includegraphics[height=0.3\textwidth]{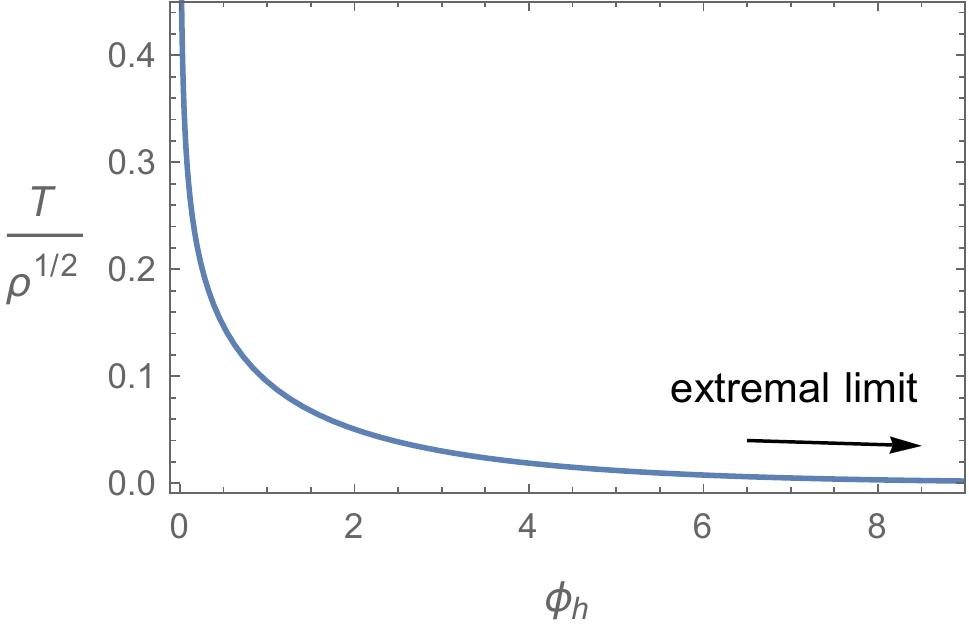}\qquad
  \includegraphics[height=0.3\textwidth]{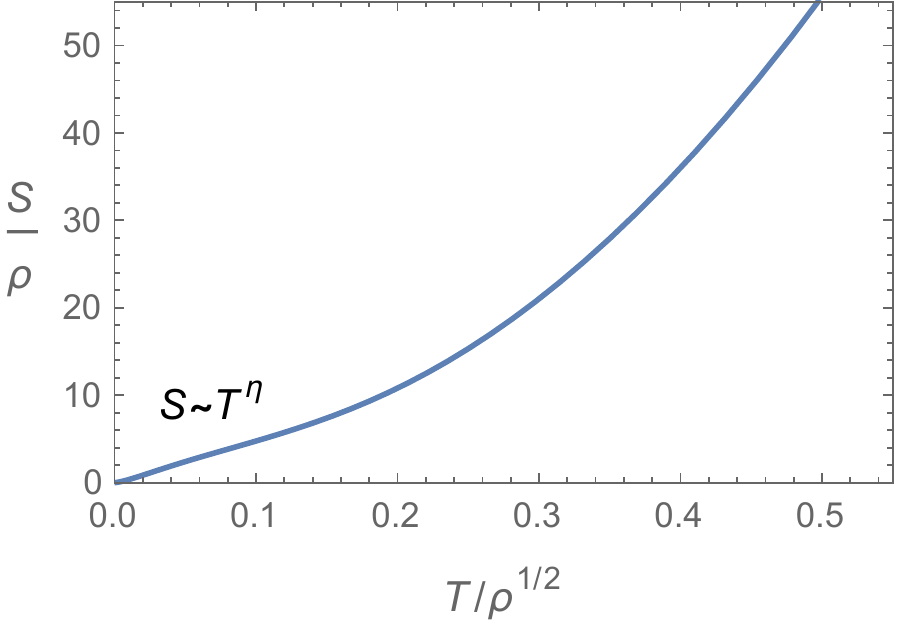}\vspace{10pt}\\
  \includegraphics[height=0.3\textwidth]{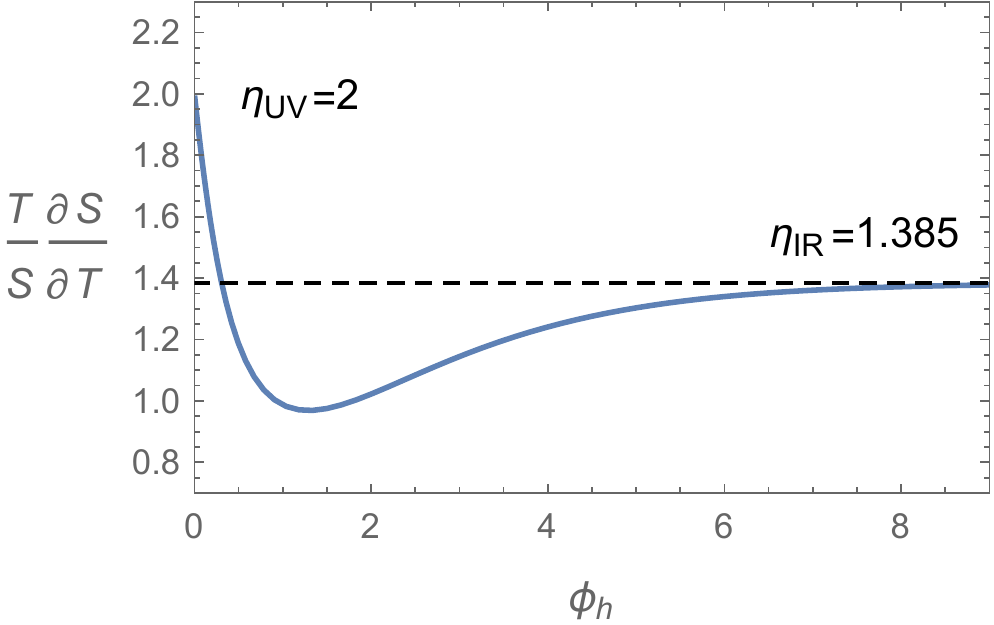}\quad
  \includegraphics[height=0.3\textwidth]{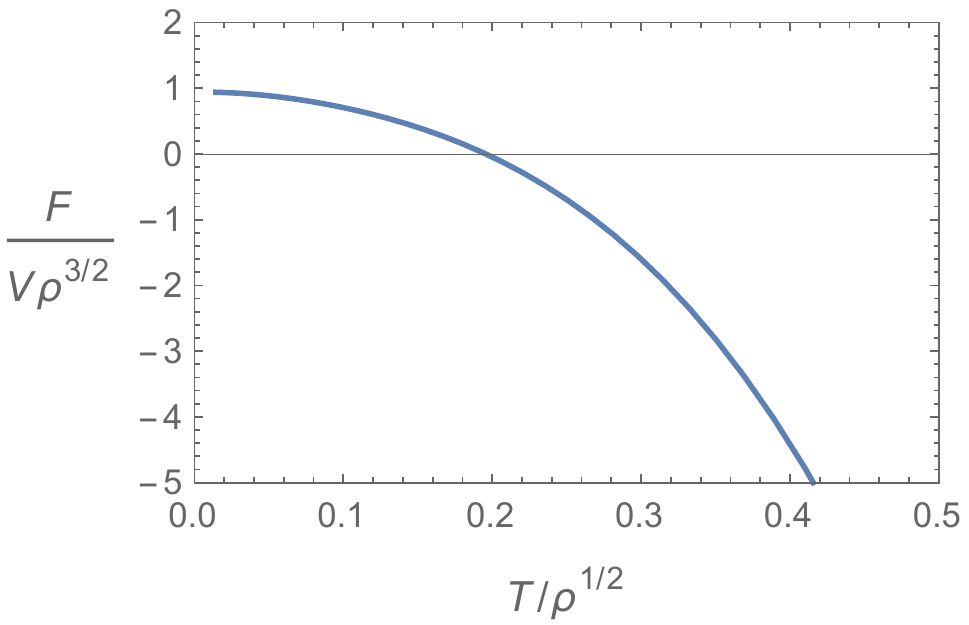}\quad
  \caption{Numerical results for physical properties of a gapless geometry. The parameters are chosen by $\gamma=0.5$ and $\delta=-0.7$. In this case, the extremal limit is at $T\to 0$.}
\label{fig:gapless}
\end{figure}

\begin{figure}
  \includegraphics[height=0.3\textwidth]{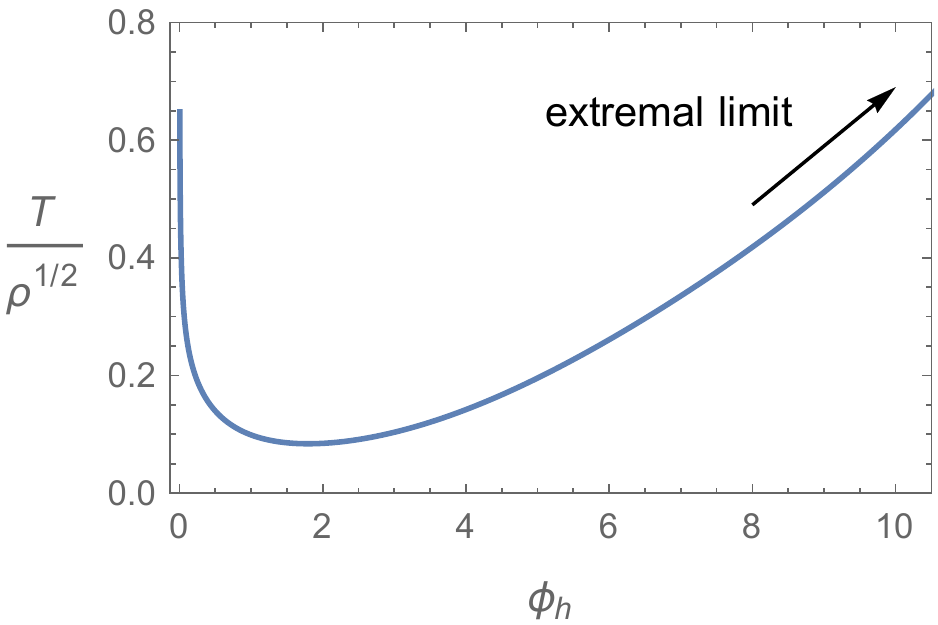}\qquad\quad
  \includegraphics[height=0.3\textwidth]{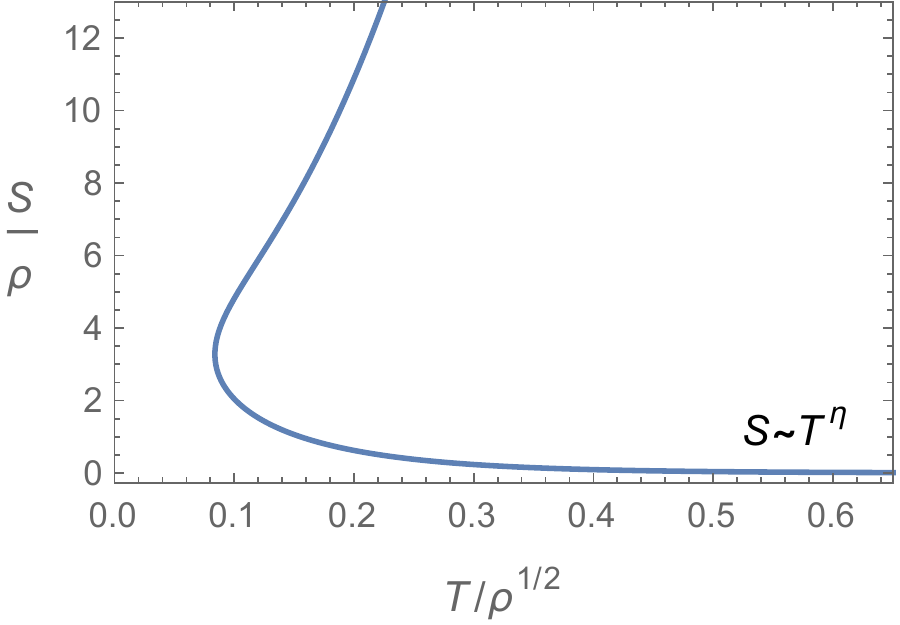}\vspace{10pt}\\
  \includegraphics[height=0.3\textwidth]{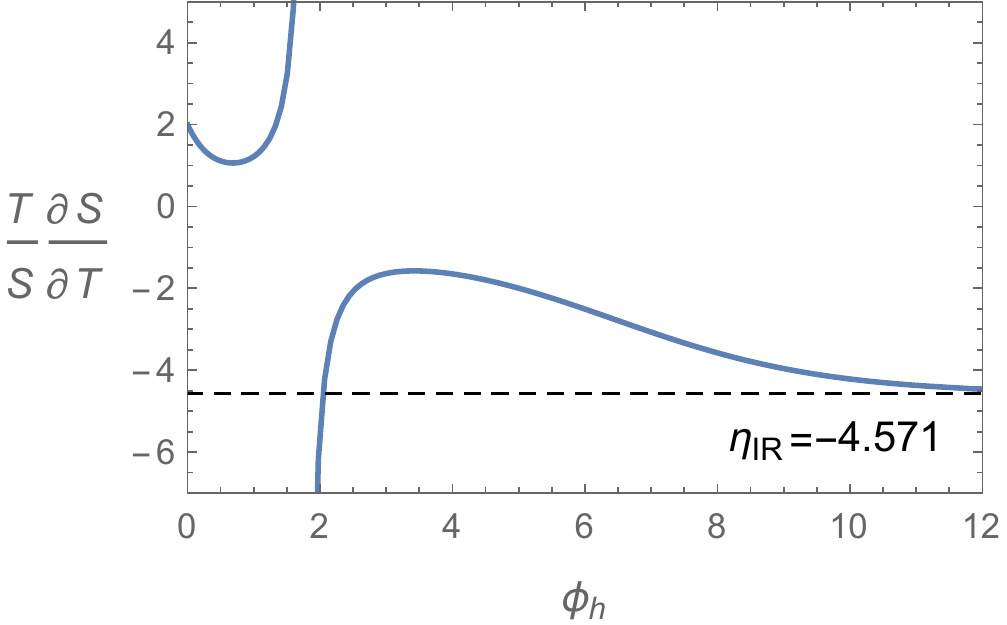}\quad
  \includegraphics[height=0.3\textwidth]{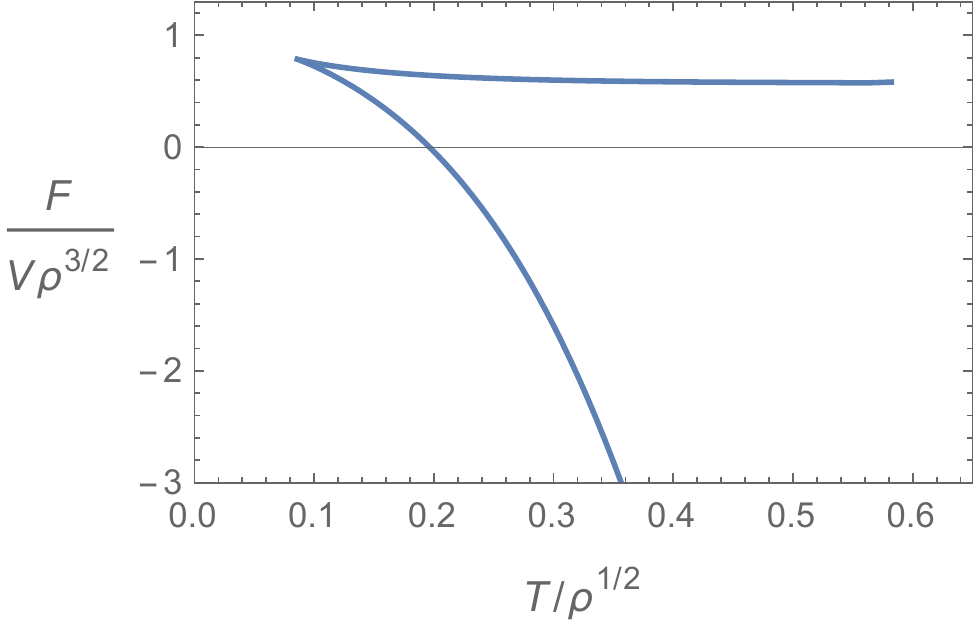}
  \caption{Numerical results for physical properties of a gapped geometry. The parameters are chosen by $\gamma=0.4$ and $\delta=-1.2$. In this case, the extremal limit is at $T\to\infty$.}
\label{fig:gapped}
\end{figure}

\begin{figure}
  \includegraphics[height=0.3\textwidth]{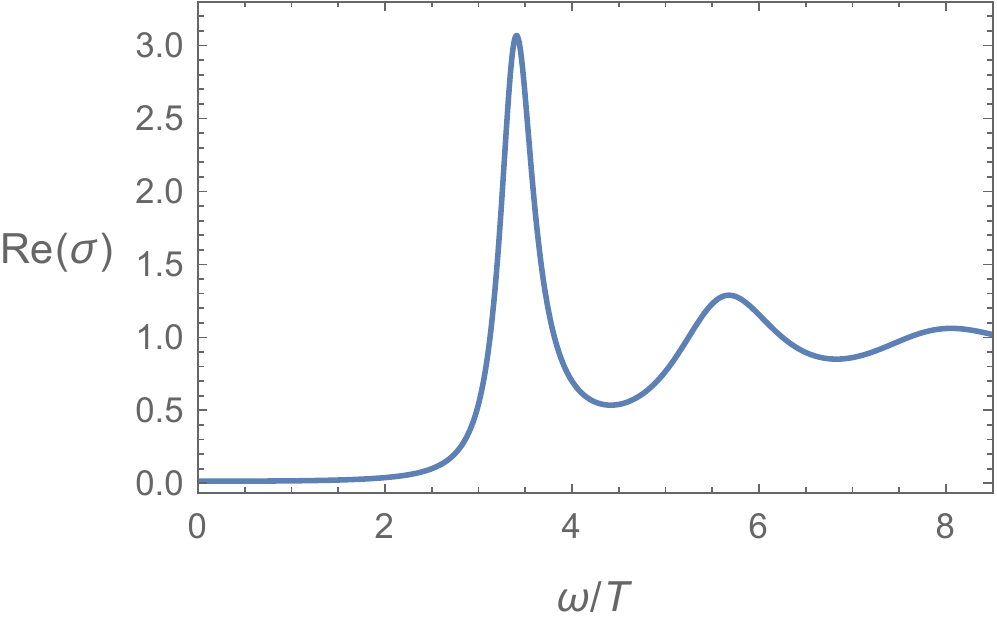}\quad
  \includegraphics[height=0.3\textwidth]{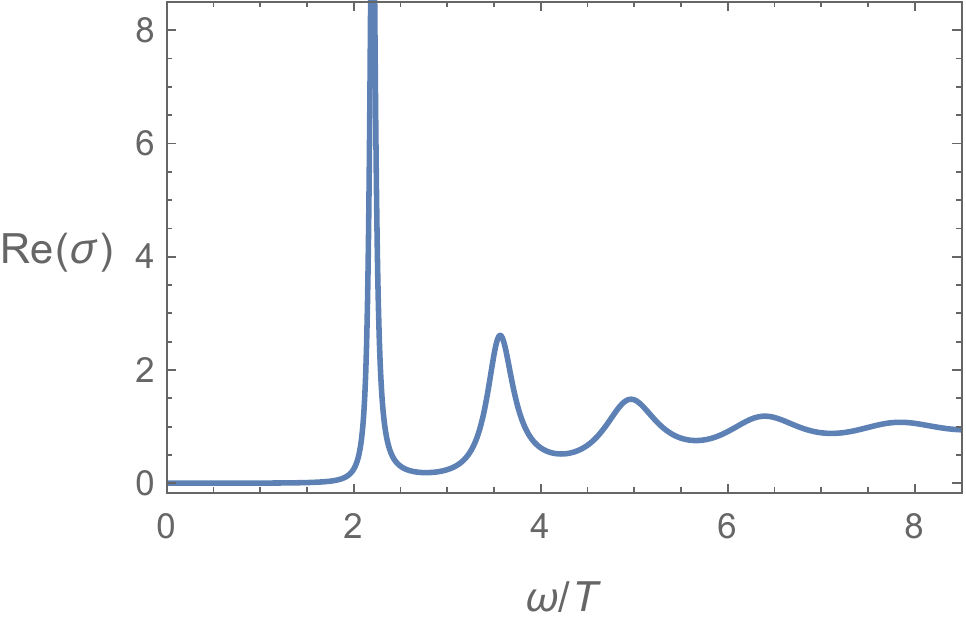}
  \caption{AC Conductivity calculated at two different temperatures: $T/\rho^{1/2}=0.538$ for the left plot, and $T/\rho^{1/2}=0.858$ for the right plot. The background is a gapped geometry with $\gamma=0.4$ and $\delta=-1.2$. As we go to the extremal limit by increasing the temperature of the unstable branch, the spectrum will have a hard gap.}
\label{fig:sigma}
\end{figure}

We compare a gapless case and a gapped case. We choose the parameters $(\gamma,\delta)$ according to figure~\ref{fig:par}, and solve the equations of motion numerically.

Figure~\ref{fig:gapless} is for a gapless case with $(\gamma,\delta)=(0.5,-0.7)$, and figure~\ref{fig:gapped} is for a gapped case with $(\gamma,\delta)=(0.4,-1.2)$. In figures~\ref{fig:gapless} and \ref{fig:gapped}, the upper-left plot shows the relation between the temperature and $\phi_h\equiv\phi(r_h)$. The upper-right plot shows the relation between the entropy and the temperature.
 The lower-left plot shows the scaling exponent in $S\sim T^\eta$ as a function of $\phi_h$, and the lower-right plot shows the relation between the free energy and the temperature.

In the gapped case, there is a minimal temperature $T_{min}$ for the black holes and there is a small black hole and large black hole branch. This is clearly visible in the left-upper plot of figure~\ref{fig:gapped}. The small black hole branch is on the right side of this figure. The extremal limit can be approached from this branch by going to the right. In this limit the temperature of the black holes  $T\to\infty$ like the usual Schwarzschild black holes. Only the left branch is stable.

The extremal limit is the ground state of the system. As we heat up the system, there will be a first-order phase transition at $T_c>T_{min}$ and the system will jump from the low temperature ground state to the black hole in the stable branch, \cite{gkmn,gkmn2}. The AC conductivity calculated from the same gapped geometry at two different temperatures is plotted in figure~\ref{fig:sigma}. It shows that as we increase the temperature of the unstable branch in order to approach the extremal limit, a hard gap will emerge in the extremal limit.

\section{The spectrum of tensor and scalar fluctuations}
\label{sec:tensor-scalar}

In previous sections, we studied systems where the current-current correlator has a gap. Here we study the spin-2 and spin-0 excitations associated to the stress tensor. Such fluctuations determine holographically the spin-2 (also known as the shear channel) and the spin-0 (bulk channel) parts of the stress-tensor two-point function.

Generally, we need to turn on the following perturbations and derive the linearized equations of motion, to calculate various transport coefficients:
\begin{equation}
g_{\mu\nu}\to g_{\mu\nu}(1+e^{-i\omega t}h_{\mu\nu}(r)),\qquad A_\mu\to A_\mu(1+e^{-i\omega t}a_\mu(r)),\qquad \phi\to\phi\,(1+e^{-i\omega t}\delta\phi(r)).
\end{equation}
The fluctuations can be classified by their spins under the spatial rotations (here $i=x$, $y$) as follows.
\begin{itemize}
  \item spin-0: $h_{tt}$, $h_{rr}$, $h_{ii}$, $h_{tr}$, $a_t$, $a_r$, $\delta\phi$. In a gauge where $\delta\phi$ is unperturbed \cite{Gubser:2008sz,DeWolfe:2011ts}, a decoupled equation for $h_{ii}$ can be obtained, responsible for the bulk viscosity.
  \item spin-1: $h_{ti}$, $h_{ri}$, $a_i$. A decoupled equation for $a_i$ can be obtained, responsible for the conductivity.
  \item spin-2: $h_{ij}$. The equation for $h_{ij}$ is responsible for the shear viscosity.
\end{itemize}

\subsection{Spin-2 fluctuation - Shear viscosity}
The spin-2 fluctuation $h_{xy}$ satisfies the Laplace's equation. The Laplace's equation $\nabla^2\Phi=0$ with the the metric \eqref{eq:BCD} is
\begin{equation}
\left(\sqrt{\frac{D}{B}}C\Phi'\right)'+\sqrt{\frac{B}{D}}C\omega^2\Phi=0.
\end{equation}
After the change of variables by
\begin{equation}
\frac{d\xi}{dr}=\sqrt{\frac{B}{D}},\qquad \tilde{\Phi}=\sqrt{C}\Phi,
\end{equation}
we can obtain a Schr\"{o}dinger equation
\begin{equation}
-\frac{d^2\tilde{\Phi}}{d\xi^2}+\tilde{V}(\xi)\tilde{\Phi}=\omega^2\tilde{\Phi}.
\end{equation}
The potential is
\begin{equation}
\tilde{V}=-\frac{DB'C'}{4B^2C}+\frac{D'C'}{4BC}-\frac{DC'^2}{4BC^2}+\frac{DC''}{2BC},
\end{equation}
where the prime is with respect to $r$. If $\tilde{V}$ goes to to infinity in both IR and UV, then the Schr\"{o}dinger equation can only have bound states that correspond to a discrete spectrum for the spectral density of the correlator. To analyze the behavior of $\tilde{V}$ in the IR, we use the hyperscaling violating geometry \eqref{eq:ztheta} as the background.

\begin{figure}
  \centering
  \includegraphics[width=0.47\textwidth]{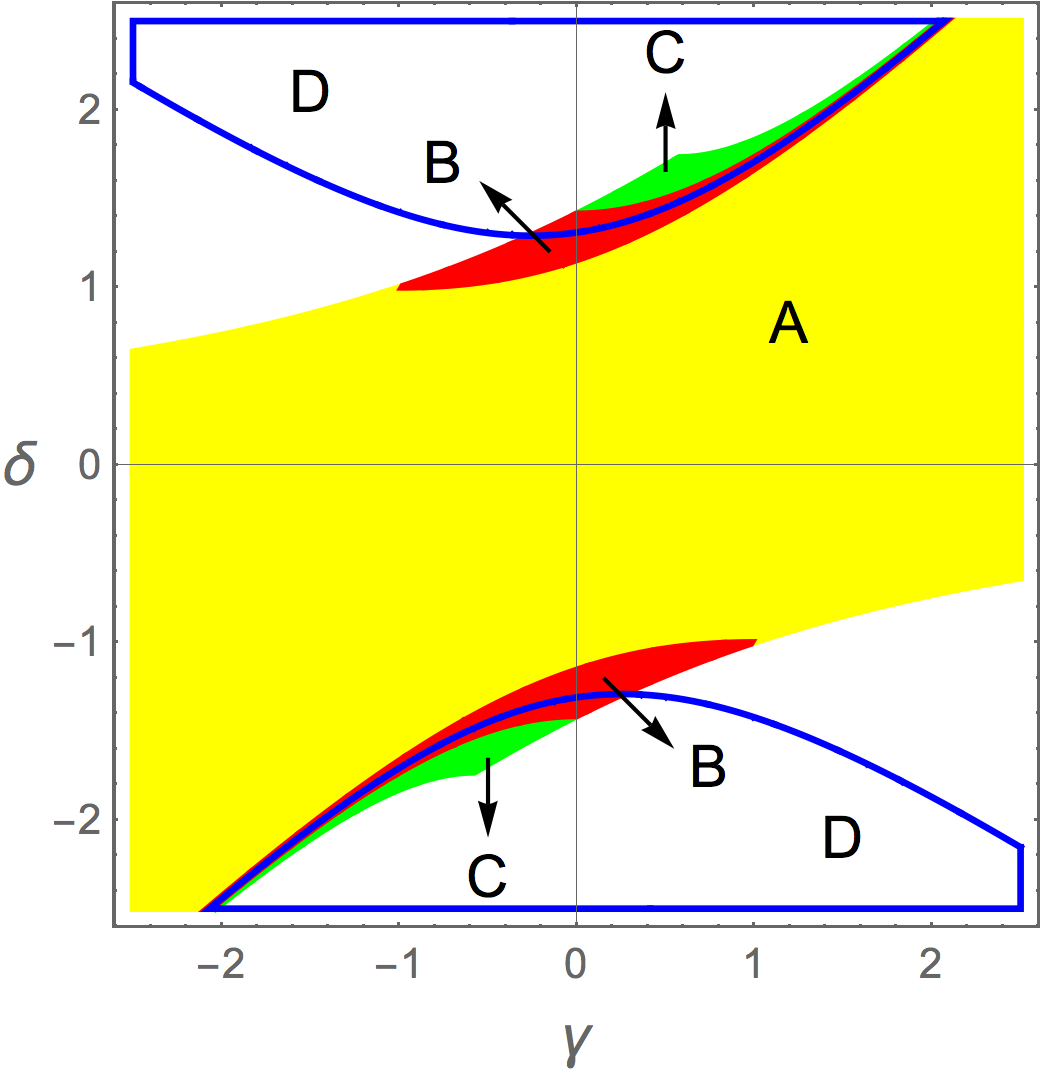}\qquad
  \includegraphics[width=0.47\textwidth]{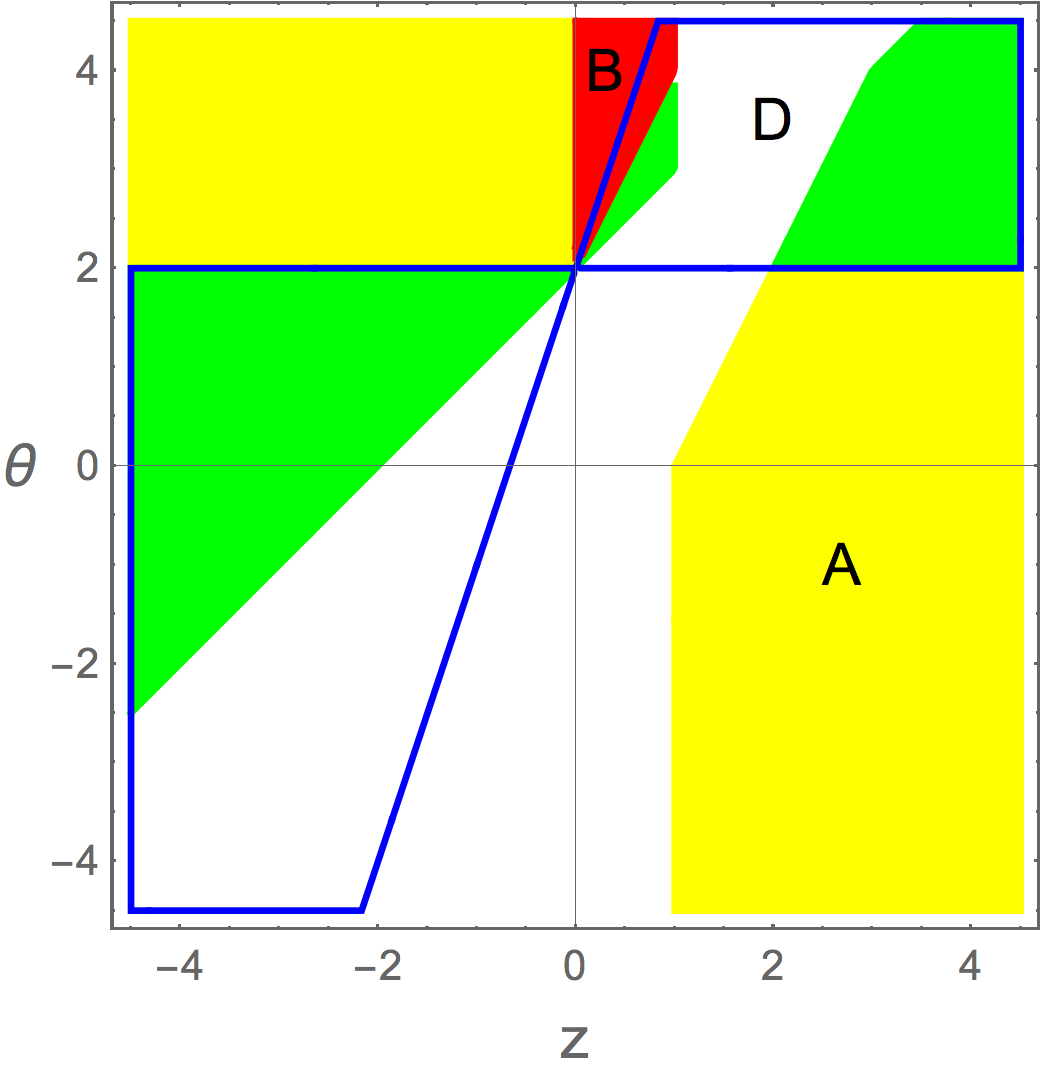}\qquad
  \caption{\label{fig:parLap} The regions A, B, and C are the parameter space constrained by the Gubser criterion. We assume that the conditions in the UV for a gapped geometry has been satisfied as explained in the end of section 3.1. In region A (yellow), the extremal limit is at $T\to 0$, and the Laplacian is gapless. In region B (red), the extremal limit is at $T\to\infty$, and the Laplacian is gapped. In region C (green), the extremal limit is at $T\to\infty$, and the Laplacian is gapless. Region D (enclosed by blue boundaries) is holographically unreliable. When both correlators are holographically reliable a gapped current-current correlator implies a gapped shear tensor correlator. }
\end{figure}

{\it IR charged solution.} We change the variable $r$ to $\xi$ by \eqref{eq:xr}, and the Schr\"{o}dinger potential is
\begin{equation}
\tilde{V}(x)=\frac{\nu_L^2-1/4}{\xi^2},
\end{equation}
where
\begin{equation}
\nu_L=\frac{(\gamma-\delta)(3\gamma+\delta)+4}{2[(\gamma-\delta)(\gamma+3\delta)+4]}=\frac{2+z-\theta}{2z}.\label{eq:nuLap}
\end{equation}
Assuming the Gubser criterion is always satisfied, we have $\nu_L>0$ when the extremal limit at $T\to 0$ ($\xi\to\infty$), and $\nu_L<0$ when the extremal limit is at $T\to\infty$ ($\xi\to 0$). In the later case, we should further impose that $|\nu|>1$ so that there is only one normalizable solution in the IR. By comparing \eqref{eq:nu} with \eqref{eq:nuLap}, we can see that $|\nu|-|\nu_L|>0$, which implies that when the Laplacian is gapped, the current-current correlator is also gapped and vice versa in the holographically reliable region. Figure~\ref{fig:parLap} is a summary of the parameter space $(\gamma,\delta)$ for the spectrum of the Laplacian.

{\it IR neutral solution.} We change the variable $r$ to $\xi$ by \eqref{eq:xr}, and the Schr\"{o}dinger potential is
\begin{equation}
\tilde{V}(x)=\frac{\nu_{L0}^2-1/4}{\xi^2},
\end{equation}
where
\begin{equation}
\nu_{L0}=\frac{\delta^2-3}{2(\delta^2-1)}=\frac{3-\theta}{2}.
\end{equation}
We observe that $\nu_0$ in terms of $\theta$ can be obtained from \eqref{eq:nuLap} with $z=1$, although the expression in terms of $\delta$ is very different. Assuming the Gubser criterion is always satisfied, we have $\nu_{L0}>0$ when the extremal limit at $T\to 0$ ($\xi\to\infty$), and $\nu_{L0}<0$ when the extremal limit is at $T\to\infty$ ($\xi\to 0$). In the later case, we should further impose that $|\nu_{L0}|>1$ so that there is only one normalizable solution in the IR. Considering all above constraints, the parameter space for the gapped Laplacian is $1<\delta^2<5/3$.

\subsection{Spin-0 fluctuation - Bulk viscosity}

To calculate the bulk viscosity, we can turn on the perturbations $h_{tt}$, $h_{rr}$, $h_{xx}$, and $a_t$ only, according to the gauge choice of \cite{Gubser:2008sz,DeWolfe:2011ts}. After eliminating $h_{tt}$, $h_{rr}$, and $a_t$, a decoupled equation for $h_{xx}$ is derived\footnote{This appearance of this equation is not unique by applying the equations of motion. It can be written in a simple form in the coordinates \eqref{eq:ansatz} with $f=g/\sqrt{h}$:
\begin{equation}
\frac{1}{f\phi'^2}(f\phi'^2h_{xx}')'+\left(\frac{\omega^2}{f^2}-\frac{r^2h(hZ)'A_t'^2}{gh'}-\frac{(g/h)'}{g/h}\frac{(r\phi')'}{r\phi'}\right)h_{xx}=0.
\end{equation}}
\begin{small}
\begin{multline}
h_{xx}''+\left(\frac{7C'}{2C}+\frac{D'}{D}-\frac{C\phi'^2}{2C'}-\frac{3C''}{C'}+\frac{2\phi''}{\phi'}\right)h_{xx}'
+\left[\frac{\omega^2B}{D}+\left(\frac{C'Z'}{C\phi'^2}-\frac{Z}{2}\right)\frac{A_t'^2}{D}+\frac{C''}{C}-\frac{D''}{2D}\right.\\
\left.+\left(\frac{CD'}{DC'}-1\right)\frac{\phi'^2}{4}+\frac{C'}{C}\left(\frac{5C'}{4C}-\frac{5D'}{4D}+\frac{\phi''}{\phi'}\right)+\frac{D'}{D}\left(\frac{3C''}{2C'}-\frac{\phi''}{\phi'}\right)
\right]h_{xx}=0.
\end{multline}
\end{small}
After the change of variables by
\begin{equation}
\frac{d\xi}{dr}=\sqrt{\frac{B}{D}},\qquad \tilde{h}_{xx}=\phi'h_{xx},\label{eq:xihxx}
\end{equation}
where the prime is with respect to $r$, we can obtain a Schr\"{o}dinger equation
\begin{equation}
-\frac{d^2\tilde{h}_{xx}}{d\xi^2}+\tilde{V}(\xi)\tilde{h}_{xx}=\omega^2\tilde{h}_{xx}.
\end{equation}
If $\tilde{V}$ goes to to infinity in both IR and UV, then the Schr\"{o}dinger equation can only have bound states that correspond to a discrete spectrum for the spectral density of the correlator. To analyze the behavior of $\tilde{V}$ in the IR, we use the hyperscaling violating geometry \eqref{eq:ztheta} as the background.

{\it IR charged solution.} We change the variable $r$ to $\xi$ by \eqref{eq:xihxx}, and the leading order of the Schr\"{o}dinger potential is
\begin{equation}
\tilde{V}(\xi)=\frac{\nu_s^2-1/4}{\xi^2},
\end{equation}
where
\begin{equation}
\nu_s=\frac{\sqrt{[(\gamma-\delta)(3\gamma+\delta)+4][(\gamma-\delta)(3\gamma+17\delta)+36]}}{2[(\gamma-\delta)(\gamma+3\delta)+4]}
=\frac{\sqrt{(z-\theta+2)(9z-\theta-6)}}{2z\,\text{sgn}(2-\theta)}.\label{eq:nus}
\end{equation}
Assuming the Gubser criterion is always satisfied, the expression inside the square root is always positive, and we have $\nu_s>0$ when the extremal limit at $T\to 0$ ($\xi\to\infty$), and $\nu_s<0$ when the extremal limit is at $T\to\infty$ ($\xi\to 0$). In the later case, we should further impose that $|\nu_s|>1$ so that there is only one normalizable solution in the IR. The three correlators responsible for the conductivity, shear viscosity, and bulk viscosity are gapped at the same time in their common holographically reliable region.

{\it IR neutral solution.} The Schr\"{o}dinger potential is the same as (B.7). Therefore, the discussion about the spectrum for the scalar fluctuation is the same as the Laplacian.

\section{Conclusion and outlook}
\label{sec:conclusion}
We have obtained two new classes of holographic matter, constructed from geometries that have correlators with discrete spectrum of excitations.

The first is a class of holographic ground-state saddle points of the bulk action (\ref{eq:action}) with $W(\phi)=0$ which implies the U(1) symmetry is intact. Such saddle points have a source of momentum dissipation due to the non-trivial axion solutions that break the translational symmetry.
The characteristic exponent $\lambda$ in (\ref{li5}) that controls the IR properties of the coupling of the axions to the rest of the degrees of freedom can be selected so that the momentum dissipation effects are irrelevant in the IR.

In  the absence of momentum dissipation, the saddle-point solutions of interest, found already in \cite{cgkkm} have a discrete spectrum for the charge charge correlators, implying the associated spectral density is a discrete sum of $\delta$-functions. Translation invariance implies that one of the $\delta$-functions is at zero frequency.

A careful analysis, done in this paper, indicates that in the presence of IR irrelevant momentum dissipation, the correlator still has a discrete spectrum, with the only difference that the zero frequency $\delta$-function has now disappeared. Therefore, these are bona-fide insulators with a hard gap and a discrete spectrum of charged degrees of freedom. They resemble in several respects Mott insulators but we suspect this is a novel class of insulators.

We have found explicit numerical solutions that connect well-defined asymptotically AdS$_4$ UV fixed points with a non-trivial charge density to the insulating fixed points discussed above.
There are however caveats in our discussion. Although explicit flows exist, it is not yet clear whether they are the dominant (in the sense of free energy) flows between the UV and IR fixed points in top-down string theory effective actions, although the M-theory example of \cite{McGreevy} is a good candidate for a stable flow. We believe that the mechanism is a wider interest and is the one expected to be at play for example in real QCD at finite but O(1) baryon density, \cite{ajkkrt,iz}.

Other characteristics of such saddle points are as follows:
\begin{itemize}

\item All non-trivial quantum critical saddle points with $z\not= 1$ have a mild IR geometrical singularity. It is a naked singularity in the presence of hyperscaling violation, while it is a much subtler singularity not captured by scalar curvature invariant in the case of Lifshitz scaling symmetry. Despite the fact that holography demands regular saddle points, singularities can be tolerated because they can arise from the neglect of some degrees of freedom (coarse graining). They should be however resolvable, as adding back the missing degrees of freedom, one should recover a regular background. A criterion for resolvability was advanced by Gubser, \cite{Gubser:2000nd}, and we always impose it on our singular saddle points.   There is however a further issue in the presence of a resolvable singularity: that the calculation of correlators should not depend on the resolution of the IR singularity. This issue crops up as
the presence (or not) of two normalizable modes, in the Sturm-Liouville problem that is relevant for the calculation of a two-point function. In such a case the solution is not unique, and a further boundary condition at the singularity is necessary in order to define the correlator, \cite{cgkkm}. This is the case where the holographic calculation is unreliable, and a resolution of the singularity is necessary in order to provide a concrete answer for the relevant correlator.

We find that in the saddle-points of interest studied in this paper, the $\delta$-function at zero frequency, is absent {only} for those bulk theories for which the charge correlator is holographically well defined (does not suffer from singularity resolution ambiguities).
This is interesting and may provide important clues for other cases where zero-frequency $\delta$-functions in transport appeared when it was not expected, \cite{dgk}.

\item In a related theme, in appendix \ref{sec:SUGRA} we have computed analytically  the conductivity of a class of black holes solutions of gauged supergravity (known as STU black holes)  dual to deformations of the ABJM theory.
    In particular, in the extremal two-charge case (where the charge density is zero), we obtain a gap and a zero-frequency $\delta$-function although none was expected due to symmetry principles.
The reason is that there is no charge density and without charge density there is no argument for a zero-frequency $\delta$-function.
Moreover, the equation for the conductivity below the gap is in the holographically non-well-defined class as it has two normalizable solutions. However here a unique solution can be chosen using analyticity in the frequency $\omega$.

\item The zero-temperature phase of the gapped solutions persists up to a finite critical temperature $T_c$. Indeed the phase diagram of candidate black-hole ground states is similar to that relevant for holographic QCD, \cite{gkmn,gkmn2,ajkkrt}. There is a minimum temperature, $T_{min}<T_c$ below which there are no black-hole solutions as can be seen in the upper left part of figure \ref{fig:gapped}. Above $T_c$ the system jumps to a plasma-like phase via a first order transition. Calculating the conductivity of this phase is an interesting problem. We expect that this will correspond to a conductor, where momentum dissipation effects may be strong.

\item In the low temperature insulating phase the change of conductivity with temperature is an effect suppressed in the large-$N$ limit. It can be calculated however by a one-loop gravity calculation. The relevant diagrams will be the one-loop correction to the bulk propagator of two gauge bosons. This calculation is in principle feasible (a related one-loop correction to holographic thermodynamics was performed recently in \cite{ajkkt}).

\item Energy transport is controlled by the stress tensor two-point function. In section \ref{sec:tensor-scalar} we analyzed the equations for shear and bulk channels. We found that for all holographically reliable geometries, when the current-current correlator is gapped, then the shear and bulk channels are also gapped.
           Therefore there is no low-energy transport in these insulators.
           On the other hand, in the supersolid class this fact is consistent with our interpretation.

\item Our analysis has used effective holographic theories following the philosophy developed in \cite{cgkkm}. Some of the solutions we find and analyze however appear in string-derived supergravities, (see \cite{McGreevy} and appendix \ref{sec:SUGRA}), namely the AdS$_4$ gauged supergravity obtained from compactification of M-theory on $S^7$ that is dual to the 3d ABJM theory, \cite{abjm}.
    It is therefore expected that they may be ubiquitous in three dimensional QFTs that are deformations of the ABJM theory at finite density.

\end{itemize}

We do expect that in codimension-one subspaces of the theory space we are studying there will be IR fixed points that are non-scaling. This is the case we expect to give rise to Arhenius-type insulators and looking for them is an interesting endeavour.

The second class of holographic saddle-points have finite density but in theories where the U(1) symmetry is broken. The action in (\ref{eq:action}) with $W(\phi)\not= 0$ is describing this class of saddle points and the universality classes of quantum critical IR fixed points were classified in \cite{gk2}.
This class contains both superfluid IR fixed points when the U(1) symmetry is spontaneously broken as well as fixed points with the U(1) symmetry explicitly  broken.

There are several possible IR fixed points but in this paper we have studied the fractionalized phases for  which the bulk mass term $W(\phi)$ flows to zero in the IR.
There is still a source of breaking of translational invariance in the form of non-trivial axion profiles.

We have found fractionalized fixed points with a current-current correlator that has a discrete spectrum. Unlike the previous case studied above and despite the presence of momentum dissipation, the presence of symmetry breaking guarantees that there is a non-trivial $\delta$-function at zero frequency due to the superfluid mode.

 Such systems are therefore superfluids (or superconductors when the U(1) is weakly gauged) in the presence of translational symmetry breaking with a discrete spectrum. This is not unlike to what is expected in supersolids with the sole exception that the translational invariance breaking is spontaneous.
Like the case of insulators mentioned above, the systems here are also separated by a first order phase transition from a liquid-like deconfined plasma phase. Moreover the temperature dependence of the low-temperature
supersolid-like phase is again a subleading effect in 1/$N$.

There are several problems that remain open in this direction.
We highlight that the fate of other phases, in particular the cohesive ones requires further study.

We also expect that the nature of the breaking of translational symmetry in the supersolid like ground states can be improved to be as required for standard supersolids, \cite{review}.
This can be done by ``injecting" skyrmion-like charged solitons in the geometry (these are baryons in holographic models of QCD that are represented by flavor gauge theory instantons, \cite{ss,ss2}). Such charge is expected to form a crystal at sufficiently high density, which is not expected to be high enough so that there is serious backreaction on the original geometry. Implementing this construction is an interesting problem.

In this respect, looking back at standard (massless) QCD at zero density, we can interpret it as a zero density superfluid, where the broken symmetry is the $SU(N_f)_A$ flavor symmetry and the superfluid is composed of the massless pions. Turning on a baryon density, the system is driven to a quantum critical AdS$_2$ phase, \cite{ajkkrt}. It is plausible that for higher density there may be a crystalline phase that would be a supersolid.

\addcontentsline{toc}{section}{Acknowledgments}
\acknowledgments

We would like to thank B. Gouteraux, C. Rosen, X. Zotos and especially Jan Zaanen for useful discussions. We also thank Gouteraux and Zaanen for constructive comments on the manuscript.

This work was supported in part by European Union's Seventh Framework Programme under grant
agreements (FP7-REGPOT-2012-2013-1) no 316165, the EU-Greece program ``Thales" MIS 375734 and was also co-financed by the European Union (European Social Fund, ESF) and Greek national funds through the Operational Program ``Education and Lifelong Learning" of the National Strategic Reference Framework (NSRF) under ``Funding of proposals that have received a positive evaluation in the 3rd and 4th Call of ERC Grant Schemes".

\newpage
\addcontentsline{toc}{section}{Appendices}
  \renewcommand{\theequation}{\thesection.\arabic{equation}}
\appendix

\section{Equations of motion}
\label{sec:eom}
In this appendix we make explicit the equations of motion from the action we are using in \eqref{eq:action}
\begin{align}
& R_{\mu\nu}-\frac{1}{2}g_{\mu\nu}R=\frac{1}{2}\left(T_{\mu\nu}^A+T_{\mu\nu}^\phi+T_{\mu\nu}^{\psi}\right),\\
& \nabla^{\mu}\left(Z(\phi)F_{\mu\nu}\right)=W(\phi)A_\nu,\\
& \nabla^2\phi-V'(\phi)-\frac{Z'(\phi)}{4}F^2-\frac{W'(\phi)}{2}A^2-\frac{Y'(\phi)}{2}\sum_{i=1}^2(\partial\psi_i)^2=0,\\
& \nabla^\mu(Y(\phi)\nabla_\mu\psi_i)=0,\label{eq:psi}
\end{align}
where
\begin{align}
T_{\mu\nu}^A &= Z(\phi)\left(F_{\mu\rho}F^\rho_{\;\;\nu}-\frac{1}{4}g_{\mu\nu}F^2\right)+W(\phi)\left(A_\mu A_\nu-\frac{1}{2}g_{\mu\nu}A^2\right),\\
T_{\mu\nu}^\phi &= \partial_\mu\phi\partial_\nu\phi-\frac{1}{2}g_{\mu\nu}(\partial\phi)^2-g_{\mu\nu}V(\phi),\\
T_{\mu\nu}^{\psi} &= Y(\phi)\sum_{i=1}^2\left(\partial_\mu\psi_i\partial_\mu\psi_i-\frac{1}{2}g_{\mu\nu}(\partial\psi_i)^2\right).
\end{align}
We take $\psi_i=k_ix_i$ ($i=1,2$), which solves \eqref{eq:psi}. The equations of motion with the ansatz
\begin{equation}
ds^2=-D(r)dt^2+B(r)dr^2+C(r)(dx^2+dy^2)
\end{equation}
are as follows. The Einstein equations are
\begin{gather}
\frac{2C''}{C}-\frac{C'}{C}\left(\frac{B'}{B}+\frac{C'}{2C}\right)+\frac{ZA_t'^2}{2D}+\frac{BWA_t^2}{2D}+BV+\frac{k^2BY}{2C}+\frac{1}{2}\phi'^2=0,\\
\frac{2C''}{C}-\frac{C'}{C}\left(\frac{B'}{B}+\frac{C'}{C}+\frac{D'}{D}\right)+\frac{BWA_t^2}{D}+\phi'^2=0,\\
\frac{2D''}{D}-\frac{2C''}{C}-\frac{D'}{D}\left(\frac{B'}{B}-\frac{C'}{C}+\frac{D'}{D}\right)+\frac{B'C'}{BC}-\frac{2ZA_t'^2}{D}-\frac{2k^2BY}{C}=0,
\end{gather}
The Maxwell equation is
\begin{equation}
\left(\frac{CZ}{\sqrt{BD}}A_t'\right)'=CW\sqrt{\frac{B}{D}}A_t,\label{eq:Maxwell}
\end{equation}
The equation for the scalar is
\begin{equation}
\phi''+\left(\frac{C'}{C}+\frac{D'}{2D}-\frac{B'}{2B}\right)+\frac{A_t'^2Z'(\phi)}{2D}+\frac{A_t^2BW'(\phi)}{2D}-BV'(\phi)-\frac{k^2BV'(\phi)}{2C}=0.
\end{equation}
Without the momentum dissipation $(k=0)$, there is a conserved charge
\begin{equation}
{\cal Q}=\frac{C}{\sqrt{BD}}\left[ZA_tA_t'-C\left(\frac{D}{C}\right)'\right].
\end{equation}
With the momentum dissipation, ${\cal Q}$ satisfies the following equation:
\begin{equation}
{\cal Q}'+k^2\sqrt{BD}Y=0.
\end{equation}

\section{The potential for the scalar}
\label{sec:potential}
In this appendix we analyze optimal choices for the scalar  potential.

We assume that the potential for the scalar $V(\phi)$ only contains exponential terms, as is the case in supergravity systems. We need only the leading exponential term in the IR to discuss the hyperscaling violating geometry. However, we need at least two exponential terms to construct an asymptotically AdS geometry. A potential with only two exponentials can be parameterized as
\begin{equation}
V_\text{2-exp}=-{6\over L^2[6\delta^2+\Delta(3-\Delta)]}\left[\Delta(3-\Delta)e^{-\delta(\phi-\phi_0)}+6\delta^2e^{{\Delta(3-\Delta)\over 6\delta}(\phi-\phi_0)}\right]
\label{eq:2exp}
\end{equation}
We have traded the four real parameters of a two-exponential potential in terms of the position of the UV fixed point, $\phi_0$, the UV $AdS_4$ scale $L$, the UV dimension $\Delta$ of the scalar operator dual to $\phi$, and the extra exponent $\delta$ that controls the  IR hyperscaling violating solution when the factor $e^{-\delta(\phi-\phi_0)}$ dominates in the IR.

The potential (\ref{eq:2exp}) (at zero charge density), has a single UV fixed point at $\phi=\phi_0$. In the IR, at finite charge density,  it has two hyperscaling-violating  fixed points.
One  corresponds to $-\delta\phi\to \infty$
with  an  IR asymptotic geometry given in \cite{cgkkm} with exponents $(\gamma,\delta)$.
The other corresponds to a hyperscaling violating geometry when $\delta\phi\to \infty$ with exponents ($\gamma,-{\Delta(3-\Delta)\over 6\delta}$).

This parametrization emerges from imposing the following constraints on the potential
\begin{itemize}
  \item $V(0)=-\frac{d(d-1)}{L^2}$. For $AdS_4$ with $L=1$, $V(0)=-6$.
  \item $V'(0)=0$.
  \item $V''(0)=m^2$. For $\Delta=1$ or $2$, $V''(0)=-2$.
\end{itemize}
By applying the above three constraints, the potential only has one parameter that is $\delta$. If we use the following potential
\begin{equation}
V(\phi)=-\frac{3}{3\delta^2+1}(2e^{-\delta\phi}+6\delta^2e^{\frac{1}{3\delta}\phi}),
\end{equation}
which has $V=-6-\phi^2+\cdots$ in the UV limit, the asymptotic behavior of $\phi$ will contain $\ln z$ terms in general. The near horizon behavior of $\phi$, $g$, and $h$ is
\begin{equation}
\phi=f_1r+f_2r^2+f_3r^3+\cdots+\log r(f_{12}r^2+f_{13}r^3).\label{eq:philog}
\end{equation}
If we choose $\Delta=2$, there are no $\ln z$ terms. If $\Delta=1$, the consistency of the equations without $\ln z$ requires $\delta=\pm 1/\sqrt{3}$. However, these values of $\delta$ cannot give the gapped geometry.

Consider a three-exponential potential
\begin{equation}
V(\phi)=c_1e^{c_2\phi}+c_3e^{c_4\phi}+c_5e^{c_6\phi}.
\end{equation}
If we set $c_6=0$ for simplicity, we have one more free parameters in addition to $\delta$. We parametrize the potential as
\begin{equation}
V(\phi)=-\frac{2(6+u)e^{-\delta\phi}}{2+(6+u)\delta^2}-\frac{(6+u)^2\delta^2e^\frac{2\phi}{(6+u)\delta}}{2+(6+u)\delta^2}+u.\label{eq:3expu}
\end{equation}
With this potential, if we plugin the near boundary expansion \eqref{eq:philog} without the log terms to the equations of motion, the leading term is
\begin{equation}
-\frac{2f_1^2(-2+(6+u)\delta^2)}{(6+u)\delta}r^2+\cdots
\end{equation}
If the near boundary expansion \eqref{eq:philog} is consistent without the log terms, the coefficient in the above term must be zero, which gives
\begin{equation}
u=\frac{2(1-3\delta^2)}{\delta^2}
\end{equation}
Substituting the above $u$ to \eqref{eq:3expu}, we obtain the potential \eqref{eq:3exp} used in our numerical calculations.

\begin{figure}
  \centering
  \includegraphics[width=0.4\textwidth]{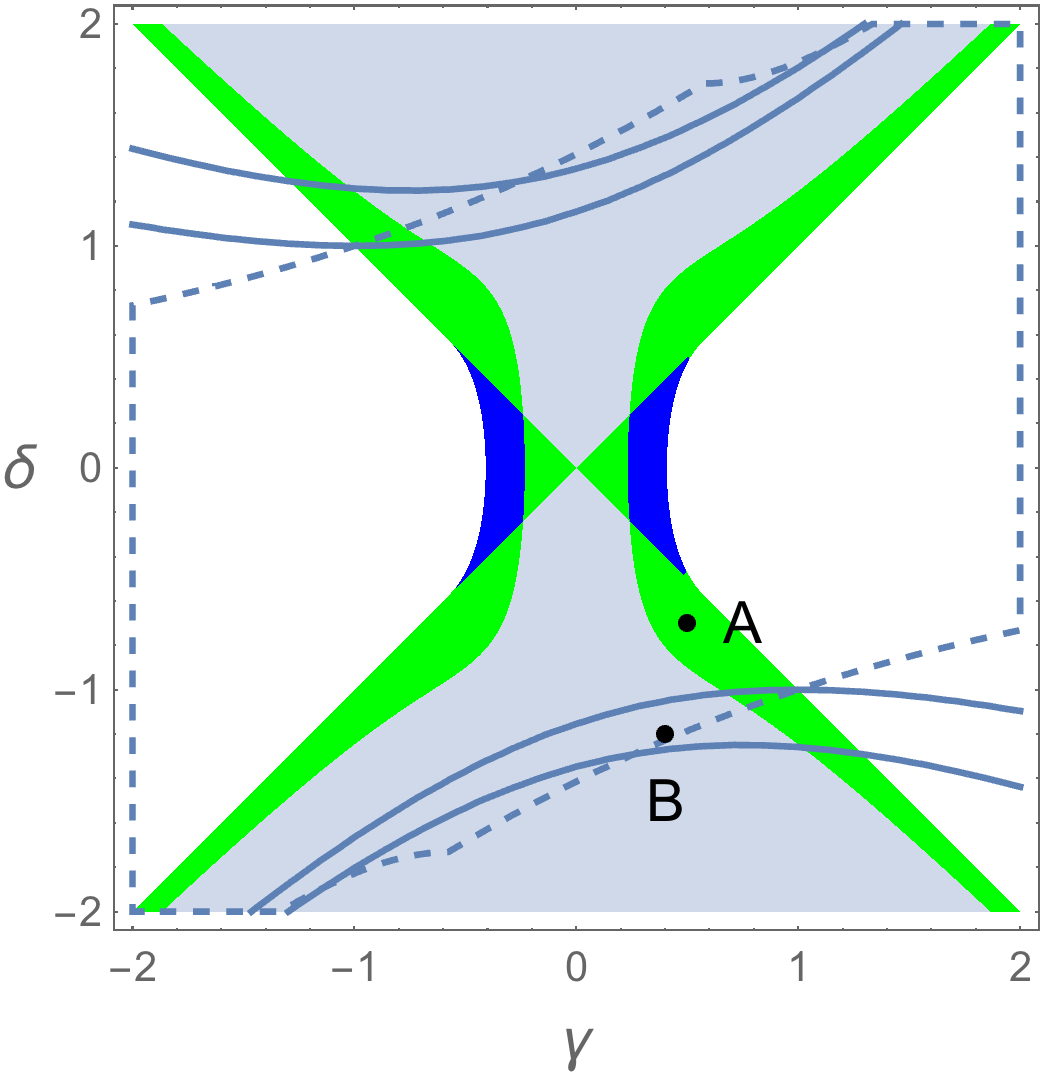}
  \caption{\label{fig:ads2} The parameter space of $(\gamma,\delta)$ under the two conditions: (1) The solution of $\phi_*$ exists; (2) $\beta_-<0$. In the light blue region, there is only one solution of $\phi_*$, and in the blue region, there are two solutions of $\phi_*$. The parameter space depends on the concrete form of $V(\phi)$ and $Z(\phi)$, and this plot is for the potential \protect\eqref{eq:3exp} and $Z=e^{\gamma\phi}$.}
\end{figure}

Note that there may also be an $AdS_2\times\mathbb{R}^2$ IR fixed point at constant $\phi=\phi_*$ \cite{gk2}, as shown in figure~\ref{fig:whole}. The condition for the existence of an IR AdS$_2$ extremum at finite charge density is
\be
\left({V'\over V}+{Z'\over Z}\right)\Big |_{\phi=\phi_*}=0~~~\to~~~{V'_*\over V_*}+\gamma=0.
\label{2}\ee
The critical exponents for the perturbation around the IR $AdS_2$ is
\be
\beta_{\pm}=\frac12\left(1\pm\sqrt{1-4\delta_1}\right),
\label{4}\ee
where
\be
\delta_1=\left({V''\over V}+{Z''\over Z}-2{V'^2\over V^2}\right)\Big |_{\phi=\phi_*}=\left({V''_*\over V_*}-\gamma^2\right)\Big |_{\phi=\phi_*}.
\label{3}\ee
If a RG-stable $AdS_2$ fixed point exists, the solution of $\phi_*$ is real, and there is an irrelevant mode associated with this $\phi_*$. For the potential \eqref{eq:3exp} and $Z=e^{\gamma\phi}$, the parameter space of $(\gamma,\delta)$ under the above condition is potted in figure~\ref{fig:ads2}.

\section{Analytic solutions from supergravity}
\label{sec:SUGRA}
Four special cases of the maximal gauged supergravity in $AdS_4$ are given in Table~\ref{tab:4charges}, regarding how they fit into the general classification. These four examples provide both the analytic understanding of some EMD systems and the numerical check of the program. The metric is
\begin{equation}
ds^2=e^{2A(\bar{r})}(-h(\bar{r})dt^2+d\mathbf{x}^2)+\frac{d\bar{r}^2}{e^{2A(\bar{r})}h(\bar{r})}.
\end{equation}
The relation between $\bar{r}$ and the $r$ in \eqref{eq:ansatz} is $e^{2A(\bar{r})}=1/r^2$. The functions $A(\bar{r})$ and $h(\bar{r})$ are given in Table~\ref{tab:4charges}. These solutions are obtained from a more general black hole with four U(1) charges found in \cite{Cvetic:1999xp}.

\begin{figure}
  \centering
  \includegraphics[height=0.33\textwidth]{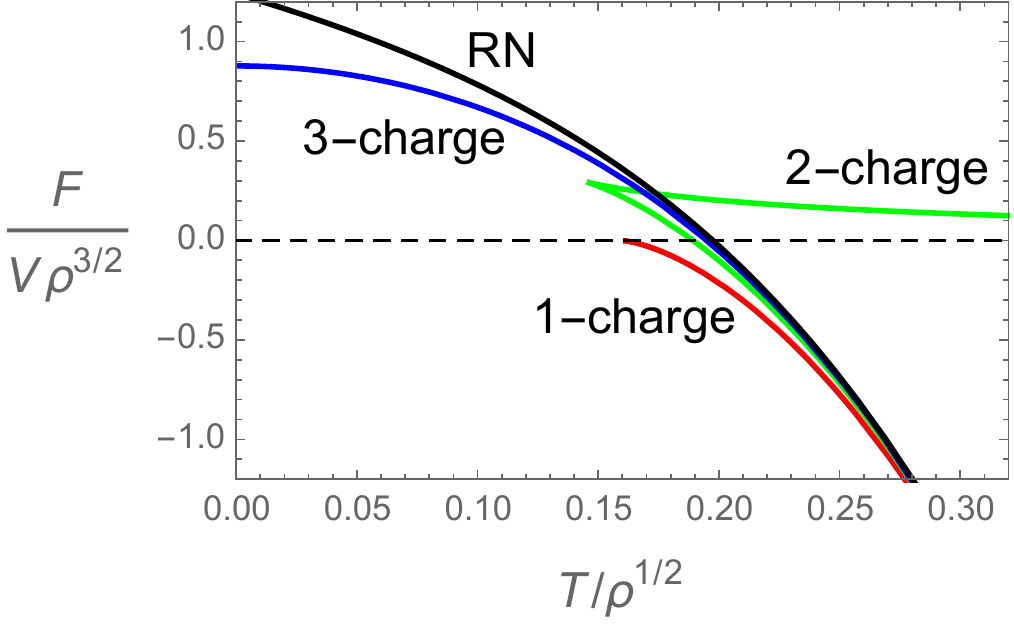}
  \caption{\label{fig:sugra} The relation between the free energy and the temperature for 1-charge ($\gamma=\sqrt{3}$, $\delta=-1/\sqrt{3}$), 2-charge ($\gamma=1$, $\delta=-1$), 3-charge ($\gamma=1/\sqrt{3}$, $\delta=-1/\sqrt{3}$), and RN black holes in $AdS_4$.}
\end{figure}

The temperature and entropy are given by
\begin{equation}
T=\frac{|h'|e^{2A}}{4\pi},\qquad
s=4\pi e^{2A(r_H)}.
\end{equation}
The relation between the free energy and the temperature is plotted in figure~\ref{fig:sugra}. There are some exceptional features for these systems. For example, if we set the horizon size $r_H=0$, then the gauge field will vanish for the 1-charge and 2-charge black holes. Consequently, the extremal geometries have to be at zero density. The geometries are significantly simplified, and it allows us to obtain exact, analytic solutions for $\sigma(\omega)$ for arbitrary $\omega$. The extremal geometries for the 1-charge and 2-charge black holes can only be at zero density, and thus they are not ``topologically" connected to the finite temperature black holes. The situation is similar to the asymptotically-$AdS_5$ case that was discussed in \cite{DeWolfe:2012uv}.

\begin{table}
\centering
\caption{\label{tab:4charges} Geometries and properties of 1, 2, 3, 4-charge black holes in AdS$_4$.}
\renewcommand{\arraystretch}{2.5}
\setlength\doublerulesep{0.2pt}
\begin{scriptsize}
\begin{tabular}{c|c|c|c|c}
  \hline\hline
  & 1-charge & 2-charge & 3-charge & 4-charge (RN)\\

  \hline
  $V(\phi)$ & $-6\cosh(\phi/\sqrt{3})$ & $-2(\cosh\phi+2)$ & $-6\cosh(\phi/\sqrt{3})$ & $-6$\\

  \hline
  $Z(\phi)$ & $e^{\sqrt{3}\phi}$ & $e^\phi$ & $e^{\phi/\sqrt{3}}$ & $1$\\

  \hline
  $e^{2A}$ & $\bar{r}^{3/2}(\bar{r}+Q)^{1/2}$ & $\bar{r}(\bar{r}+Q)$ & $\bar{r}^{1/2}(\bar{r}+Q)^{3/2}$ & $(\bar{r}+Q)^2$\\

  \hline
  $h$ & $1-\dfrac{\bar{r}_H^2(\bar{r}_H+Q)}{\bar{r}^2(\bar{r}+Q)}$ & $1-\dfrac{\bar{r}_H(\bar{r}_H+Q)^2}{\bar{r}(\bar{r}+Q)^2}$
  & $1-\dfrac{(\bar{r}_H+Q)^3}{(\bar{r}+Q)^3}$ & $1-\dfrac{\bar{r}(\bar{r}_H+Q)^4}{\bar{r}_H(\bar{r}+Q)^4}$\\

  \hline
  $A_t$ & $\dfrac{\sqrt{Q}\bar{r}_H}{\sqrt{\bar{r}_H+Q}}\Bigl(1-\dfrac{\bar{r}_H+Q}{\bar{r}+Q}\Bigr)$ & $\sqrt{2Q\bar{r}_H}\Bigl(1-\dfrac{\bar{r}_H+Q}{\bar{r}+Q}\Bigr)$ & $\sqrt{3Q(\bar{r}_H+Q)}\Bigl(1-\dfrac{\bar{r}_H+Q}{\bar{r}+Q}\Bigr)$
  & $\dfrac{2\sqrt{Q}(\bar{r}_H+Q)}{\sqrt{\bar{r}_H}}\Bigl(1-\dfrac{\bar{r}_H+Q}{\bar{r}+Q}\Bigr)$\\

  \hline
  $\phi$ & $\dfrac{\sqrt{3}}{2}\ln\Bigl(1+\dfrac{Q}{\bar{r}}\Bigr)$ & $\ln\Bigl(1+\dfrac{Q}{\bar{r}}\Bigr)$ & $\dfrac{\sqrt{3}}{2}\ln\Bigl(1+\dfrac{Q}{\bar{r}}\Bigr)$ & 0\\

  \hline
  $(\gamma,\delta)$ & $\gamma=\sqrt{3}$, $\delta=-\dfrac{1}{\sqrt{3}}$ & $\gamma=1$, $\delta=-1$ & $\gamma=\dfrac{1}{\sqrt{3}}$, $\delta=-\dfrac{1}{\sqrt{3}}$ & $\gamma=0$, $\delta=0$\\

  \hline
  $(z,\theta)$ & $z=1$, $\theta=-1$ & $z=1$, $\theta\to \infty$ & $z\to\infty$, $-\theta/z=1$ & $z\to\infty$, $\theta=0$\\

  \hline
  $(\kappa,\zeta)$ & $\kappa=\sqrt{3}$, $\zeta=-1$ & $\bar{\kappa}=2$, $\bar{\zeta}=-1$ & $\kappa=\sqrt{3}$, $\zeta=1$ & \\

  \hline
  $(\alpha,\beta)\,^\text{a}$  & $\alpha=-\dfrac{4}{3}$, $\beta=0$ & $\alpha=-2$, $\beta=0$ & $\alpha=-4$, $\beta=0$ & \\

  \hline
  $\tau$ & $-\dfrac{1}{2\sqrt{3}}$ & 0 & $\dfrac{1}{2\sqrt{3}}$ &\\

  \hline
  $T$ & $\dfrac{(3\bar{r}_H+2Q)\sqrt{\bar{r}_H}}{4\pi\sqrt{\bar{r}_H+Q}}$ & $\dfrac{3\bar{r}_H+Q}{4\pi}$
  & $\dfrac{3\sqrt{\bar{r}_H(\bar{r}_H+Q)}}{4\pi}$ & $\dfrac{(\bar{r}_H+Q)(3\bar{r}_H-Q)}{4\pi\bar{r}_H}$\\

  \hline
  $S$ & see$\,^\text{b}$ & see$\,^\text{b}$ & $\propto T$ & const\\

  \hline
  & Exceptional & Exceptional$\,^\text{c}$ & IR charged, gapless & IR charged, gapless\\
  \hline\hline
\end{tabular}
\end{scriptsize}
\flushleft\footnotesize{$^\text{a}$ They are defined in eq.~\eqref{eq:hor}.\\
$^\text{b}$ The scaling exponent depends on the ensemble (fixing $\rho$ or $\mu$). See figures~\ref{fig:2-charge} and \ref{fig:1-charge}.\\
$^\text{c}$ For the $r_H=0$ geometry, the conductivity has a gap. See \eqref{eq:sigma-2}.}
\end{table}

\begin{figure}
  \centering
  \includegraphics[width=0.47\textwidth]{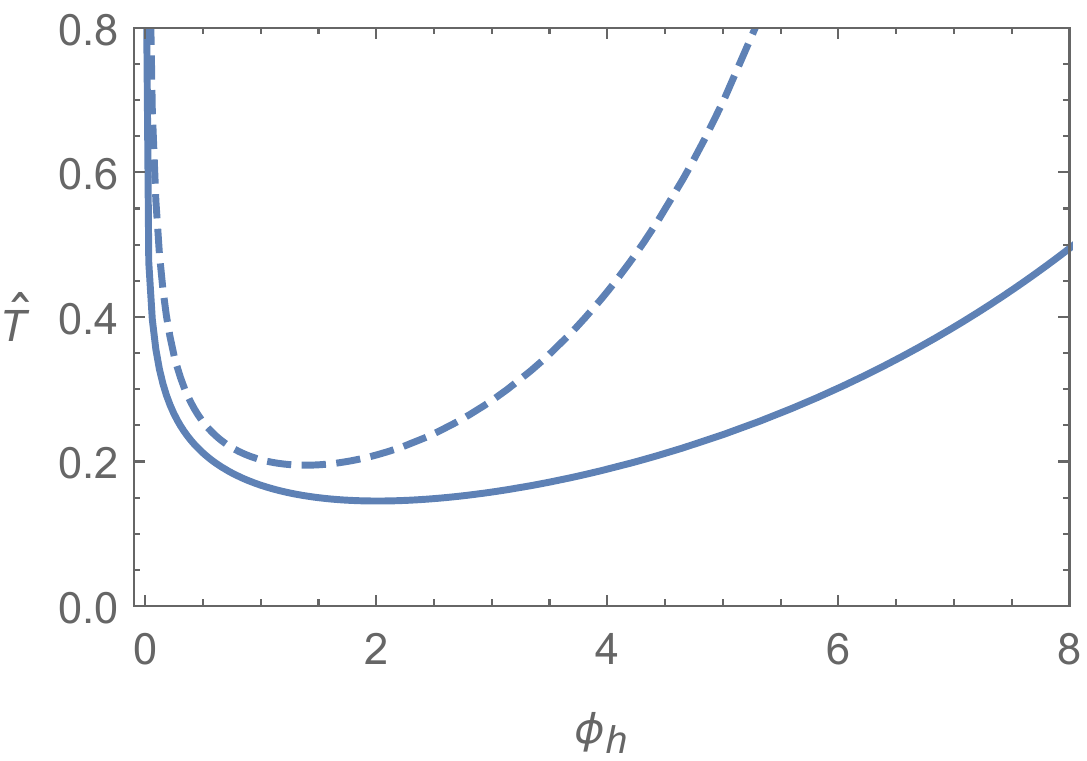}\qquad
  \includegraphics[width=0.47\textwidth]{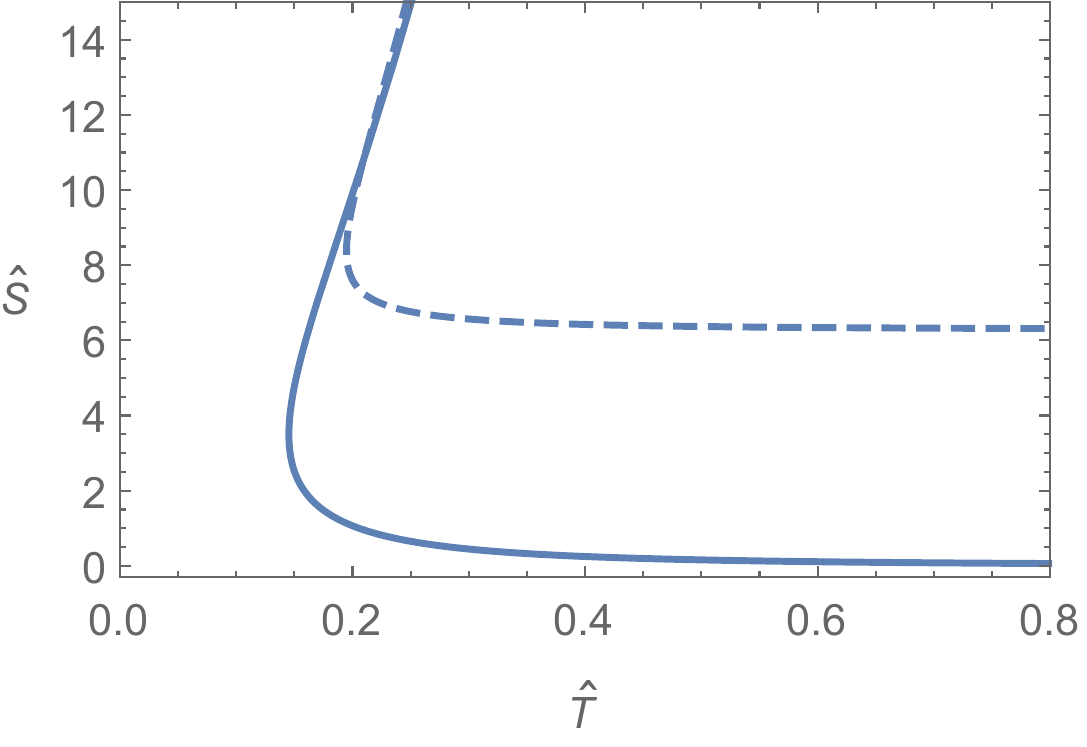}\qquad
  \caption{\label{fig:2-charge} 2-charge black hole. The solid line is for the canonical ensemble (fix $\rho$), for which $\hat{T}=T/\sqrt{\rho}$ and $\hat{S}=S/\rho$; the dashed line is for the grand canonical ensemble (fix $\mu$), for which $\hat{T}=T/\mu$ and $\hat{S}=S/\mu^2$.}
\end{figure}

\begin{figure}
  \centering
  \includegraphics[width=0.47\textwidth]{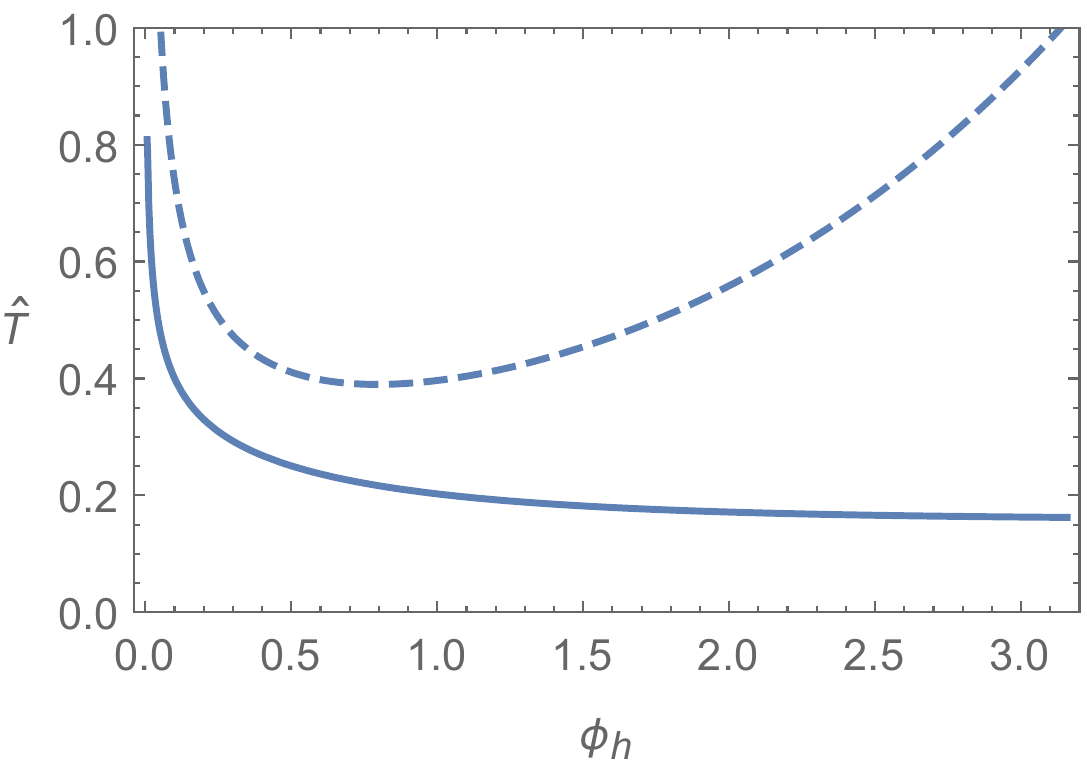}\qquad
  \includegraphics[width=0.47\textwidth]{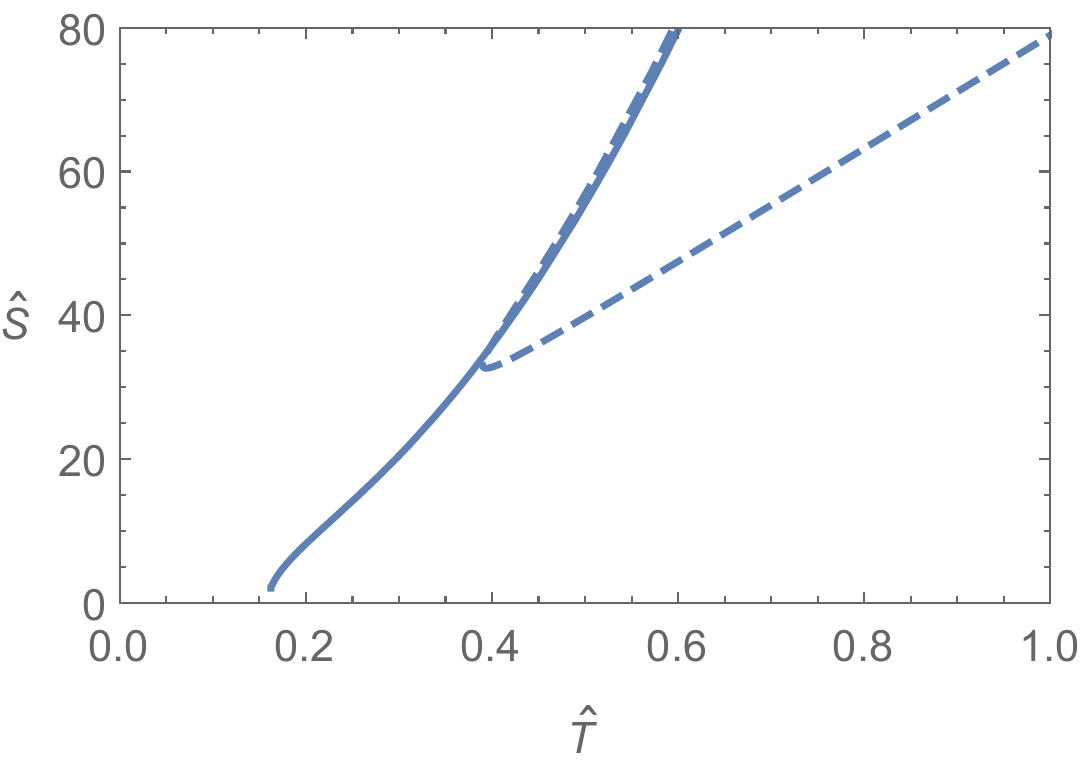}\qquad
  \caption{\label{fig:1-charge} 1-charge black hole. The solid line is for the canonical ensemble (fix $\rho$), for which $\hat{T}=T/\sqrt{\rho}$ and $\hat{S}=S/\rho$; the dashed line is for the grand canonical ensemble (fix $\mu$), for which $\hat{T}=T/\mu$ and $\hat{S}=S/\mu^2$.}
\end{figure}

{\it 2-charge black hole in $AdS_4$.}
To obtain the conductivity, we perturb the system \eqref{eq:LVZ} around the solution to \eqref{eq:eoms} by $\delta A_x=a_x(\bar{r})e^{-i\omega t}$, and obtain the equation for $a_x$:
\begin{equation}
a_x''+\frac{2}{\bar{r}+Q}a_x'+\frac{\omega^2}{\bar{r}^2(\bar{r}+Q)^2}a_x=0,
\end{equation}
where the prime stands for derivative with respect to $\bar{r}$. By the usual redefinitions this equation can be mapped to a Schr\"odinger equation with  a constant potential equal to $\tilde{V}=Q^2/4$.
The general solution for $a_x$ is
\begin{equation}
a_x=C_1\left(\frac{\bar{r}}{\bar{r}+Q}\right)^\frac{Q+\sqrt{Q^2-4\omega^2}}{2Q}+C_2\left(\frac{\bar{r}}{\bar{r}+Q}\right)^\frac{Q-\sqrt{Q^2-4\omega^2}}{2Q}.
\end{equation}
There are two linearly-independent solutions. When $\omega^2>Q^2/4$, the first solution describes the in-falling wave with the $\omega\to\omega+i\epsilon$ prescription. When $\omega^2<Q^2/4$, both solutions are real and normalizable. As the background extremal solution is singular, this situation seems to be in the holographically non-well defined class we have discussed earlier: we need to impose an extra boundary condition at the singularity in order to obtain a unique solution.

\begin{figure}
  \centering
  \includegraphics[width=0.47\textwidth]{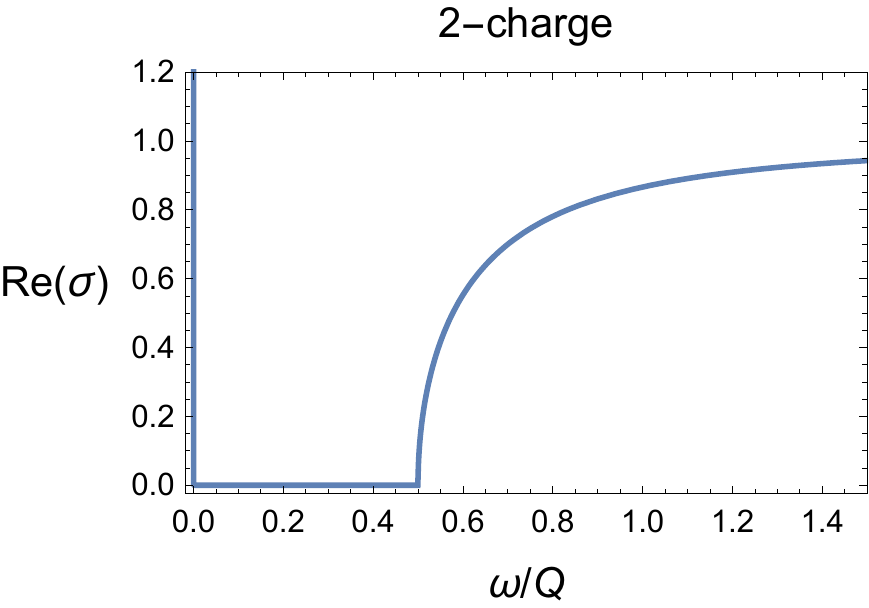}\qquad
  \includegraphics[width=0.47\textwidth]{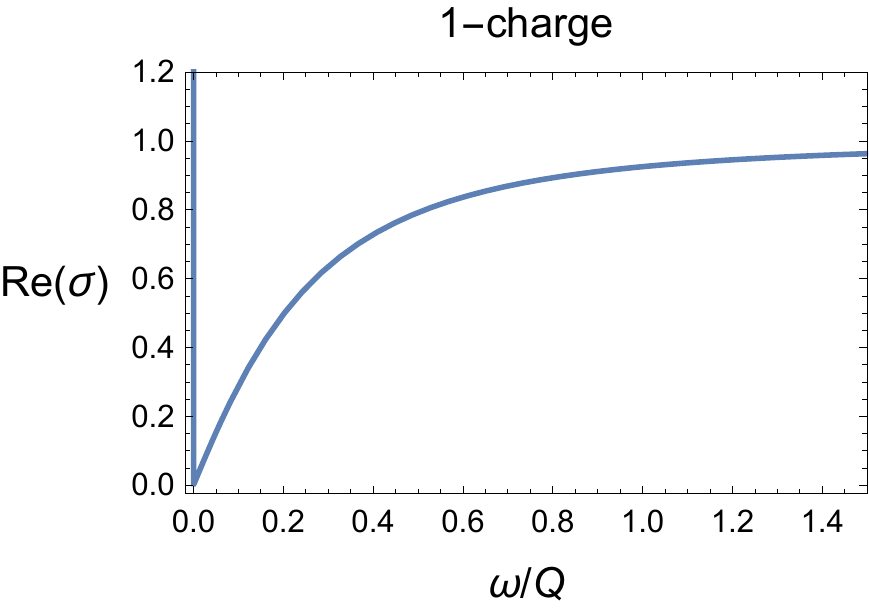}\qquad
  \caption{\label{fig:12-charge} The left plot is the AC conductivity calculated from the 2-charge black hole in AdS$_4$ at extremality, and the right plot is the AC conductivity calculated from the 1-charge black hole in AdS$_4$ at extremality. Both of them have a $\delta$-function at $\omega=0$.}
\end{figure}

However, in this case, analyticity (in $\omega$) of the correlator makes the solution unique. Indeed, if we analytically continue the $\omega^2>Q^2/4$ solution to $\omega^2<Q^2/4$, the solution for $a_x$ is unambiguous for all $\omega$. The conductivity is\footnote{We assume $\omega>0$. It is crucial to shift the pole to the lower half $\omega$-plane: $\sqrt{-\omega^2}\to\sqrt{-(\omega+i\epsilon)^2}=\omega\sqrt{-1-i\epsilon}=-i\omega$.}
\begin{equation}
\sigma(\omega)=\left.\frac{i(Q+\sqrt{Q^2-4\omega^2})}{2\omega}\right|_{\omega\to\omega+i\epsilon}.\label{eq:sigma-2}
\end{equation}
From (\ref{eq:sigma-2}) we observe that the spectrum is gapped and continuous above the gap.\footnote{This is very reminiscent to the situation studied in \cite{ihqcd,ihqcd2} in Einstein-Dilaton gravity with a potential that asymptotes at strong coupling to the Liouville potential of non-critical string theory and the dilaton solution to the well-known linear dilaton solution of string theory.} The conductivity has a $\delta$-function at $\omega=0$. The AC conductivity as a function of $\omega$ is plotted in figure~\ref{fig:12-charge}.
We should note that very few exact solutions for the conductivity at arbitrary $\omega$ are available for extremal geometries; another example is the 1-charge black hole in $AdS_5$ \cite{DeWolfe:2012uv}.

{\it 1-charge black hole in $AdS_4$.}
The equation for $a_x$ is
\begin{equation}
a_x''+\frac{2}{\bar{r}+Q}a_x'+\frac{\omega^2}{\bar{r}^3(\bar{r}+Q)}a_x=0.
\end{equation}
The effective Schr\"odinger potential is given by
\begin{equation}
\tilde{V}=\frac{3Q^2\bar{r}}{16(Q+\bar{r})}.
\end{equation}
The solution with the in-falling wave in the IR is
\begin{equation}
a_x=C_1\sqrt{\frac{\bar{r}}{\bar{r}+1}}\,H_1^{(1)}\Bigl(\omega\sqrt{\frac{\bar{r}+1}{\bar{r}}}\Bigr),
\end{equation}
where $H_n^{(1)}$ is the Hankel function.

From this solution we obtain the conductivity as
\begin{equation}
\sigma(\omega)=i\left.\frac{H^{(1)}_2(2\omega)}{H^{(1)}_1(2\omega)}\right|_{\omega\to\omega+i\epsilon}.
\end{equation}
As $\omega\to 0$,
\begin{equation}
\sigma=\frac{i}{\omega}+(\pi+2i\gamma+2i\ln\omega)\omega+\cdots.
\end{equation}
The spectrum is gapless, and the conductivity has a $\delta$-function at $\omega=0$. The AC conductivity as a function of $\omega$ is plotted in figure~\ref{fig:12-charge}.

\section{DC conductivity at zero density}
\label{sec:zero-density}
Although we focus on finite density systems, it is illuminating to examine zero density systems at extremality. Consider a zero density density system whose background is described by
\begin{equation}
S=\int d^{4}x\sqrt{-g}\left[R-\frac{1}{2}(\partial\phi)^2-V(\phi)\right],\label{eq:action0}
\end{equation}
and the dynamics of the gauge field is given by a $Z(\phi)F^2$ term, where $V(\phi)$ and $Z(\phi)$ are defined by \eqref{eq:Vexp}. At finite temperature, the DC conductivity is finite and can be calculated from \eqref{eq:sigmaDC} with the IR neutral geometry. The result is
\begin{equation}
\sigma_\text{DC}=Z_h\sim r_h^{\bar{\kappa}\gamma}=r_h^{-2\gamma\delta},
\end{equation}
where $r_h\to\infty$ is the extremal limit.

\begin{itemize}

\item When $\gamma\delta>0$, $\sigma_\text{DC}\to 0$ in the extremal limit, so the system is an insulator.

   \item  When $\gamma\delta<0$, $\sigma_\text{DC}\to\infty$ in the extremal limit, so the system is a conductor.

\end{itemize}
For the extremal limit ($T\to 0$) and the DC limit
($\omega\to 0$) to be compatible the conductivity should have a $\delta$-function at $\omega=0$ when $\gamma\delta>0$, while the conductivity should not have a $\delta$-function at $\omega=0$ when $\gamma\delta<0$. We will examine whether this is true for the gapped geometries in the following.

To obtain the conductivity, we perturb the system \eqref{eq:action0} around the background metric \eqref{eq:BCD} by $\delta A_x=a_x(r)e^{-i\omega t}$, and obtain the equation for $a_x$:
\begin{equation}
\frac{1}{fZ}(fZa_x')'+\frac{\omega^2}{f^2}a_x=0,\label{eq:ax0}
\end{equation}
where $f=\sqrt{D/B}$. As noted before, we can ignore the $\omega^2$ term when we discuss the $\omega\to 0$ limit of the conductivity for extremal geometries at $T\to\infty$, while we need to do the asymptotic match for extremal geometries at $T\to 0$. We will discuss the extremal geometries at $T\to\infty$. In terms of the Schr\"{o}dinger coordinate $\xi$ in \eqref{eq:xr}, the general solution of $a_x$ in the IR neutral geometry is
\begin{align}
a_x=C_1\sqrt{Z}\xi^{1/2-\nu_0}+C_2\sqrt{Z}\xi^{1/2+\nu_0},\qquad \xi\to 0\label{eq:ax0sol},
\end{align}
where $\n_0$ is defined in (\ref{li2}). From \eqref{eq:ax0}, we observe that there is a radially conserved quantity at $\omega=0$:
\begin{equation}
\Pi=fZa_x'\,,\qquad \partial_r\,\Pi=0\,.
\end{equation}
If $\Pi\to 0$ as $r\to\infty$ (the IR), the conductivity does not have a $\delta$-function at $\omega=0$. If $\Pi$ is constant as $r\to\infty$, the conductivity has a $\delta$-function at $\omega=0$ and its coefficient is proportional to $\Pi(\omega=0)$. That there is only one normalizable solution for $a_x$ requires $|\nu_0|>1$. There are two cases as follows:

\begin{itemize}
\item {\bf Case 1}: $\nu_0>1$, for which we have $\delta^2-2\gamma\delta-1>0$. The solution for $a_x$ is $a_x\sim\sqrt{Z}\xi^{1/2+\nu_0}$
The radially conserved quantity $\Pi$ evaluated at the IR is
\begin{equation}
\Pi=fZa_x'\to\text{constant}.
\end{equation}

\item {\bf Case 2}: $\nu_0<-1$, for which we have $\delta^2-2\gamma\delta-1<0$. The solution for $a_x$ is $a_x\sim\sqrt{Z}\xi^{1/2-\nu_0}$. The radially conserved quantity evaluated in the IR is
\begin{equation}
\Pi=fZa_x'\sim r^{\delta^2-2\gamma\delta-1}\to 0.
\end{equation}
\end{itemize}

We expect that the insulating phase corresponds to the case 1, and the conducting phase corresponds to the case 2. We will show that this is true after we excluded the holographically unacceptable region $|\nu_0|<1$ and have imposed Gubser's criterion.

{\it Gapped, conducting phase.} The condition that the extremal geometries are at $T\to\infty$ limit of the small black-hole branch is $\delta^2>1$. The condition that the system is a conductor is $\gamma\delta<0$. These two inequalities imply  $3\delta^2-2\gamma\delta-3>0$. Consequently,
\begin{equation}
\nu_0+1=\frac{3\delta^2-2\gamma\delta-3}{2(\delta^2-1)}>0.
\end{equation}
If we further impose that there is only one normalizable solution, which translates to $|\nu_0|>1$, we finally obtain $\nu_0>1$. According to the discussion above, this corresponds to the case 1. The conductivity has a $\delta$-function at $\omega=0$ in the gapped, conducting phase. Note that this is not always true without the constraint $|\nu_0|>1$.

{\it Gapped, insulating phase.} The condition that the extremal geometries are at $T\to\infty$ is $\delta^2>1$. The condition that the system is an insulator is $\gamma\delta>0$. These two inequalities give $\delta^2+2\gamma\delta-1>0$. Consequently,
\begin{equation}
\nu_0-1=-\frac{\delta^2+2\gamma\delta-1}{2(\delta^2-1)}<0.
\end{equation}
If we further impose that there is only one normalizable solution, which is $|\nu_0|>1$, we obtain $\nu_0<-1$. According to the discussion above, this corresponds to the case 2. The conductivity does not have a $\delta$-function at $\omega=0$ in the gapped, insulating phase as expected from general principles.

The above conclusion can be obtained in another way. The equations for $a_x$ at $\omega=0$ can be solved by
\begin{equation}
a_x=\tilde{C}_1+C_2\int_0^r\frac{dr'}{f(r')Z(\phi(r'))}.\label{eq:axsol}
\end{equation}
In the IR, the first constant term corresponds to the first term in \eqref{eq:ax0sol}, and the second term corresponds to the second term in \eqref{eq:ax0sol}. In the UV, we have $a_x=\tilde{C}_1+C_2r+\cdots$, and thus the conductivity is
\begin{equation}
\sigma=\frac{C_2}{i\omega \tilde{C}_1}.
\end{equation}
We can see that if only the first solution is normalizable, we have $C_2=0$, and thus there is no $\delta$-function at $\omega=0$; if only the second solution is normalizable, we have $C_2\neq 0$, and thus there is a $\delta$-function at $\omega=0$.

For the finite density systems with axions, we can obtain a solution for $\lambda_1$ similar to \eqref{eq:axsol}. The first term in $\Pi$ is always dominant in the IR. Equation~\ref{eq:lambda1} at $\omega=0$ in the IR can be significantly simplified to be $(fZ_2\lambda_1')'=0$, which can be solved as
\begin{equation}
\lambda_1=\tilde{D_1}+D_2\int\frac{dr}{fZ_2}.\label{eq:lambda1sol}
\end{equation}
This is the counterpart of \eqref{eq:axsol} for finite density systems with axions, but the difference is that \eqref{eq:lambda1sol} is valid only for the IR geometry.

\end{document}